\documentclass[
    reprint,
    superscriptaddress,
    aps,
    prx,
    longbibliography,
    nofootinbib,
]{revtex4-1}

\usepackage{graphicx}
\usepackage{amsmath}
\usepackage{amssymb}
\usepackage[
    colorlinks=true,
    urlcolor=blue,
    linkcolor=blue,
    citecolor=blue,
    breaklinks
]{hyperref}
\usepackage[normalem]{ulem}
\usepackage{xcolor}

\newcommand{\tr}[0]{\mathrm{tr}}
\newcommand{\sgn}[0]{\mathrm{sgn}}
\newcommand{\euler}[0]{\mathrm{e}}
\mathchardef\mhyphen="2D

\begin{document}

\title{Universality classes of thermalization for mesoscopic Floquet systems}

\author{Alan Morningstar}
\affiliation{Department of Physics, Princeton University, Princeton, NJ 08544, USA}
\affiliation{Department of Physics, Stanford University, Stanford, CA 94305, USA}

\author{David A. Huse}
\affiliation{Department of Physics, Princeton University, Princeton, NJ 08544, USA}

\author{Vedika Khemani}
\affiliation{Department of Physics, Stanford University, Stanford, CA 94305, USA}

\date{\today}

\begin{abstract}
We identify several distinct phases of thermalization that describe regimes of behavior in isolated, periodically driven (Floquet), mesoscopic quantum chaotic systems. In doing so, we also identify a new Floquet thermal ensemble---the ``ladder ensemble"---that is qualitatively distinct from the ``featureless infinite-temperature" state that has long been assumed to be the appropriate maximum-entropy equilibrium ensemble for driven systems. The phases we find can be coarsely classified by (i) whether or not the system irreversibly exchanges energy of order $\omega$ with the drive, i.e., \textit{Floquet thermalizes}, and (ii) the \textit{Floquet thermal ensemble} describing the final equilibrium in systems that do Floquet thermalize. These phases are representative of regimes of behavior in mesoscopic systems, but they are sharply defined in a particular large-system limit where the drive frequency $\omega$ scales up with system size $N$ as the $N\to\infty$ limit is taken: we examine frequency scalings ranging from a weakly $N$-dependent $\omega(N) \sim \log N$, to stronger scalings ranging from $\omega(N) \sim \sqrt{N}$ to $\omega(N) \sim N$. We show that the transition where Floquet thermalization breaks down happens at an extensive drive frequency and, beyond that, systems that do not Floquet thermalize are distinguished based on the presence or absence of rare resonances across Floquet zones. We produce a thermalization phase diagram that is relevant for numerical studies of Floquet systems and experimental studies on small-scale quantum simulators, both of which lack a clean separation of scales between $N$ and $\omega$. A striking prediction of our work is that, under the assumption of perfect isolation, certain realistic quench protocols from simple pure initial states can show Floquet thermalization to a novel type of Schrodinger-cat state that is a global superposition of states at distinct \emph{temperatures}. Our work extends and organizes the theory of Floquet thermalization, heating, and equilibrium into the setting of mesoscopic quantum systems.
\end{abstract}

\maketitle

\section{Introduction\label{sec:intro}}

Breakthrough experimental developments in building isolated quantum systems have led to significant recent progress in quantum statistical mechanics.
This has fueled advances in our understanding of fundamental questions surrounding the process of \emph{thermalization} and its various exceptions in isolated many-body systems~\cite{DAlessio-Rigol2016_review, Nandkishore-Huse2015_review,Abanin-Serbyn2019_review,Alet-Laflorencie2018_review,Serbyn-Papic2021_review,Moudgalya-Regnault2021_review,Chandran-Moessner2022_review}.

In the common case of a system governed by a time-independent Hamiltonian, the system thermalizes if, at late times, probability distributions of local observables are indistinguishable from those in a relevant thermal ensemble. The appropriate thermal ensemble is determined by the principle of entropy maximization, constrained by the conservation laws of the system.  
The Eigenstate Thermalization Hypothesis (ETH)~\cite{Jensen-Shankar1985_statistical, Deutsch1991_quantum,Srednicki1994_chaos,Rigol-Olshanii2008_thermalization,Srednicki1996_thermal,Srednicki1999_approach, DAlessio-Rigol2016_review} posits conditions for thermalization on individual eigenstates of the dynamics, and empirically these conditions hold in examples of thermalizing systems~\cite{Rigol2009_breakdown,Rigol2009_quantum,Beugeling-Haque2014_finite,Kim-Huse2014_testing,Mondaini-Rigol2016_eigenstate,Mondaini-Rigol2017_off_diagonal}.

Upon the addition of a periodic drive of frequency $\omega$, i.e., making the system ``Floquet", the Hamiltonian and eigenstates of the stroboscopic dynamics gain a periodic time dependence.
The drive breaks the conservation of energy and the appropriate long-time maximum-entropy equilibrium is assumed to be a \emph{featureless} ``infinite temperature'' state~\cite{Lazarides-Moessner2014_equilibrium, DAlessio-Rigol2014_long-time}.
Exceptions to this ``heat death" are possible~\cite{prosen1998time, DAlessio-Polkovnikov2013_manybody}, notably in many-body localized (MBL) or integrable Floquet systems~\cite{Ponte-Abanin2015_mbl, Lazarides-Moessner2015_fate, gritsev2017integrable, claeys2017breaking}, in which case the system may thermalize to a generalized periodic Gibbs ensemble~\cite{Lazarides-Moessner2014_periodic, Russomanno-Santoro2012_steady} and/or realize novel  ordered phases such as the discrete time-crystal~\cite{Khemani2016, else_floquet_2016, CVS, SachaReview, Khemani2019} or the anomalous Floquet insulator~\citep{Rudner13, Titum16}. Heating can also be suppressed for a time exponential in the drive frequency, a transient phenomenon called Floquet prethermalization~\cite{Abanin-Huveneers2015_exponentially,Mori-Saito2016_rigorous, Kuwahara-Saito2016_floquet, Abanin-Huveneers2017_effective, Abanin-Huveneers2017_rigorous, RubioAbadal-Bloch2020,DeRoeck-Verreet2019_very_slow,Luitz-Khemani2020,Ho-DeRoeck2020_rigorous_without,Else-Nayak2017_time_translation,Lerose-Silva2019_kapitza,Machado-Yao2020_long_range,Collura-Lerose2022_confinement,Else-Nayak2017_prethermal_zero}. All of these results on Floquet thermalization and its exceptions were obtained in works aimed at the limit where the drive frequency $\omega$ is finite and the number of degrees of freedom in the system $N$ is infinite.

However, as we show in this paper, this limit provides an incomplete description of thermalization in chaotic Floquet systems. In particular, in \textit{mesoscopic} systems where $N$ is finite, there are other regimes of thermalization captured by thermal ensembles that are \emph{qualitatively distinct} from a featureless infinite-temperature state.
\begin{figure}
    \centering
    \includegraphics[width=1.05\linewidth]{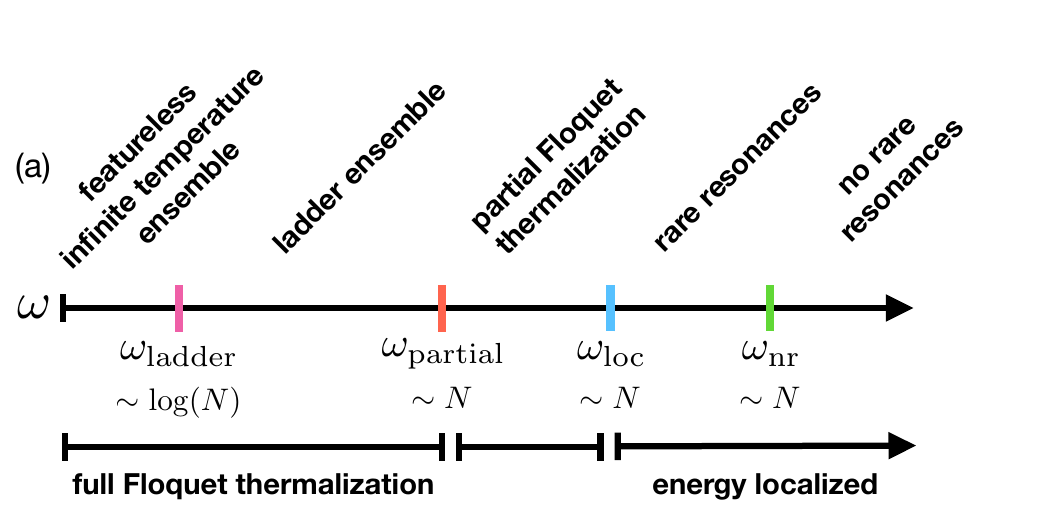}\\
    \vspace{0.2in}
    \includegraphics[width=1.0\linewidth]{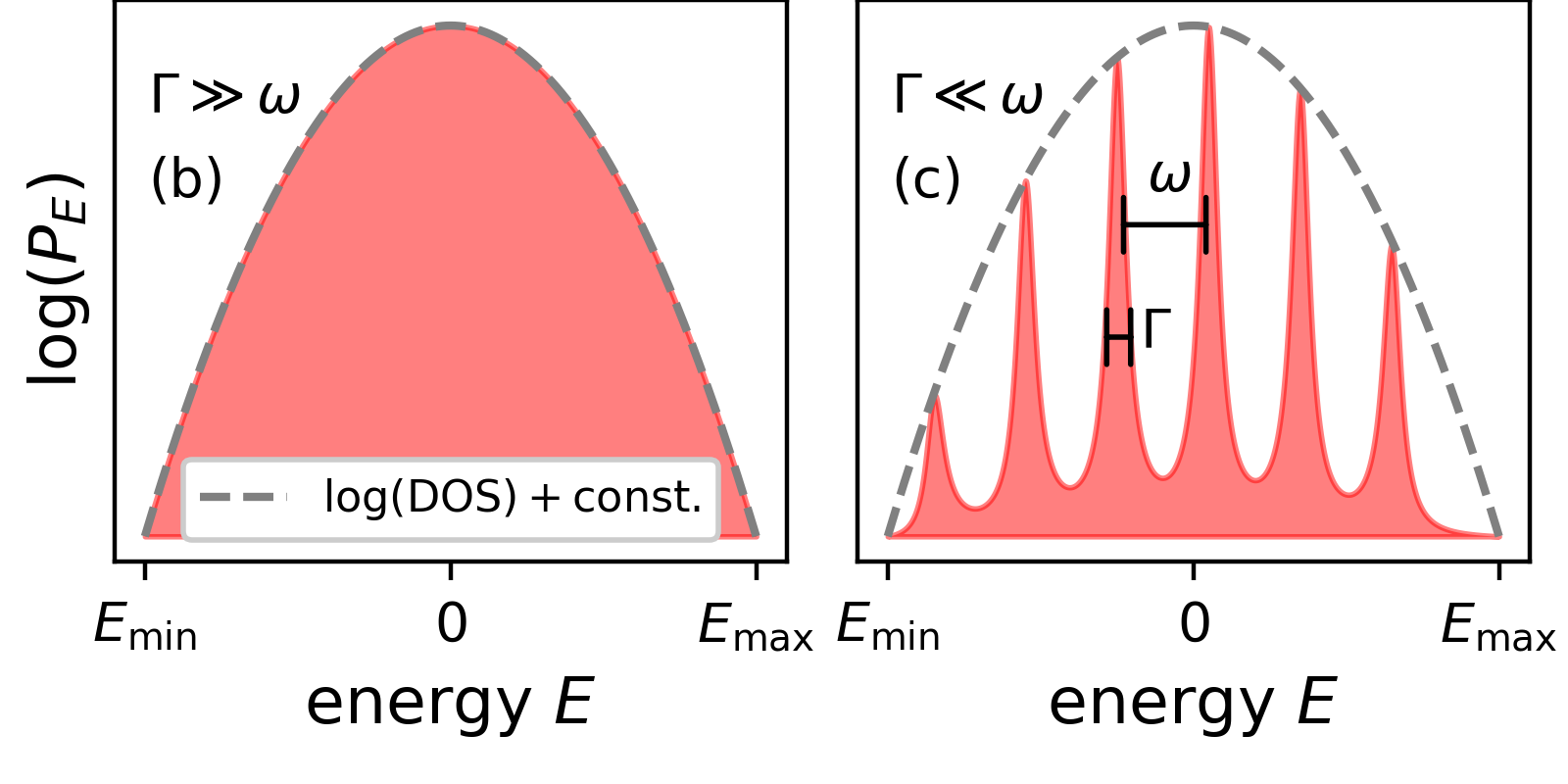}
    \caption{Distinct regimes of thermalization in mesoscopic Floquet systems. (a) A sketch of the different regimes of thermalization, listed in Sec.~\ref{sec:summary_results}. The various frequencies marked along the axis depend on the system size $N$ as shown, and also on (quasi)energy, so the line shown is a cut through the full 2D phase diagram of Fig.~\ref{fig:phase_diagram_one_way_two_way}. 
    (b) and (c):  A depiction of the (b) ``featureless infinite-temperature"  and (c) ``ladder" ensembles. These are characterized by the probability distribution of energy defined by an effective Hamiltonian. In the ladder ensemble, the system has a drive frequency $\omega$ that is large enough so that, while the system does Floquet thermalize, it maintains an approximate conservation of energy modulo $\omega$. The weights of the peaks follow the density of states (DOS) at those energies, $P_E \propto \mathrm{DOS}(E)$.}
    \label{fig:ladder}
\end{figure}
These regimes, and the crossovers between them, occur at drive frequencies that depend on the system size $N$, so to study them we allow for a drive frequency $\omega \propto \Omega(N)$ that is scaled up with $N$. We examine frequency scalings ranging from a weakly $N$-dependent $\Omega(N) = \log N$, to stronger scalings ranging from $\Omega(N) = \sqrt{N}$ to $\Omega(N) = N$. The distinctions between the different regimes we obtain can be made sharp in a particular large-$N$ limit~\cite{Altshuler-Levitov1997_quasiparticle,Burin2015_long_range,Tikhonov-Mirlin2018_power_law,Gopalakrishnan-Huse2019_instability,Bulchandani-Gopalakrishnan2021_onset}, discussed later, where $N \to \infty$ is taken with $\omega/\Omega(N)$ held constant, i.e., a large-$\omega$ limit is taken at the same time.\footnote{The width of the crossovers [in the control parameter $\omega/\Omega(N)$] scales as $\Omega(N)^{-1}$ and sharpens up as the limit is taken.} This limit may appear non-standard when compared to conventional thermodynamic limits studied in many-body physics, but is standard in studies of mesoscopic systems where interaction strengths and/or other parameters are often taken to scale with $N$~\cite{Altshuler-Levitov1997_quasiparticle}.

One motivation for studying the different possible regimes of thermalization in mesoscopic Floquet systems is that many settings---for instance, experiments on near-term quantum simulators---allow controlled access to ``intermediate-scale"~\cite{Preskill2018_nisq} many-body quantum systems where there is not a clean separation of scales between $N$ and $\omega$ (measured in units of a microscopic energy scale). This is also true for numerical studies of Floquet phenomena, which are limited to small system sizes~\cite{Morningstar-Vidal2021_simulation}. There is currently a major gap in the literature in theoretically and systematically addressing thermalization, Floquet heating, and equilibrium in this setting, which we hope to bridge with this present work.  

 \subsection{Summary of results \label{sec:summary_results}}

We identify a number of distinct regimes of thermalization, and crossovers between them, that can occur in mesoscopic Floquet systems. In this work, we are only considering isolated chaotic many-body systems subject to periodic driving, focusing on the delocalization of energy across Floquet zones; in particular, all the crossovers we consider are between chaotic regimes with different degrees of energy conservation, so the physics of many-body localization or integrability will play no role in our discussion. Ref.~\cite{Seetheram-Refael2018_absence} instead considers the fate of finite-size driven integrable systems, with an $N$-dependent driving amplitude.
The different regimes of thermalization we find are summarized below and in Fig.~\ref{fig:ladder}(a), and explained in detail later: 

\begin{itemize}
    \item At the smallest frequencies, the system irreversibly exchanges energy with the drive and Floquet thermalizes to a conventional \textbf{featureless infinite-temperature} state, i.e., the relevant Floquet thermal ensemble is a uniform distribution over all states [Fig.~\ref{fig:ladder}(b)]. 

    \item At larger frequencies, beyond $\omega = \omega_\mathrm{ladder}(N) \sim \log N$ (or, in some physically relevant cases discussed later, $\omega_\mathrm{ladder}(N) \sim \sqrt{N}$), the system instead Floquet thermalizes to a \textbf{ladder ensemble}: In this regime, energy conservation is not completely destroyed, but downgraded to an approximate conservation of energy modulo $\omega$. Thus, energy becomes delocalized across a ``ladder" of narrow  energy windows that are spaced  by $\omega$. This is the relevant maximum-entropy Floquet thermal ensemble under the constraint of conservation of energy modulo $\omega$ [Fig.~\ref{fig:ladder}(c)]. While in some cases this ensemble can have the same average energy as the infinite temperature ensemble, the distribution of energy is distinct.
    
    \item At yet larger frequencies beyond $\omega = \omega_\mathrm{partial}(N) \sim N$, the system only partially thermalizes across the rungs of the ladder. We call this the regime of \textbf{partial Floquet thermalization}.  In this regime, a version of Floquet heating may still occur for many initial states.
    
    \item Finally, at the largest frequencies, beyond $\omega = \omega_{\rm loc}(N) \sim N$, the system does not exchange energy of order $\omega$ or more with the drive, i.e., it does not Floquet thermalize and instead becomes \textbf{energy localized}. In this regime, there exists an extensive energy, defined by a quasilocal effective Hamiltonian, that is approximately conserved for all times. While the extensive many-body bandwidth furnishes an upper bound for $\omega_{\rm loc}(N)$, we find that energy localization sets in at a smaller scale, and distinct non-trivial energy-localized regimes exist that can be distinguished by the presence or absence of isolated Floquet many-body resonances in rare states.

\end{itemize}

A few points are of note. First, complete Floquet thermalization occurs for frequencies less than $\omega_\mathrm{partial}(N)\sim N$, in the sense that the system effectively exchanges energy with the drive. In most of this regime, i.e., between the two scales $\omega_\mathrm{ladder}(N)\sim\log(N)$ and $\omega_\mathrm{partial}(N)\sim N$, the ladder ensemble is the relevant description of the final thermal equilibrium. In contrast, the regime in which the system thermalizes to a featureless infinite-temperature state is parametrically smaller, extending only up to $\omega_\mathrm{ladder}(N)\sim\log(N)$. Second, while we have only focused on the frequency dependence of the different regimes in the discussion above [and in Fig.~\ref{fig:ladder}(a)], there is also a strong dependence on (quasi)energy which we explore below. In Fig.~\ref{fig:phase_diagram_one_way_two_way}, we map out the full two-parameter phase diagram of the different types of thermalization mentioned above, and one important message of our work is that sometimes the thermalization of Floquet systems needs to be state or energy-resolved instead of uniformly averaging over all Floquet eigenstates or initial states. 

Finally, a notable consequence of our results is that under a suitable quench protocol, isolated systems in pure states can thermalize to novel \emph{Schrodinger cat states of temperature}, i.e., superpositions of states at globally different energy \textit{densities} [Fig.~\ref{fig:ladder}(c)]. Although the coherences of such states are notoriously fragile, the ladder-like distribution of energy is a stable signature of this novel form of Floquet thermalization.

The rest of this paper is organized as follows: In Sec.~\ref{sec:theory} we set up our theoretical understanding of the different regimes of thermalization that occur in mesoscopic Floquet many-body quantum systems. We support our theoretical reasoning with numerical evidence using a concrete model in Sec.~\ref{sec:num}. In Sec.~\ref{sec:experiment} we explore some of the prospects for studying the physics discussed in this work experimentally, and show that indeed experimental studies seem to be accessible on some near-term platforms for quantum simulation. Finally, we summarize and discuss our findings in Sec.~\ref{sec:conclusions}.

\section{Theory\label{sec:theory}}

\subsection{Setup and review of Floquet heating}

For the purpose of discussion, we consider $N$ qubits evolving under a time-periodic Hamiltonian $H(t) = H_0 + g_\omega(t) V_0$, where $g_\omega$ is an $O(1)$-valued periodic function of time that time-averages to zero, with period $T=\frac{2\pi}{\omega}$. $H_0$ is a quantum chaotic Hamiltonian that is a sum of one- and two-body terms, and $V_0$ couples the system to the drive, also consisting of a sum of one- and two-body terms. A characteristic microscopic energy scale of $H$ is set to one here. 
We are generally interested in behavior at $\omega \gg 1$, although in practice $\omega \gtrapprox \omega_0$ can be a more accurate condition, where $\omega_0$ is $O(1)$ and depends on the specific system.
Both $H_0$ and $V_0$ are traceless, so the energy corresponding to infinite temperature is zero. $\frac{1}{N 2^N}\tr(H_0^2)$, $\frac{1}{N 2^N}\tr(V_0^2)$, and $\frac{1}{N 2^N}\tr([H_0, V_0]^2)$ are all of order one, so the only small parameters present are $1/\omega$ and $1/N$. The stroboscopic dynamics are governed by the Floquet unitary $U_F = \mathcal{T} \exp\left(-i \int_0^T H(t) dt\right)$ that time-evolves the system by one period. The Floquet unitary defines the Floquet Hamiltonian $H_F$ via $U_F \equiv e^{-i H_F T}$. The quasienergies $\theta$ are defined such that the eigenvalues of $U_F$ are $\euler^{-i\theta T}$, so $\theta$ is only defined modulo $\omega$ and is strictly conserved by the dynamics, and this may be the only such strict conservation law. The specific model we use for later numerical demonstrations is given in Sec.~\ref{sec:num}, but our results are more general.

The process of Floquet heating entails a system resonantly exchanging energy with the drive in quanta of size $\sim \omega$. In this work, we will ideally consider frequencies $\omega$ that are large compared to the microscopic energy scale of $H$, which is set to $1$ here (the regime when $\omega$ is comparable to the local energy scales leads to rapid heating, but the $\omega \sim O(1)$ boundary between these two regimes is system dependent).
In this high-frequency regime, absorbing a quanta $\sim \omega$ of energy requires a high-order process involving $O(\omega)$ local energy moves, which occurs at a rate that is exponentially suppressed in $\omega$. Because these processes can happen anywhere in the system, the system as a whole exchanges ``photons" with the drive at a rate~\cite{Abanin-Huveneers2015_exponentially,Mori-Saito2016_rigorous, Kuwahara-Saito2016_floquet, Abanin-Huveneers2017_effective, Abanin-Huveneers2017_rigorous} 
\begin{align}
    \Gamma \sim N \euler^{-\omega/\omega_0},
\end{align}
with some microscopic (order one) $\omega_0$ that may, in general, depend on the energy density.\footnote{In one dimension there is a logarithmic correction such that the heating rate is bounded by $\Gamma \sim N \euler^{-(\omega \log \omega) / \omega_0}$~\cite{Cao2021_statistical,Sels-Polkovnikov2022_dilute,Parker-Altman2019_hypothesis}.  We focus on the general case in higher dimensions for our discussion and numerical studies below, but the results are qualitatively the same in one dimension.} This is the behavior in the drive frequency range $\omega_0\lessapprox \omega\ll N\omega_0$, and, in this regime, there is an effective (``prethermal") quasi-local Hamiltonian $H_\mathrm{eff}$ that captures the dynamics of the system on timescales shorter than $t\sim \Gamma^{-1}$.  $H_\mathrm{eff}$ can be obtained perturbatively, and represents the most optimal quasi-local truncation of a high-frequency Magnus expansion for $H_F$~\cite{Kuwahara-Saito2016_floquet}. The leading term in the expansion is the time-averaged Hamiltonian, $H_0$ (see App.~\ref{app:Heff}). 

The time-scale $\Gamma^{-1}$ sets the crossover time between the prethermal regime with dynamics governed by $H_\mathrm{eff}$ (which has an extensive conserved energy), and the regime of \textit{Floquet thermalization}, where the system thermalizes \emph{across} different Floquet zones due to a resonant drive-mediated coupling between states separated in energy by $\omega$ and therefore becomes delocalized in energy~\cite{DAlessio-Polkovnikov2013_manybody}. The slow thermalization across Floquet zones is reflected in the non-perturbative, non-local character of $H_F$. The difference between $U_F=e^{-iH_FT}$ and $e^{-iH_{\rm eff}T}$ is the thermalization process across Floquet zones visible on times $t> \Gamma^{-1}$.

\subsection{Floquet thermal ensembles and nonstandard large-$N$ limits\label{sec:scaling_up}}

We now discuss the featureless infinite temperature and ladder ensembles for Floquet thermalization, the crossover between these, and how this crossover sharpens in a particular large-$N$ limit. 

For a system that absorbs/emits energy slowly enough, an eigenstate of $U_F$ with eigenvalue $e^{-i\theta T}$ will be supported on eigenstates of $H_\mathrm{eff}$ near a ``ladder'' of energies  that differ in steps of $\omega$, i.e., with energies 
\begin{align}
    E_{\rm eff} = \theta \pm n \omega
\end{align}
with $n\in \mathbb{Z}$. However, due to the nonzero heating rate, each of the ``rungs'' of the energy ladder will have an energy uncertainty $\sim\Gamma \sim N \euler^{-\omega/\omega_0}$ (see Fig.~\ref{fig:ladder}(c) for a depiction).
In the commonly prioritized limit of finite $\omega$ and $N\to\infty$, the rate $\Gamma$ grows with $N$ and eventually becomes larger than $\omega$ (when $N \sim \omega \euler^{\omega/\omega_0}$). 
For $N$ well in excess of this, the ladder is not resolvable as the width of the rungs exceeds the spacings between rungs, and the energy conservation (even modulo $\omega$) is fully lost.  The resulting equilibrium is then ``infinite temperature" in a strict sense, because the relevant Floquet thermal ensemble is a uniform distribution over all energy eigenstates, as shown in Fig.~\ref{fig:ladder}(b).  

The strict infinite-temperature property of Floquet thermalized states (and eigenstates of $U_F$) can break down to various degrees when $\Gamma \ll \omega$. This can occur in systems with finite $N$ and $\omega$, or in large $N$ systems where we allow $\omega$ to scale up with $N$ in such a way that some behavior characteristic of finite-size systems is retained in the limit. For example, consider $\omega_{\rm ladder}(N) = \omega_0 \log N$: If $\omega$ is scaled up with $N$ faster, so that $\omega \gg \omega_{\rm ladder}(N)$,  then $\Gamma\ll\omega$ at large enough $N$.  In this regime, the distributions of energy in the  eigenstates of $U_F$ have significant weight only near a well-resolved ladder of energies with spacing $\omega$, as shown in Fig.~\ref{fig:ladder}(c). In other words, the energy defined by $H_\mathrm{eff}$ is conserved modulo $\omega$ to a precision of $\sim\Gamma$.

This ``energy ladder" is a maximum-entropy Floquet thermal ensemble that is distinct from ``infinite temperature", and notably it sets in already at a frequency scale $\omega_\mathrm{ladder} = \omega_0 \log N$ that is only weakly dependent on $N$. If we consider the rescaled frequency $\nu = \frac{\omega}{\omega_{\rm ladder}(N)} = \frac{\omega}{\omega_0 \log N}$, then the crossover from the featureless infinite temperature ensemble to the ladder ensemble happens near $\nu=1$. This crossover in the rescaled variable $\nu$ becomes sharp in the large-$N$ limit as can be seen from the behavior of $\frac{\Gamma}{\omega} = \frac{N \euler^{-\omega/\omega_0 }}{\omega} = \frac{N^{1-\nu}}{\nu \omega_0 \log(N)}$ near $\nu=1$: it diverges with $N$ if $\nu < 1$ and approaches zero if $\nu > 1$.

The ladder ensemble sets in at $\omega \sim \log(N)$ and extends to parametrically larger frequency scalings,  $\omega = \omega_{\rm partial}(N) \sim N$. First consider $\omega \sim N^\alpha$, with $0 < \alpha < 1$. As long as $\alpha < 1$, consecutive rungs on the ladder have different energies but the same energy \emph{density} as $N\to\infty$, i.e., the spacing in energy density between the rungs tends to zero. Each Floquet quasienergy $\theta$ corresponds to populating a ladder of energies $E_\mathrm{eff}~ \mathrm{mod}~\omega =\theta$ that spans across all energy densities. In particular, the ladder contains a subset of rungs that have the same energy (and entropy) \textit{density} as infinite temperature in the $N\to \infty$ limit, $E_{\rm eff}^\infty/N = \frac{1}{N2^N}\mbox{Tr}[{H_{\rm eff}}] = 0$. Thus the final equilibrium is one where the \emph{average} energy density corresponds to infinite temperature, but the \emph{distribution} of energy is markedly distinct from the uniform distribution and instead concentrated near a ladder of well-spaced energies. 

Finally, we have the case of $\omega \propto N$ ($\alpha = 1$). In this case, the frequency $\omega$ is extensive and corresponds to transitions between different \textit{energy densities} $e$; thus we denote $\Delta e \equiv \frac{\omega}{N}$ in the rest of this paper. Since $\omega$ is extensive in this case, it is not generally true that the system thermalizes to the same average energy density as infinite temperature in the $N\to\infty$ limit, even when it does equilibrate across Floquet zones (Floquet thermalizes). This is because the final energy distribution resides on a ladder of different energy densities, and the infinite temperature energy density ($e=0$) is not generally one of them (see Fig.~\ref{fig:sign_entropy} for a demonstration).

This brings us to an important point: Floquet thermalization (also referred to as Floquet heating) refers to reaching the appropriate equilibrium ensemble with energies distributed either according to the uniform infinite temperature ensemble or the appropriate ladder ensemble. The ladder ensemble is always a distinct ensemble from infinite temperature, and need not even have the same average energy density as infinite temperature. \emph{Thus, Floquet thermalization does not imply that the system thermalizes to infinite temperature, even on average.}

Some comments are in order before concluding this section. When defining the ladder ensemble, we considered energies defined according to $H_\mathrm{eff}$, the most optimal quasi-local truncation of $H_F$. In this case, the energy defined by $H_\mathrm{eff}$ is conserved modulo $\omega$ to a precision of $\sim\Gamma$ set by the heating rate. However, if $H_\mathrm{eff}$ is not chosen optimally---for instance, if energy is defined with respect to the leading term $H_0$---then the precision is lower and the width of the rungs is accordingly broader,  as discussed in App.~\ref{app:Heff}.  Likewise, if we consider thermalization of generic initial states  (instead of eigenstates of $U_F$), then the energy uncertainty of the initial state also contributes to the broadening of the rungs. A typical product initial state has energy uncertainty $\propto \sqrt{N}$, and hence requires $\omega \sim N^\alpha$ with $\alpha > \frac{1}{2}$ to resolve the rungs of the ladder for large enough $N$. In this case, the crossover from the featureless infinite temperature ensemble to the ladder ensemble happens at $\omega_{\rm ladder}(N)\sim \sqrt{N}$ and it does not sharpen up in the large-$N$ limit.

In sum, in this section we've discussed different Floquet thermal ensembles that are relevant for the different ways in which a system can Floquet thermalize.  This has been in the context of mesoscopic systems, but also using the notion of nonstandard large-$N$ limits, with $N$-dependent frequencies, that help us sharply distinguish the different regimes.
As also mentioned, beyond a certain scale $\omega_{\rm loc} \propto N$, Floquet thermalization does not occur and the effect of the drive is perturbative for almost all eigenstates.  In those cases, energy is localized to the microcanonical window of states near the initial energy, which is conserved. 

In the next section, and in much of the rest of the paper, our goal is to map out a phase diagram delineating various thermal and non-thermal Floquet regimes. In order to do this, we discuss the structure present \textit{at} the scaling $\omega \propto N$, where both Floquet thermalization and various partially-thermalizing or non-Floquet-thermal phases exist in the large-$N$ and large-$\omega$ limit. Note that at this scale, we are no longer probing the crossover between the infinite temperature and ladder ensembles (which happens at parametrically smaller frequencies); instead, our goal is to study different degrees of thermalization to the appropriate ladder ensemble.

\subsection{Thermalization phase diagram\label{sec:floquet_heating_resonances}}
Our goal is to map out the various types of thermalization that can occur in mesoscopic Floquet systems and, relatedly, in Floquet systems in nonstandard large-$N$ limits. In the former, these are regimes of behavior with smooth crossovers between them, while in the latter they are idealized sharp phases representative of the finite-$N$-and-$\omega$ regimes. To do this we fix $\omega \propto N$ and consider the matrix elements of the drive operator $V_0$ between eigenstates of $H_0$ separated in energy density by $\Delta e$, and the relevant density of states of $H_0$, in order to argue for or against interzone thermalization and Floquet resonances. Note that the rungs of the ladder represent different Floquet zones, so thermalization to the ladder ensemble corresponds to interzone thermalization. We use the leading order $H_0$ as our definition of energy because $H_\mathrm{eff}\to H_0$ as $N,\omega \to\infty$, and because it is the simplest option, e.g., using $H_0$ results in a notion of energy that is not $\omega$ dependent, as $H_\mathrm{eff}$ is (also see the discussion in App.~\ref{app:Heff}). We also corroborate our findings numerically with the full dynamics generated by $U_F$ for our model system in Sec.~\ref{sec:num}.

We refer to the inverse energy level spacing of $H_0$ as the density of states (DOS), so that it is the number density, in energy (not in energy and volume), of many-body energy levels. At large $N$, the DOS of $H_0$ is of the form
\begin{align}
    D(e) = \exp[N s(e)]~,\label{eq:DOS}
\end{align}
up to subexponential factors, where $s(e)$ is the entropy density at energy density $e$.

A similar form is motivated for the matrix elements of $V_0$, which are suppressed exponentially in $\omega$~\cite{Mondaini-Rigol2017_off_diagonal,Mallayya-Rigol2019_heating,Mori2022_heating}, in accordance with the exponentially slow heating rate. As we are considering $\omega \propto N$ now, the matrix elements are exponentially small in $N$ for a fixed $\Delta e$. Therefore, we assume that the leading behavior of the size of the matrix elements is
\begin{align}
    V(e_1, e_2) = \exp[-N f(\Delta e, \bar{e})]~,\label{eq:ME}
\end{align}
where $e_1$ and $e_2$ are the energy densities of the two eigenstates of $H_0$, and $f$ is a function of the difference and mean
\begin{align}
    \Delta e = \frac{\omega}{N} = |e_2 - e_1|\ \text{and}\ \bar{e} = \frac{e_1+e_2}{2}.
\end{align}
Note that the form of Eq.~\ref{eq:ME} matches to the ETH~\cite{DAlessio-Rigol2016_review} and to the heating behavior discussed above, with $f(\Delta e, \bar{e}) \cong \frac{1}{2}(s(\bar{e})+\frac{\Delta e}{\omega_0})$ for small $\Delta e$.  However, here we are dealing with matrix elements between extensively different energies, so we allow a more general form for $f$.

We have now completed the essential setup for understanding the main features of the thermalization phase diagram, which we explain here and support with numerical evidence in Sec.~\ref{sec:num}. The presence or absence of thermalization and/or resonances between Floquet zones should be governed by comparing the relevant energy level spacing with the relevant matrix elements or rates for those processes. When the frequency is extensive, this comparison is nontrivial because the exponents in Eqs.~\ref{eq:DOS} and \ref{eq:ME} are both $\propto N$.\footnote{In one dimension there is a correction such that $\omega \propto N/\log N$ (a slightly subextensive frequency) is instead the scaling needed to have $(1/D)$ and $V$ scale similarly~\cite{Cao2021_statistical,Sels-Polkovnikov2022_dilute,Parker-Altman2019_hypothesis}.  We proceed with the $\omega \propto N$ scaling presented above, which appears to be correct for systems in more than one dimension, and is consistent with the numerics for an all-to-all quantum dot model that we study below.  The finite-size regimes in our phase diagram are all still present for the modified scaling in one dimension, although the precise scaling of $\omega$ needed to remain in any one of the phases as one takes the large-$N$ limit is slightly altered by the $1/(\log{N})$ factor.} 

\begin{figure*}
    \centering
    \includegraphics[width=0.31\linewidth]{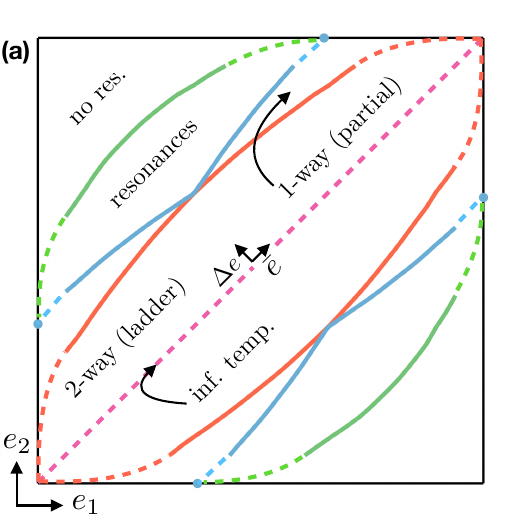}
    \includegraphics[width=0.62\linewidth]{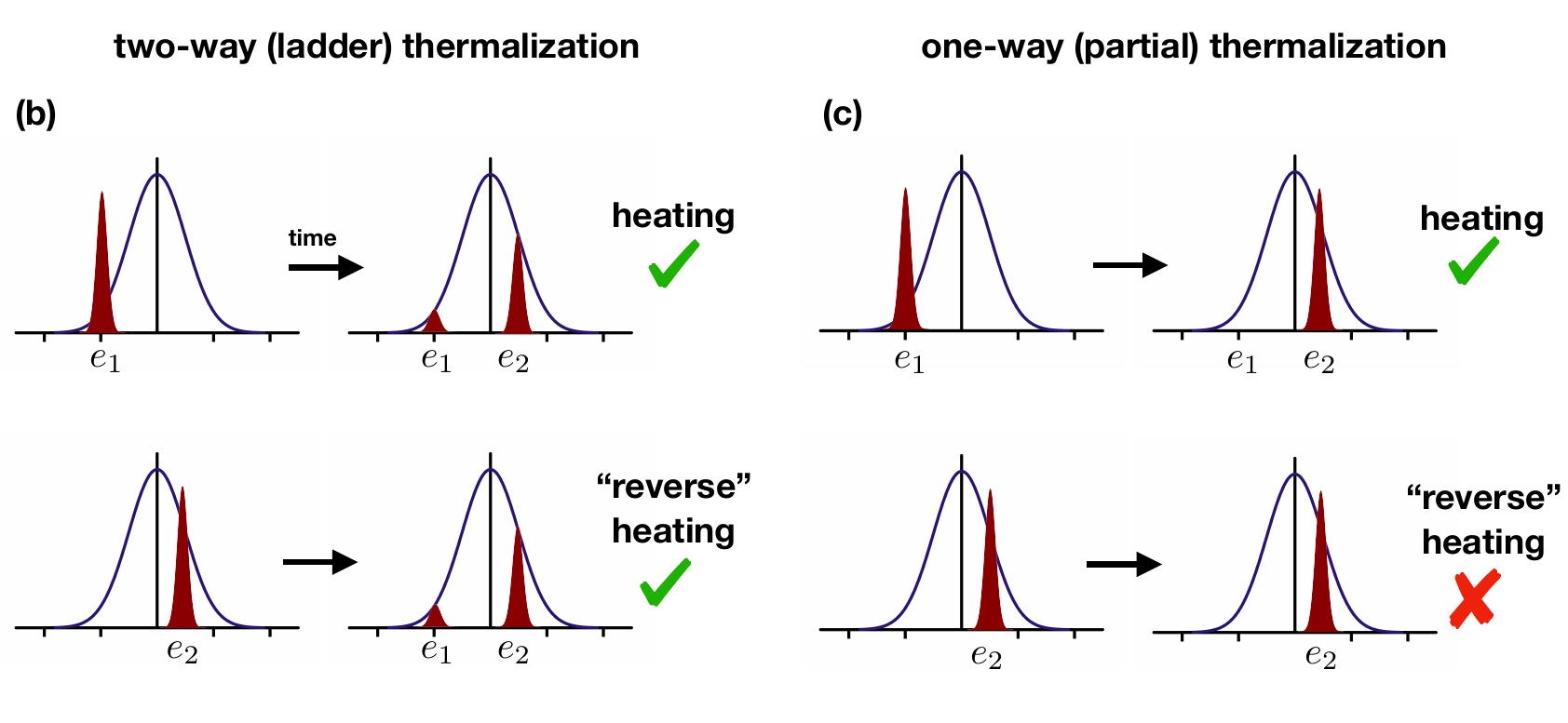}
    \caption{Phases of thermalization in Floquet systems. (a) The phase diagram in the two-zone approximation. The solid lines are obtained from $N\in\{12,14\}$ data, and the dashed curves are sketched extrapolations. The outer black square marks the edges of the spectrum of $H_0$. A point on the diagram specifies two energy densities, $e_1$ and $e_2$, which can also be labelled by $\bar{e}$ and $\Delta e$ (i.e., the frequency $\omega = N\Delta e$ is implied by the point, and is constant along lines of constant $\Delta e$). The phase that that point belongs to indicates the extent to which the Floquet dynamics mixes those two energy densities. The pink dashed line denotes $\Delta e = 0$, which contains the commonly considered case of finite $\omega$ and $N\to\infty$. The distinct behaviors are as described in Fig.~\ref{fig:ladder}(a) and in the main text. ``resonances" and ``no res." indicate energy localized phases with and without rare Floquet resonances. (b) and (c) Depictions of two-way and one-way thermalization. In two-way thermalization, initial states at either energy density thermalize to have the correct equilibrium populations at those energies. In one-way thermalization, states at a high density of states do not decay into a window of energy at a much lower density of states, so there is heating but not the ``reverse heating" needed to produce full Floquet thermalization.
    }
    \label{fig:phase_diagram_one_way_two_way}
\end{figure*}

Since the frequency and many-body bandwidth of $H_0$ are both extensive in $N$, the number of rungs in the relevant ladder ensemble is finite. To simplify the analysis, we truncate the ladder to two rungs and work in a \emph{two-zone approximation}, where we consider Floquet thermalization/resonances involving two narrow windows of energy density near $e_1$ and $e_2$ (this implies the frequency via $\omega = N|e_1-e_2|$, and the quasienergy via $\theta = Ne_1 \mod \omega$), and we discuss the minor modifications that come with considering the rest of the Floquet zones in the energy-density ladder in App.~\ref{app:two_zone}. We find four distinct phases, ordered from the most thermalizing to the least: ``two-way (full) thermalization", ``one-way (partial) thermalization", and two energy localized phases with ``isolated resonances", and with ``no resonances". The phase with two-way Floquet thermalization contains a sliver of vanishing relative size that is thermalization to the infinite-temperature ensemble or to a ladder ensemble with zero average energy density. The phase diagram is shown in Fig.~\ref{fig:phase_diagram_one_way_two_way}; we will explain the ideas behind each phase below, and support the diagram with numerical evidence later on. 

\subsubsection{Two-way (full) thermalization}

In the phase with two-way Floquet thermalization, states at $e_1$ and $e_2$ get fully mixed by the presence of the drive in the sense that in the $N\to\infty$ limit, any initial state at either energy density will thermalize to occupy states at both energy densities with weight proportional to $D(e_1)$ and $D(e_2)$, respectively. Another way of saying this is that the Floquet eigenstates of $U_F$ will all have weight on both rungs of the corresponding energy ladder, with weights proportional to the DOS.  

This phase occurs when the rate for transitioning from one energy to the other is larger than the level spacing at the initial energy, a symmetric condition given by $\lim_{N\to\infty} V(e_1,e_2)^2 D(e_1) D(e_2) = \infty$. Using Eqs.~\ref{eq:DOS} and \ref{eq:ME} this translates to
\begin{equation}
    \label{eq:two_way_condition}
    2 f(\Delta e, \bar{e}) < s(e_1)+s(e_2)~.
\end{equation}
Intuitively, this says that each energy has a high enough density of states that states at the other energy ``see" them as an effective continuum and transition to them at a Fermi Golden Rule (FGR) rate~\cite{Mallayya-Rigol2019_heating,Ikeda-Polkovnikov2021_fgr,Mori2022_heating,Rakcheev-Lauchli2020_estimating,Micklitz-Huse2022_emergence}. If we label $e_2$ as the energy closer to zero with higher DOS, so that $D(e_2) \geq D(e_1)$, then FGR is valid for transitions from $e_1$ to $e_2$, so that a system initialized at $e_1$ thermalizes to acquire weight at both $e_1$ and $e_2$ in proportion to the DOS at the two energies. We call this process ``heating" since the $e_2$ is at a higher (absolute value of) temperature than $e_1$. Due to the heating process, the resultant broadening of the levels at the lower DOS is larger than the level spacing there, thus both DOS act as a continuum. As a result, a state initialized at $e_2$ would also thermalize to acquire a small weight on $e_1$ that is proportional to $D(e_1)$, a process we called ``reverse heating" [see Fig.~\ref{fig:phase_diagram_one_way_two_way}(b)]. In general, since the DOS is exponentially larger (in $N$) for $e_2$, the system thermalizes on average to $e_2$, the energy with the higher entropy density, when $N$ is large.

A special case occurs when $D(e_1)=D(e_2)$. This occurs in models with $D(e) = D(-e)$ when both energies are situated symmetrically about zero, so that $e_2 = -e_1 = \frac{\omega}{2N}$. This corresponds to Floquet eigenstates at eigenvalue $e^{-i \theta T} = e^{i \frac{\omega}{2} \frac{2\pi}{\omega}} = -1$, i.e., with quasienergy $\theta =\pi/T$. In this case, the final thermal equilibrium is an equal-weight superposition of two different energy densities, one corresponding to positive temperature and the other to negative temperature. Because the final equilibrium is a superposition of states in two well-separated narrow energy windows, and those states look locally like they are described by the Boltzmann distribution, this final local Floquet equilibrium is approximately described by
\begin{eqnarray}
\rho = \frac{1}{2Z}\left(e^{-\beta H_0} + e^{\beta H_0}\right) = \frac{1}{Z} \cosh(\beta H_0),
\end{eqnarray}
where the inverse temperature $\beta$ is set by $e_1$. This is a new type of Floquet ETH ensemble distinct from infinite temperature: even though the average energy corresponds to that of infinite temperature, the distribution is bimodal with weight at positive and negative temperature states, and almost no weight at zero energy [see Fig.~\ref{fig:sign_entropy}(a)].   

A consequence of this is that if we start with a simple initial \emph{pure} state at energy $E_1$ and drive it with frequency $2|E_1|$, the state will thermalize to spontaneously become a type of pure \emph{Schrodinger-cat state of temperature}, i.e., a global superposition of thermal pure states at two different temperatures! While the coherences of such states are extremely sensitive to imperfect isolation, the ladder-like distribution of energy is a much more robust signal of this type of Floquet thermalization. We explore the prospects of realizing these novel equilibrium distributions experimentally in Sec.~\ref{sec:experiment}. 

\subsubsection{One-way (partial) thermalization}

The phase with one-way, or partial, Floquet thermalization is subtly different in that the higher entropy density appears as a continuum to the lower, but the converse is not true. The condition for this phase is the validity of Fermi's Golden Rule (FGR) in one direction but not the other, which translates to
\begin{equation}
    \label{eq:one_way_condition}
    s(e_1) + s(e_2) < 2f(\Delta e, \bar{e}) < 2s(e_2)~,
\end{equation}
if we label the higher DOS as $e_2$. This is equivalent to $V(e_1,e_2)^2 D(e_1) D(e_2)$ vanishing with $N$ but $V(e_1,e_2)^2 D(e_2)^2$ diverging. Under this condition, a system initialized at $e_1$ would heat to thermalize to $e_2$, as in the two-way phase, but a state initialized at $e_2$ would generally not thermalize to acquire a small weight on $e_1$ that is proportional to $D(e_1)$ [Fig.~\ref{fig:phase_diagram_one_way_two_way}(c)]. Such a subtle entropically suppressed ``reverse heating" could be unimportant if $N$ is large, due to $D(e_1) \ll D(e_2)$. In that case, the union of the phases with one-way and two-way Floquet thermalization is the more physically important concept.

\subsubsection{Energy localized}

We now move on to the two remaining phases where Floquet thermalization fails, so they are energy localized. This sets in when 
\begin{equation}
    \label{eq:loc_condition}
    s(e_1),\ s(e_2) < f(\Delta e, \bar{e})~.
\end{equation}
This condition means that typical states no longer couple to other Floquet zones under the dynamics. Notably, our arguments show transparently why the boundary of the energy localized regime occurs at an extensive frequency, because both $V(e_1, e_2)$ and $D(e)$ scale exponentially with $N$.  
In particular, this rules out other seemingly plausible scalings such as $\omega_{\rm loc} \propto \sqrt{N}$ which could be argued for, for instance, by comparing the frequency to the standard deviation in energy of the DOS~\cite{DAlessio-Rigol2014_long-time}.

The energy localized phase is further divided into two phases distinguished by the presence or absence of rare Floquet many-body resonances~\cite{Bukov-Polkovnikov2016_heating}, by which we mean rare Floquet eigenstates $U_F$ that are superpositions of eigenstates of $H_{\rm 0}$ in different Floquet zones, while most eigenstates are energy localized. 
Unlike the matrix elements responsible for many-body resonances in many-body localized systems, which can have very broad distributions~\cite{Serbyn-Abanin2015_criterion,Morningstar-Huse2021_avalanches,Garratt-Chalker2021_local,Garratt-Roy2022_resonant}, the distribution of the matrix elements of $V_0$ between energy eigenstates of $H_0$ at $e_1$ and $e_2$ should be Gaussian and thus well characterized by a single value, $V(e_1,e_2)$, because the energy eigenstates of $H_0$ are thermal. Thus the expectation of any rare Floquet many-body resonances is heralded by comparing the size of these matrix elements to the smallest energy gap $E_2 - E_1 - \omega$, where $E_{1,2}$ are within the energy windows we've been discussing. The minimum gap scales as $[D(e_1) D(e_2)]^{-1}$, so such rare resonances are expected to be present in typical samples in the energy-localized phase with isolated resonances, the condition for which is
\begin{equation}
    \label{eq:resonance_condition}
    s(e_1),\ s(e_2) < f(\Delta e, \bar{e}) < s(e_1)+s(e_2)~.
\end{equation}
When $f(\Delta e, \bar{e})$ is even larger than $s(e_1)+s(e_2)$, the system is in the ``no resonances" phase, which also contains the trivial regime of $e$ and $\omega$ such that both $(e+\Delta e)$ and $(e-\Delta e)$ are outside of the many-body bandwidth of the system, and $H_\mathrm{eff}$ is good for describing the dynamics of all states for all times (there is no nonperturbative effect of the periodic drive).

\subsubsection{Crossovers}

The relevant quantities governing the crossovers between the phases discussed in this section all have the form 
\begin{equation}
\euler^{N [a s(e_1) + b s(e_2) - c f(\Delta e, \bar{e})]}
\label{eq:crossover_scaling}
\end{equation}
for integers $a$, $b$, and $c$ (up to subexponential corrections). This is because all the conditions derived in Eqs.~\ref{eq:two_way_condition}-\ref{eq:resonance_condition} were obtained by comparing powers of matrix elements and DOS, which have the exponential forms shown in Eqs.~\ref{eq:ME} and \ref{eq:DOS}.

This implies that as we tune through the crossovers by varying, e.g., $\Delta e= \omega/N$, their widths sharpen up as $\propto N^{-1}$, which superficially looks like a transition with critical exponent $\nu =1$.  However, to the best of our knowledge this scaling does not correspond to any diverging length scale and associated critical exponent $\nu$, and thus these sharp crossovers are not like continuous phase transitions in that sense.

In this and the previous sections, we have explained our theoretical understanding of the idealized phases that represent different regimes of Floquet thermalization and many-body resonances that occur in isolated mesoscopic periodically driven systems. In the next section, we corroborate these ideas with numerical evidence to obtain the phase boundaries for the different phases plotted in the phase diagram in Fig.~\ref{fig:phase_diagram_one_way_two_way}(a). 

\section{Numerical evidence\label{sec:num}}

As a model system we take $N$ qubit degrees of freedom evolving under the time-periodic Hamiltonian
\begin{align}
    H(t) = H_0 + \sgn[\cos(\omega t)] V_0,
\end{align}
where
\begin{align}
    &H_0 = H_Z + H_X\ 
    &V_0 = H_Z - H_X\\
    &H_Z = \sum_{i<j} J_{ij} Z_i Z_j + \sum_i h_i Z_i\ 
    &H_X = g \sum_i X_i.
\end{align}
Equivalently, the dynamics are governed by the Floquet unitary
\begin{align}
    U_F = \euler^{-i \frac{H_Z T}{2}} \euler^{-i H_X T} \euler^{-i \frac{H_Z T}{2}},
\end{align}
where $T=\frac{2\pi}{\omega}$ is the period of the drive. We take the couplings to be random variables and average over realizations. For each realization, the $\frac{N(N-1)}{2}$ couplings $J_{ij}$, and the $N$ longitudinal fields $h_i$, are first sampled from a standard normal distribution, then shifted and scaled such that 
\begin{align}
    \langle J_{ij} \rangle = \langle h_i \rangle = 0\ \text{and}\ \frac{1}{2} \langle J_{ij}^2 \rangle = \langle h_i^2 \rangle = \frac{1}{N}.
\end{align}
The transverse field is $g=1$ unless otherwise stated. At strict infinite temperature ($\rho \propto I$), this model satisfies $\langle H_Z^2 \rangle = \langle H_X^2 \rangle = N$.~\footnote{Without the $h_i$ fields we would have $\langle H_Z \rangle = N-1$, so we include them to reduce this potential source of finite-size effects. These terms also break the Ising symmetry.}

We choose to study this all-to-all quantum dot model because the main effects we are examining involve long time scales that are beyond what the Thouless time would be in an alternative geometrically local model, so on the relevant timescales those systems would also behave effectively as a quantum dot. Thus the lack of geometric locality in our model is not important.

\begin{figure}
    \centering
    \includegraphics[width=0.9\linewidth]{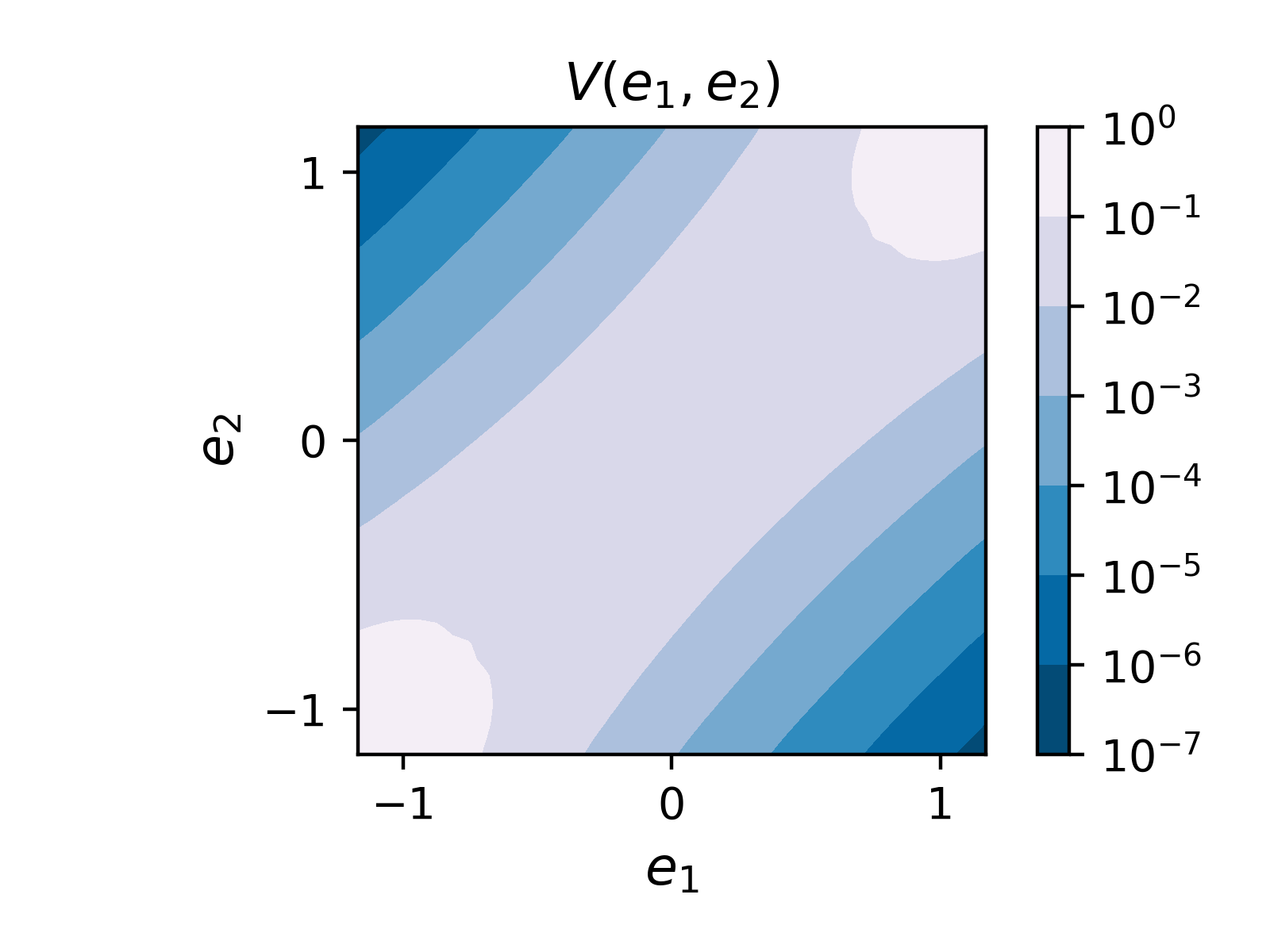}
    \caption{Matrix elements of $V_0$ between eigenstates of $H_0$ at different energy densities. $V(e_1,e_2)$ is obtained by averaging the absolute value of the matrix elements over states and samples. This data is for samples with $N=14$ qubits.}
    \label{fig:V_contours}
\end{figure}

In Fig.~\ref{fig:V_contours} we show contours of $V(e_1,e_2)$ for our model system with $N=14$ qubits. The evenly spaced, rather straight, contours indicate the expected exponential suppression in $\omega$, and a weak dependence of $f(\Delta e, \bar{e})$ on $\bar{e}$. 

\subsection{Floquet thermalization when $e_2 = -e_1 = \frac{\Delta e}{2}$}

\begin{figure}
    \centering
    \includegraphics[width=0.9\linewidth]{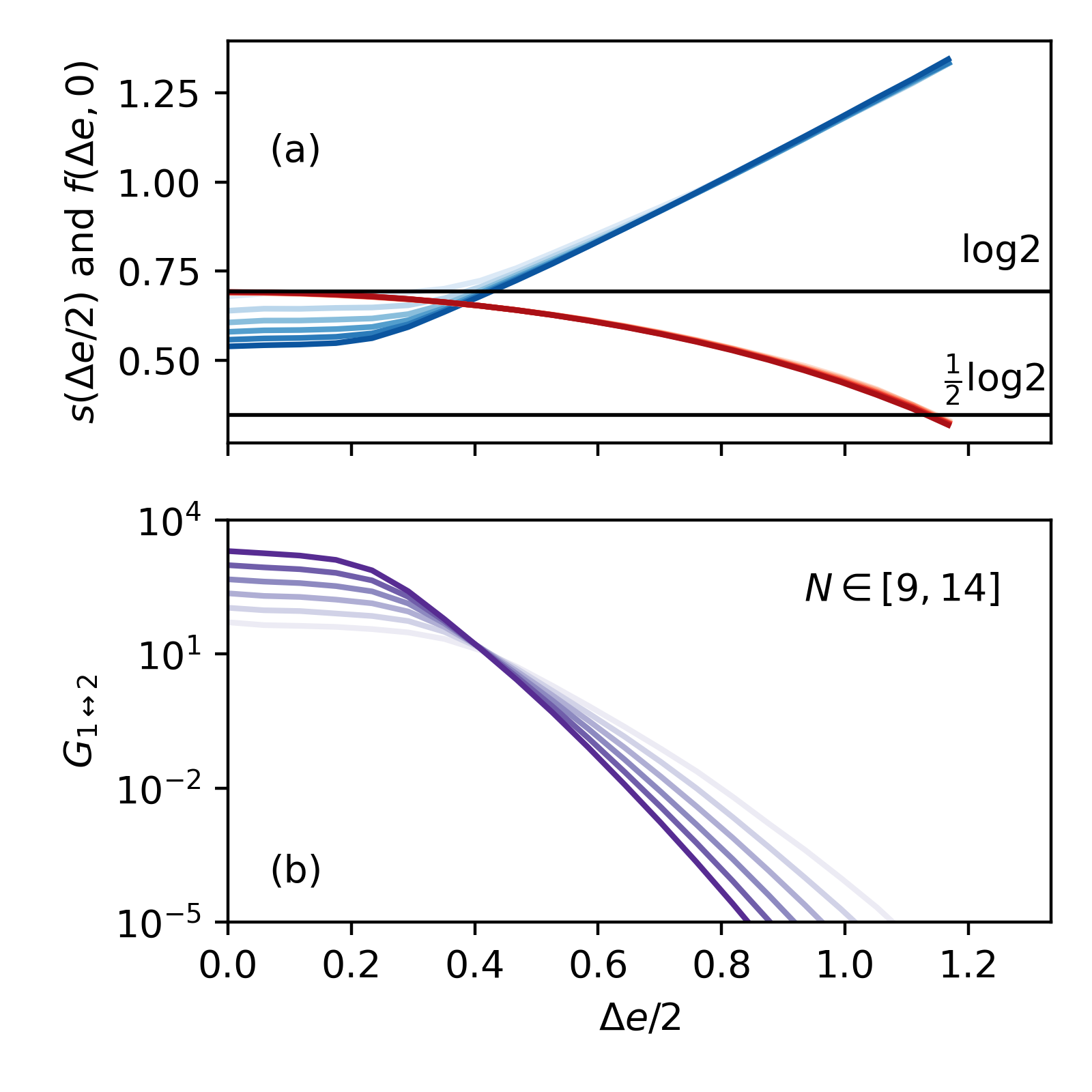}
    \caption{Crossover between Floquet thermalization and energy localization along $e_2=-e_1=\frac{\Delta e}{2}$: linear response. (a) The entropy density $s$ and a slice of the matrix element scaling function $f$ for $e_2=-e_1=\frac{\Delta e}{2}$. $s$ is shown in red and $f$ in blue. System sizes $N$ are $9\mhyphen 14$ (light to dark). The horizontal black lines are the $N\to \infty$ expectations $s(0) = \log 2$ and $f(0, 0) = \frac{\log 2}{2}$ from random matrix theory. The right edge of the plot corresponds to the realization-averaged edge of the spectrum. (b) The quantity $G_{1\leftrightarrow 2} = V_{12}^2 D_1 D_2$ for the same data as panel (a).}
    \label{fig:s_f_G_vs_e}
\end{figure}

As the most straightforward case of a crossover between the phases discussed in Sec.~\ref{sec:floquet_heating_resonances}, and shown in Fig.~\ref{fig:phase_diagram_one_way_two_way}, we take $e_2=-e_1$ and tune $\Delta e = \omega/N$ to see when the thermalization across zones (Floquet heating) turns on/off. In this case $D(e_1)=D(e_2)$ and we imagine a system initialized at energy density $e_1=-\frac{\Delta e}{2}$ and driven at frequency $\omega = N \Delta e$. In order to determine if the system will exchange energy with the drive and equally populate states at $e_2=+\frac{\Delta e}{2}$ in its final equilibrium, i.e., becomes delocalized in energy, we want to test the condition Eq.~\ref{eq:two_way_condition}. Note that the equality of the DOS at $e_1, e_2$ means that there is no one-way thermalization phase to discuss for this line-cut through the phase diagram. We test the condition Eq.~\ref{eq:two_way_condition} in the two following ways:

First, we include finite-size corrections to Eqs.~\ref{eq:DOS} and \ref{eq:ME} (see App.~\ref{app:estimate_s_f}), and extract $s\left(\frac{\Delta e}{2}\right)$ and $f(\Delta e, 0)$ from data at $N\in [9,14]$. When these two quantities cross, that indicates a change in the condition of Eq.~\ref{eq:two_way_condition}. The result is shown in Fig.~\ref{fig:s_f_G_vs_e}(a). There we see a crossover at $\frac{\Delta e}{2} \approxeq 0.4$ at large $N$. This corresponds to $\omega \approxeq 0.8 N$. For context, the energy difference between the ground state and highest excited state is, on average, $\omega \approxeq 2.7 N$ so energy localization sets in at frequencies significantly less than the many-body bandwidth. The main finite-size effect appears to be the existence of a flat region of width $\propto \frac{1}{N}$ in $f(\Delta e, 0)$ near $\Delta e = 0$. This is simply the width-$O(1)$ window of frequencies, for which a photon can be absorbed by $O(1)$ degrees of freedom in $O(1)$ time, that exists before the widely studied exponential-in-$\omega$ dependence is incurred, and we can also see this plateau in the middle of Fig.~\ref{fig:V_contours}. 

Second, we consider the related, but more direct, quantity 
\begin{align}
    \label{eq:G12}
    G_{1\leftrightarrow 2} = V(e_1, e_2)^2 D(e_1) D(e_2).
\end{align}
This quantity corresponds to the condition Eq.~\ref{eq:two_way_condition} in that $G_{1\leftrightarrow 2}$ increases (decreases) exponentially with $N$ when the condition is met (not met), modulo finite-size corrections to Eqs.~\ref{eq:DOS} and \ref{eq:ME}. In Fig.~\ref{fig:s_f_G_vs_e}(b), we directly compute the quantity $G_{1\leftrightarrow 2}$ for various system sizes. The finite-size crossings indeed appear to be consistent with our earlier analysis shown in Fig.~\ref{fig:s_f_G_vs_e}(a), and some drift is expected due to the aforementioned finite-size effects.

\begin{figure}
    \centering
    \includegraphics[width=0.9\linewidth]{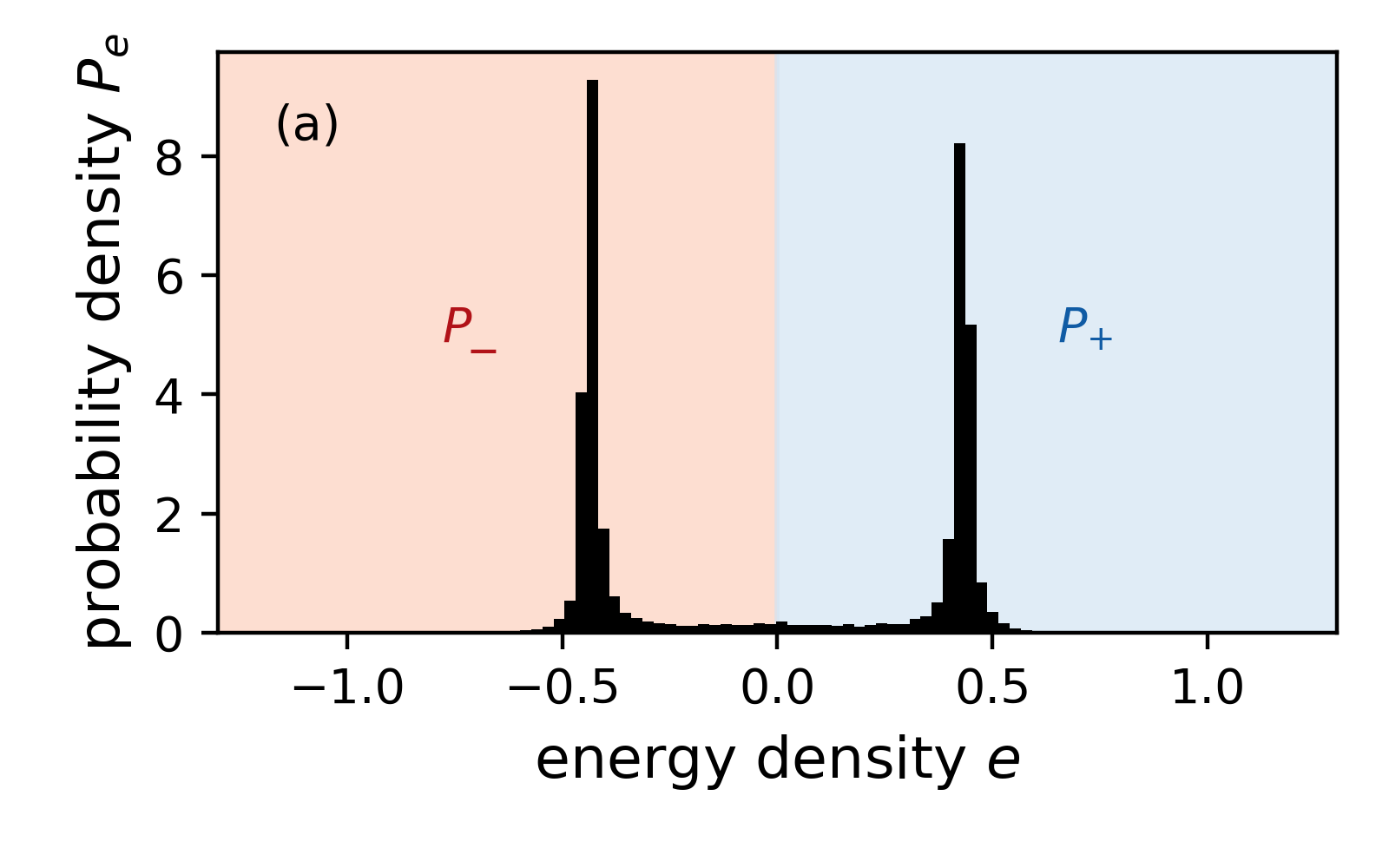}\\
    \includegraphics[width=0.9\linewidth]{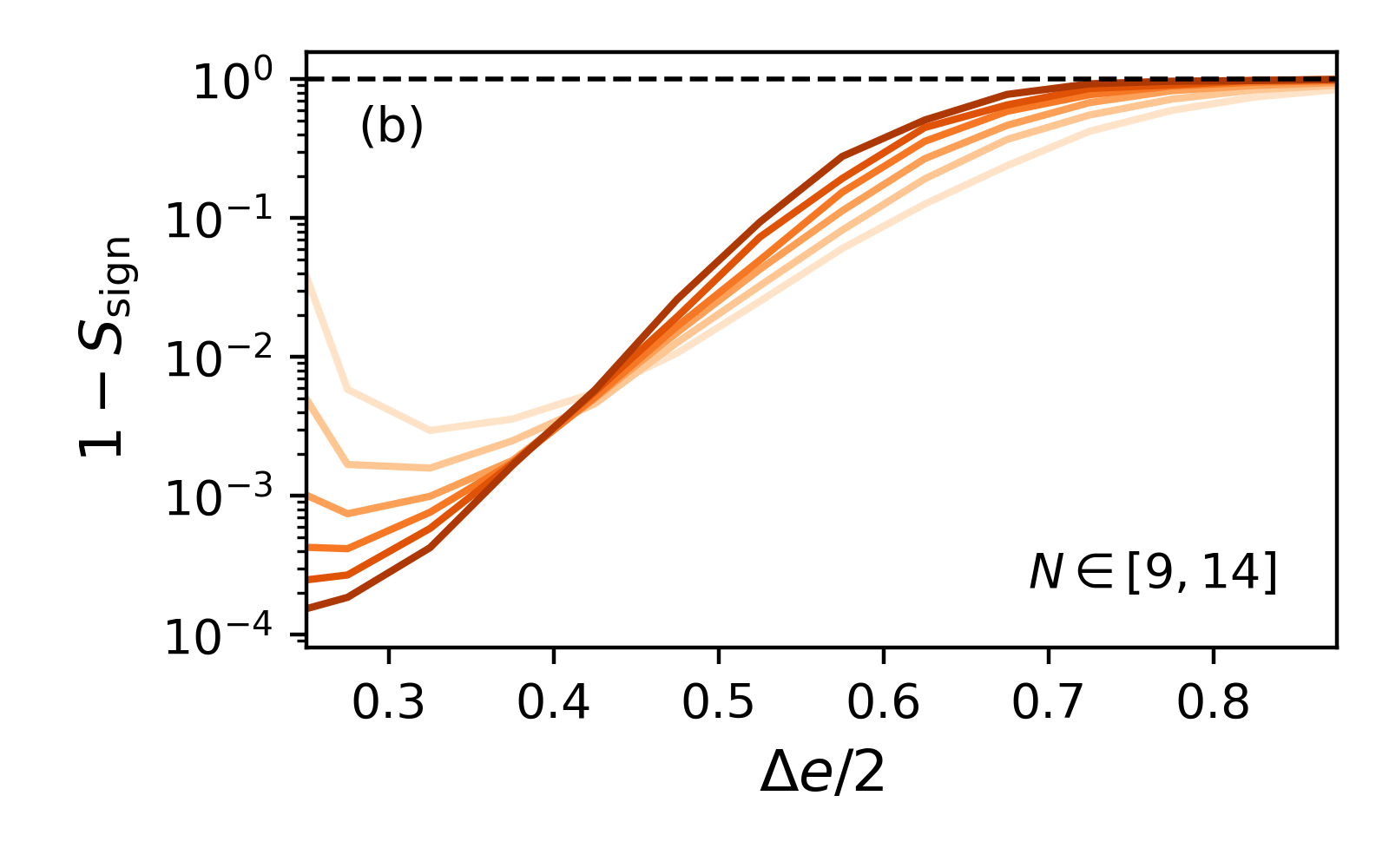}
    \caption{Crossover between Floquet thermalization and energy localization along $e_2=-e_1=\frac{\Delta e}{2}$: full Floquet dynamics. (a) An example of the probability distribution $P_e$ over energy density (with respect to $H_0$), for the eigenstate of $U_F$ closest to $\theta = \pi/T$ in a single sample with $N=13$ and $\Delta e = 0.8$. The shaded red and blue halves indicate the probability of the sign, $P_\pm$. (b) The complement of the entropy of the sign of the energy. The result is averaged over samples of $U_F$ and the two eigenstates on either side of $\theta = \pi/T$ within each sample. System sizes $N=9\mhyphen 14$ correspond to light to dark curves.}
    \label{fig:sign_entropy}
\end{figure}

An independent way to corroborate the existence of this sharp crossover between the two-way Floquet thermalizing phase and the energy-localized phase with only isolated resonances (see Fig.~\ref{fig:phase_diagram_one_way_two_way} along the line $e_2=-e_1$ where the one-way phase vanishes) is to examine the eigenstates of the Floquet operator $U_F$ near $\theta=\pi/T$, which corresponds to the ladder of energy densities $e\in [...,-\frac{\omega}{2N}, +\frac{\omega}{2N},...]$ that we have been considering. Note that since $\omega \propto N$, the number of rungs on the ladder is finite and constant in $N$. In the phase where the system freely (but slowly) exchanges energy with the drive, the eigenstates of $U_F$ will have energy distributions that have weight on all rungs of the energy ladder, with the weight on each rung set by the density of states there (as in Fig.~\ref{fig:ladder}). In contrast, when the system cannot Floquet thermalize, the eigenstates will be well-localized (in energy) on a single rung. As a measure of this crossover we compute the entropy of the sign of the energy,
\begin{align}
    S_\mathrm{sign} = -P_- \log_2 (P_-) -P_+ \log_2 (P_+),\label{eq:sign_entropy}
\end{align}
for eigenstates of $U_F$ with eigenvalue near $-1$ ($\theta=\pi/T$), where $P_\pm$ is the probability of $\mathrm{sgn}(E) = \pm 1$. This quantity tends to zero in the energy localized phase, and to one bit for states at quasienergy $\theta = \pi/T$ in the two-way Floquet thermalizing phase. In Fig.~\ref{fig:sign_entropy}(a) we show the distribution of energy density for the Floquet eigenstate closest to $\theta = \pi/T$ in a single sample of $U_F$ at $\omega = 0.8 N$. The sign entropy is shown in Fig.~\ref{fig:sign_entropy}(b), where we again see a finite-size crossing near $\frac{\Delta e}{2}\approxeq 0.4$ at accessible sizes, consistent with the picture developed around Fig.~\ref{fig:s_f_G_vs_e}. 

Note that the widths of the peaks in Fig.~\ref{fig:sign_entropy}(a) are not set by the thermalization rate $\Gamma$, as in Fig.~\ref{fig:ladder}, because we are using $H_0$ rather than the optimal local $H_\mathrm{eff}$ here (App.~\ref{app:Heff}). The peaks can be made much narrower by including corrections to $H_0$~\cite{Bukov-Polkovnikov2016_heating}.

\begin{figure}
    \centering
    \includegraphics[width=1\linewidth]{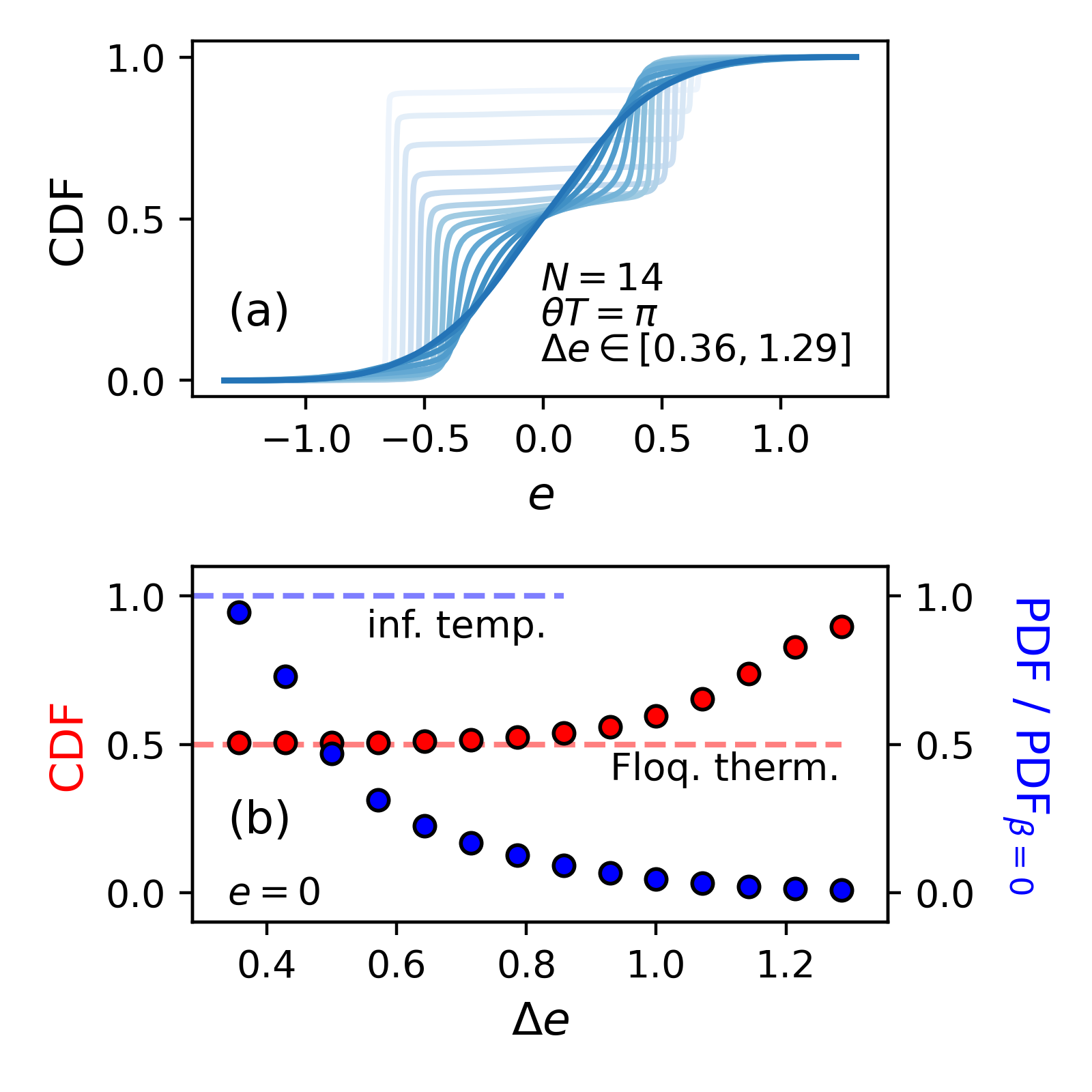}
    \caption{
    Distributions of energy density in Floquet eigenstates at $\theta = \pi / T$. (a) The cumulative distribution function (CDF), averaged over 10 samples of $U_F$ and 10 eigenstates near $\theta = \pi/T$ in each sample. We flip the sign of $e$ as needed for each state before averaging over states so that if the states are imbalanced in energy, the average over states is too. $\Delta e \in [0.36, 1.29]$ are evenly spaced and correspond to dark-to-light curves. All data is for $N=14$ qubits. (b) The CDF (red) and normalized probability density function (PDF) (blue) at $e=0$. The CDF at $e=0$ takes a value $1/2$ (dashed red line) when the distribution is balanced about $e=0$. This happens when the system Floquet thermalizes. The normalized PDF at $e=0$ takes the value $1$ (dashed blue line) when the system thermalizes to the featureless infinite temperature ensemble.}
    \label{fig:E_distributions}
\end{figure}

Finally, in Fig.~\ref{fig:E_distributions} we further examine the distribution of energy density in Floquet eigenstates at $\theta = \pi / T$ as we tune $\Delta e=\omega/N$: In (a) we plot the cumulative distribution function (CDF) of $e$ for $\Delta e \in [0.36, 1.29]$. This range of $\Delta e$ is chosen to span the infinite-temperature regime, the regime of the ladder ensemble, and the energy localized regime. Importantly, we average over 10 samples and the 10 states closest to $\theta = \pi/T$ in each sample, but we flip the sign of $e$ when needed so that states that are imbalanced about $e=0$ are all imbalanced towards negative $e$. We do this so that we don't generate balanced distributions by averaging over oppositely imbalanced states, which would hide signs of energy localization. In Fig.~\ref{fig:E_distributions}(a), we see that at low frequencies the distribution follows an infinite-temperature curve; at moderate frequencies the distribution is still balanced about $e=0$, but is peaked at positive and negative $e$ and not $e=0$, characteristic of the ladder ensemble; at high frequencies the distribution becomes imbalanced about $e=0$, indicating energy localization. In the lower panel, Fig.~\ref{fig:E_distributions}(b), we show the value of the CDF at $e=0$ in red, and the probability density function (PDF) at $e=0$ normalized by the infinite-temperature ($\beta=0$) value in blue. We compute the PDF as the slope of the CDF. In that panel we see that the PDF at $e=0$ begins to be reduced relative to the infinite-temperature value already at $\Delta e \approxeq 0.4$ (for $N=14$), whereas the CDF deviates from the balanced value of $1/2$ around $\Delta e \approxeq 0.8$, in agreement with our earlier finding that energy localization begins to set in near that point (Figs.~\ref{fig:s_f_G_vs_e} and \ref{fig:sign_entropy}). This demonstrates that even in small numerically accessible systems of size $N=14$ the regimes of applicability for the infinite-temperature and ladder ensembles are distinguishable: Floquet thermalization to the ladder ensemble extends significantly beyond the regime in which the system thermalizes to a featureless infinite-temperature state.

\subsection{Floquet heating the ground state}

\begin{figure}
    \centering
    \includegraphics[width=0.9\linewidth]{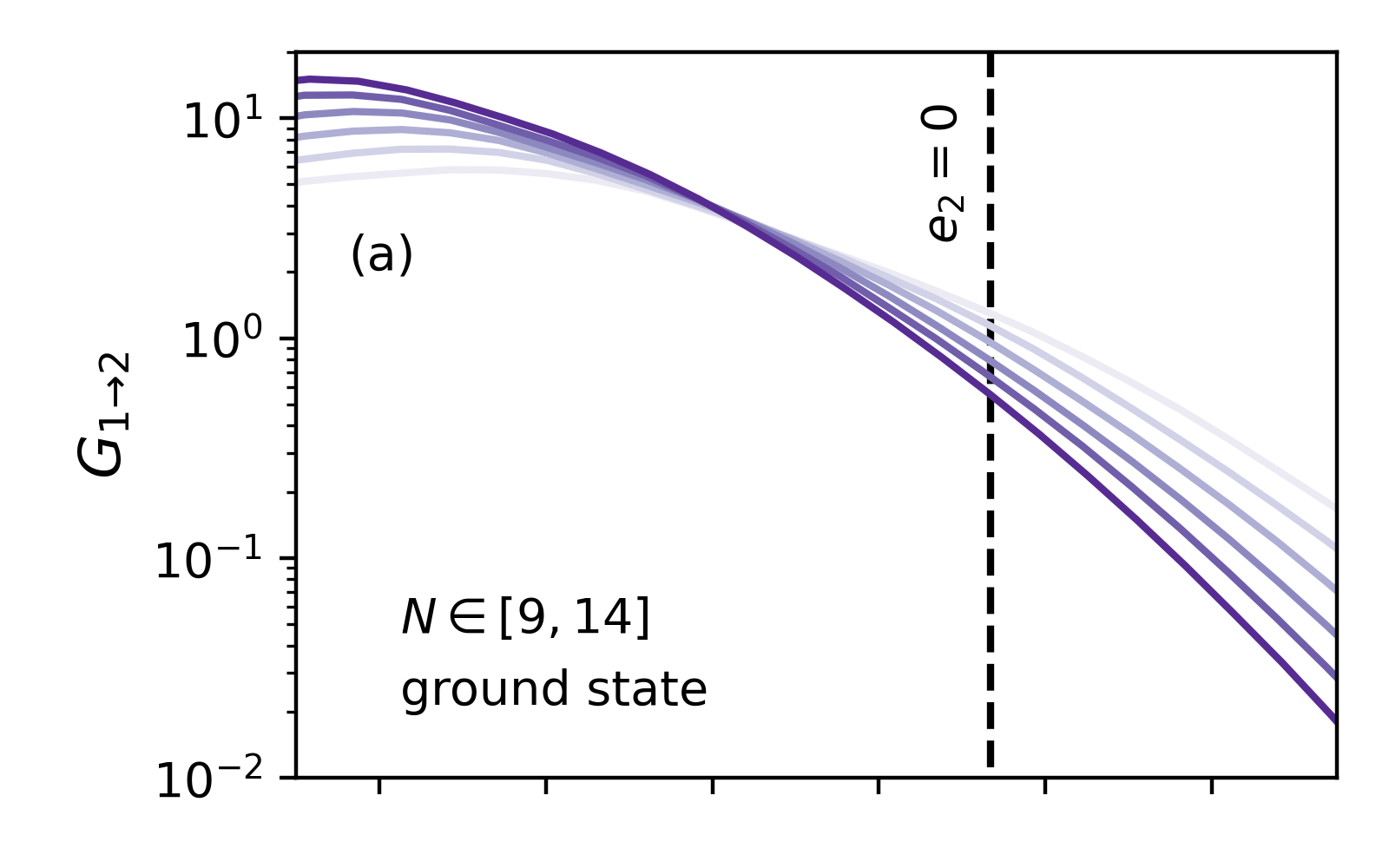}\\
    \vspace{-0.3cm}
    \includegraphics[width=0.9\linewidth]{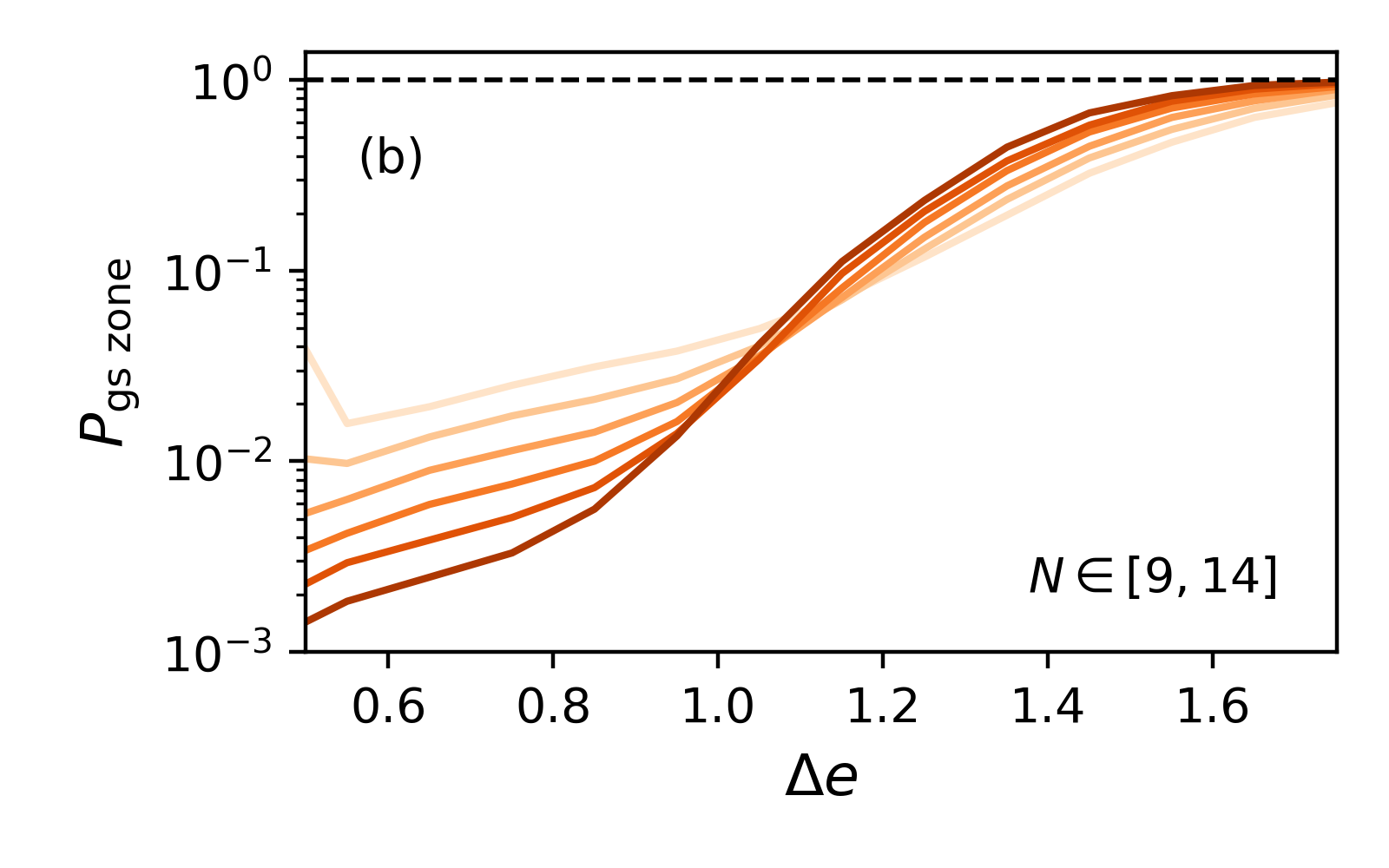}
    \caption{Ground state heating crossover. (a) The quantity $G_{1\to 2}$ (see main text) indicating the validity of Fermi's Golden Rule for heating from the ground state energy density $e_1=e_\mathrm{gs}$ to $e_2=e_\mathrm{gs} + \Delta e$. Darker curves correspond to larger $N$. The vertical dashed line marks the point at which $e_2$ is at the center of the spectrum. The horizontal axis is shared with the bottom panel. (b) The probability that the ground state of $H_0$ stays within its Floquet zone ($e\in [e_\mathrm{gs}, e_\mathrm{gs} + \frac{\Delta e}{2}]$) indefinitely, under time evolution by $U_F$.}
    \label{fig:gs_heating}
\end{figure}

Another scenario of interest is one where the system is initialized in the ground state of $H_0$, and driven at a frequency $\omega = N \Delta e$ so as to potentially cause heating to higher energy densities ($e_1 = e_\mathrm{gs}$ and $e_2 = e_\mathrm{gs} + \Delta e$). Here, too, there is a sharp ``heating crossover'', in the limit of large $N$, as a function of $\Delta e$, but it is of a slightly different nature than the earlier case of $e_2 = -e_1 = \frac{\Delta e}{2}$. In this crossover, the ground state will not serve as a continuum to transition to, from higher energy densities, because of its vanishing entropy density, so the question is whether or not states at $e_2$ appear to be a continuum to the ground state at $e_1$. This question is closely related to the very recently investigated ``Emergence of Fermi's Golden Rule"~\cite{Micklitz-Huse2022_emergence}, where a single state coupled to a pseudocontinuum produced by a finite system was considered. The corresponding crossover is from the one-way Floquet thermalizing phase directly to the energy-localized phase with no resonances, so in Fig.~\ref{fig:gs_heating}(a) we examine the quantity 
\begin{align}
    \label{eq:G23}
    G_{1\to 2} = V(e_1, e_2)^2 D(e_2)^2,
\end{align}
which corresponds to, e.g., the right-side condition in Eq.~\ref{eq:one_way_condition}. There we see an indicated crossover, which sharpens up with increasing $N$, at about $\Delta e \approxeq 0.96$ for the largest system sizes we can access. We also note that the quantity $G_{1\to2}$ corresponds (up to an $O(1)$ factor) to the quantity $\gamma$ from Ref.~\cite{Micklitz-Huse2022_emergence}, which is the unitless tuning parameter of a universal scaling function for the emergence of FGR. That emergence occurs over an $O(1)$ scale in $\log\gamma \sim 2[\log V(e_1,e_2) + \log D(e_2)]\propto N$, consistent with the aforementioned statement that the width of the crossovers at $\omega \propto N$ studied in this work are asymptotically $\propto N^{-1}$.

Again as an independent check of the sharpening crossover in Fig.~\ref{fig:gs_heating}(a), we also compute the probability that the ground state of $H_0$ will remain indefinitely within the initial Floquet zone under the full dynamics of $U_F$. We define the boundary of the ground state's Floquet zone as $E = E_\mathrm{gs} + \frac{\omega}{2}$. This is shown in Fig.~\ref{fig:gs_heating}(b), and we see that there is a finite-size crossing in agreement with our analysis of $G_{1\to 2}$.

\subsection{The full phase diagram}

So far, we have considered two cuts through the full $(e_1,e_2)$-plane, and examined the crossovers that occur along them while the (extensive) frequency of the drive is tuned. These crossovers separate different degrees of thermalization that can occur in Floquet systems. We now expand our analysis to the full 2D phase diagram in this plane, with all of the phases discussed in Sec.~\ref{sec:theory}. A notable feature of the phase diagram is the strong dependence of phase boundaries on the Floquet quasienergy (set by $e_1, e_2$), an issue that is mostly ignored in conventional studies of Floquet systems that typically do not explore the dependence on quasienergy.

The phase diagram is shown in Fig.~\ref{fig:phase_diagram_one_way_two_way}. The solid lines and dots on that diagram are obtained from $N\in\{12,14\}$ data, and the dashed lines are sketched extrapolations. The boundary between the phases with two-way and one-way Floquet thermalization marks the point at which $G_{1\leftrightarrow 2}$ goes from increasing with $N$ to decreasing. Similarly, the boundary between the one-way Floquet thermalizing phase and the energy-localized phase with isolated resonances marks where $G_{1\rightarrow 2}$ goes from increasing to decreasing with $N$. The union of the two-way and one-way phases are where full or partial interzone thermalization occurs, e.g., Floquet heating, so the blue boundary in Fig.~\ref{fig:phase_diagram_one_way_two_way} is the most experimentally relevant. The middle and edge of this boundary are the crossovers observed in Figs.~\ref{fig:s_f_G_vs_e} and \ref{fig:gs_heating}. Lastly, the boundary between having isolated resonances and not having Floquet resonances is marked by the quantity $G_r=V(e_1,e_2) D(e_1) D(e_2)$ increasing/decreasing with $N$, because $G_r$ corresponds to the condition on the right side of Eq.~\ref{eq:resonance_condition}.

\section{Experimental considerations \label{sec:experiment}}
Since one inspiration for our work is that many experimental platforms for exploring isolated quantum many-body physics are mesoscopic in size, in this section we discuss some experimental considerations and simulate an experiment that could be performed on such small near-term platforms. 

Ideally (in theory), we are studying features of systems that occur at large $N$-dependent drive frequencies. Floquet thermalization rates (when Floquet thermalization does occur) at those frequencies are exponentially suppressed in $\omega\propto \Omega(N)$, where $\Omega(N)$ is an increasing function of $N$, and this slowness is one of the main obstacles for any experimental realizations of the physics discussed in this work. However, as we show in this section, for small systems it may be possible to observe some nontrivial phenomena on experimentally realistic timescales. Other obstacles for potential experiments are that the resolution with which the energy distribution can be probed can be limited by both the initial states that can be prepared and the type of measurements that are possible. Platforms with an intermediate number (tens) of qubits, atoms, ions etc., long coherence times (in units of the relevant interaction time), and ``site-resolved" measurement capabilities would be the most appropriate for experimental explorations of some of the physics presented in this paper.

As an example, here we explore the specific goal of probing the kind of ladder-like Floquet thermal equilibrium shown in Fig.~\ref{fig:ladder}(c) and Fig.~\ref{fig:sign_entropy}(a), where the average energy density corresponds to infinite temperature, but the system actually relaxes to a linear combination of positive and negative temperature states, with almost no weight on states with zero energy density---a situation quite different from the conventional featureless infinite-temperature ensemble. For the purpose of demonstration, we continue to use the $N$-qubit model detailed in Sec.~\ref{sec:num}.

The system must be in the Floquet thermalizing phase for this to work, so first we choose a value for $\Delta e = \frac{\omega}{N}$ such that the system will eventually exchange energy with the drive. The initial state will be a pure state with energy density $-\frac{\Delta e}{2}$, and must have energy density uncertainty $\ll \Delta e$. We then allow the system to evolve for long enough times that Floquet thermalization occurs and the system populates a ladder of distinct energy densities. In this situation, while $\langle H_0 \rangle / N$ converges to zero, the variance of the energy density is $O(1)$ rather than $\propto \frac{1}{N}$, as is the case in the featureless infinite-temperature ensemble. Ideally, this can be observed via few-body energy-density correlations; some variation of the system size would be needed in order to differentiate a small $O(1)$ value from a value that is vanishing with $N$. However, a simpler protocol would be of more practical value, and that is what we discuss below. 

Projectively measuring the global energy $H_0$ is not realistically possible, so the distribution of $H_0$ (not to mention some ideal $H_\mathrm{eff}$) cannot be probed exactly. However, if a substantial part of the Hamiltonian can be measured, e.g., $H_X = g\sum_i X_i$ in our model (see Sec.~\ref{sec:num}), then this simple piece can act as a good proxy for the global energy in some respects. For example, in our model, extensive energy fluctuations can be detected through long-range $XX$ correlations in the system. Better yet, one can look at the full distribution (``full counting statistics"~\cite{Wei-Zeiher2021_kpz,levitov-Lesovik1993_charge,Groha-Calabrese2018_fcs,Gopalakrishnan-Khemani2022_xxz}) of measurement outcomes of the total $X$ magnetization $X_\mathrm{tot} = \sum_i X_i$, which is a realistic observable in many platforms that can take simultaneous snapshots of all degrees of freedom in the system. [Alternatively, one could measure all $Z$ operators and thus look at that component of the energy. But we will focus on the $X$ part of the energy, since one-qubit operators can generally be measured with higher fidelity.]

\begin{figure}
    \centering
    \includegraphics[width=0.9\linewidth]{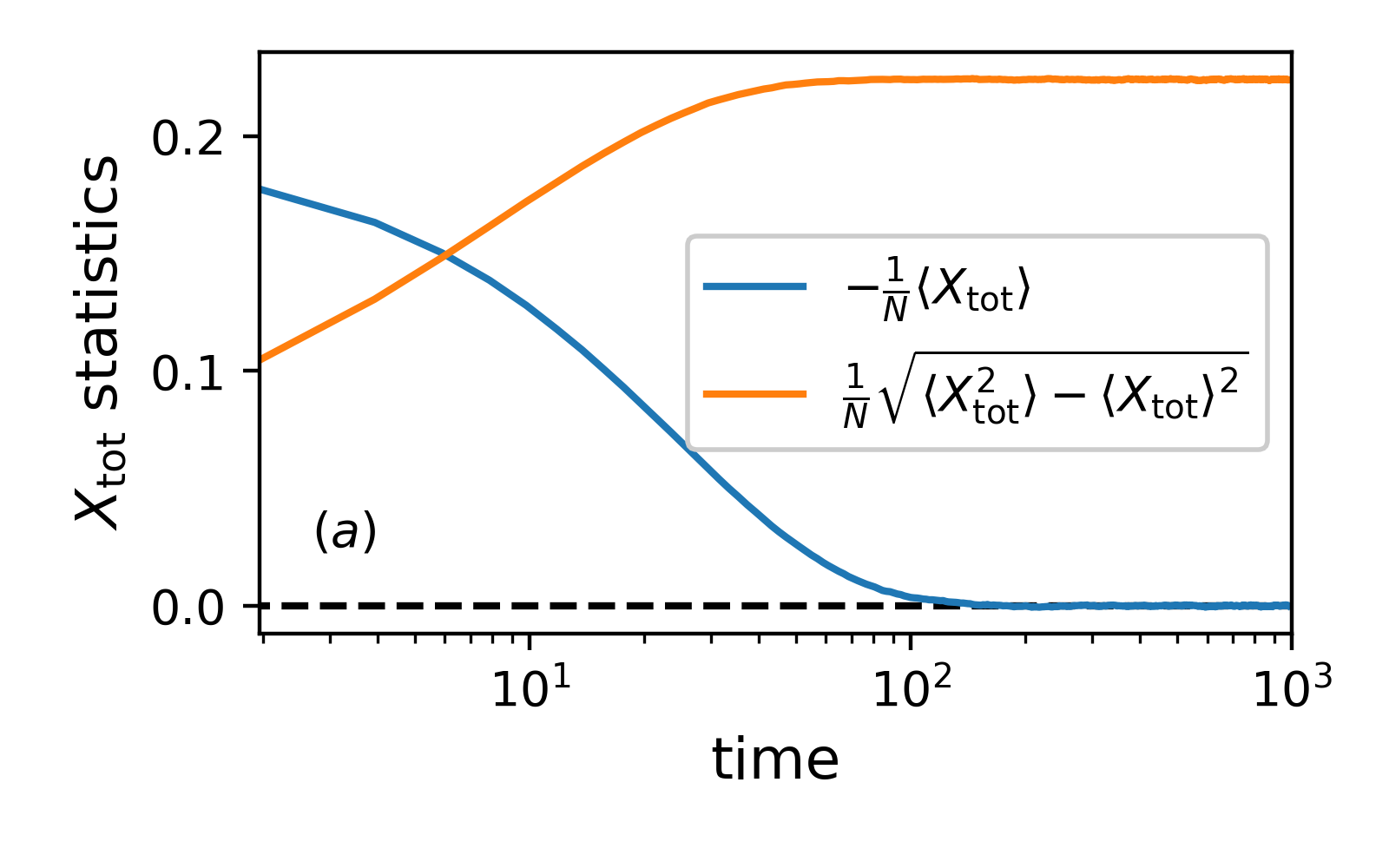}\\
    \vspace{-0.5cm}
    \includegraphics[width=0.9\linewidth]{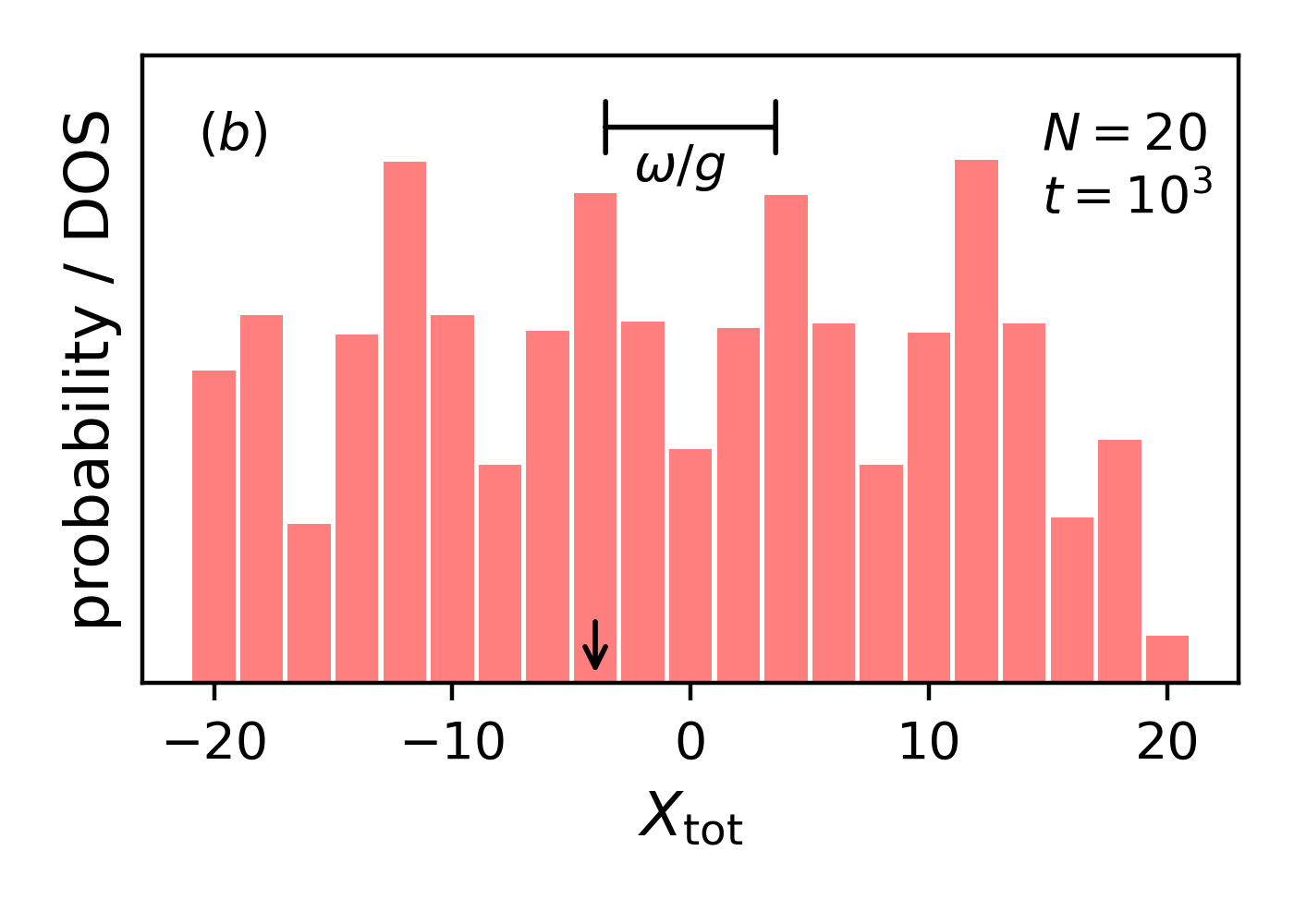}
    \caption{The distribution of measurements of $X_\mathrm{tot}$. The data is for one sample of the system described in the main text of Sec.~\ref{sec:experiment}, with $N=20$. (a) The time evolution of the (negative) mean and standard deviation of $X_\mathrm{tot}/N$. Time is measured in microscopic units, not Floquet cycles. (b) The full distribution of outcomes when measuring $X_\mathrm{tot}$ at time $t=10^3$. The initial state is an eigenstate of $X_\mathrm{tot}$ with eigenvalue indicated by the small black arrow.} 
    \label{fig:experiment}
\end{figure}

The above discussion suggests that we should set $g > 1$ so that $H_X$ is a larger component of $H_0$, so here we set $g=2$, which turns out to be sufficient for the present purposes. We consider the unitary evolution of the initial pure state $|\psi(0)\rangle = |+\rangle^{N_+} |-\rangle^{N - N_+}$, i.e., $N_+$ qubits in the $X=+1$ state and $N-N_+$ in the $X=-1$ state. This initial state has energy density $e_1 = -g \left(1-\frac{2N_+}{N}\right)$ and energy density uncertainty $\frac{1}{\sqrt{N}}$ (the prefactor is 1). We drive the system at frequency $\omega = 2N|e_1|$ to induce transitions to $e_2 = -e_1$ and other energies in the ``ladder". In Fig.~\ref{fig:experiment} we show data for a system with $N=20$ qubits, initialized in the aforementioned state with $N_+ = 8$, and driven at the frequency $\omega = 16$. Importantly, this frequency is not so large that we should worry about coupling the state space of the idealized model to other ``non-model" states that would exist in a real experimental system.
For the above values of $N$, $N_+$, and $\omega$, the initial state has $\frac{\sigma_E}{2|E|} \approxeq 0.28$, so we can hope to resolve an energy-ladder-like equilibrium at long times by measuring $X_i$ on all of the qubits simultaneously. We limit ourselves to the times $t \le 10^3$ so as to not require times far beyond what may be experimentally feasible on near-term platforms. The top panel (a) shows the time evolution of the mean and standard deviation of $X_\mathrm{tot}$, both normalized by $N$ based on the expectation of extensive fluctuations. The bottom panel (b) shows the full quantum distribution, at $t=10^3$, of $X_\mathrm{tot}$ normalized by the DOS of $X_\mathrm{tot}$, so that the state $\rho\propto I$ would result in a flat graph. The distribution clearly shows a ladder of spacing $\frac{\omega}{g}$ that develops, which corresponds to an energy ladder of spacing $\omega$. 

From the above numerical experiment, we conclude that some aspects of the more complete description of Floquet thermalization that we have provided in this work should be experimentally accessible on near-term quantum simulation platforms.

\section{Summary and discussion \label{sec:conclusions}}

In this paper we present a more complete picture of Floquet thermalization and its absence in many-body quantum systems than was previously established, thus advancing our understanding of fundamental aspects of isolated periodically driven systems. By allowing the drive frequency $\omega$ to scale up with $N$, we identify a variety of phases, separated by crossovers that sharpen as $N\to\infty$; these are representative of different regimes of thermalization that can be present in mesoscopic Floquet systems. We find a new Floquet thermal ensemble, the ladder ensemble, which is qualitatively distinct from the featureless infinite-temperature state that is achieved in the conventionally studied case where $\omega$ remains finite as $N\to \infty$. In fact, we show that the conventional infinite-temperature ensemble is valid only in a vanishing fraction of the phase diagram where Floquet thermalization does occur. Thus, we show that Floquet thermalization does \emph{not} imply that the system thermalizes to infinite temperature, even on average.   Two of the phases we detail host full or partial Floquet thermalization (delocalization in energy), but their equilibrium distribution over energy is qualitatively different from an infinite-temperature state. The other two phases fail to Floquet thermalize and thus are localized in energy, but are distinguished by the presence or absence of rare Floquet many-body resonances in typical samples. We present a phase diagram delineating these various regimes, and the phase boundaries show strong dependence on Floquet quasienergy, or the energy of an initial state, a feature that is mostly ignored in conventional studies. 

We also explored the feasibility of experimentally observing our findings, and found that some interesting types of Floquet thermalization elucidated in this work should be realizable on near-term quantum simulation platforms. In particular, we simulated an experiment in which a small Floquet system thermalizes on an accessible time scale to a superposition of thermal states at a set of evenly spaced and extensively different energies.

One of our primary motivations for studying Floquet thermalization in mesoscopic systems is that many physically relevant settings only have access to small or moderate-size systems, and our theoretical understanding of thermalization and equilibrium in this setting is still quite incomplete, as illustrated by our findings. These settings include numerical experiments limited to small sizes, and experiments on near-term quantum simulators operating in the intermediate-scale regime~\cite{Preskill2018_nisq}. For example, recent works have studied Trotter approximations for digital quantum simulation~\cite{Heyl-Zoller2019_trotter, Sieberer-Zoller2019_kicked, Kargi-Landford2021_performance}, and identified thresholds as a function of the Trotter step size separating regimes with controllable and uncontrolled Trotter errors. These trotterization protocols are simply Floquet evolutions with frequency controlled by the step size, and our work elucidates the theoretical basis behind these topical studies: the observed thresholds are simply Floquet heating thresholds that occur at an extensive frequency, $\omega\propto N$, a scaling that was not identified in these works.

A second motivation for our work draws on the fact that finite-size and -time crossovers between fully- and partially- or non-thermalizing regimes have become increasingly important in a number of contexts, most notably in the study of many-body localization,  but also in, e.g., integrability breaking~\cite{Bulchandani-Gopalakrishnan2021_onset}. 
Here, recent work has led to the understanding that the numerically and experimentally observed crossovers between thermalizing and MBL-like regimes have distinct physics from that of the asymptotic phase transition~\cite{Crowley-Chandran2021,Morningstar-Huse2021_avalanches, Long-Chandran2022_jacobi}. Indeed, an MBL-to-thermal phase transition may not even exist in the standard thermodynamic limit in higher dimensions or with power-law interactions, even though a crossover is clearly observed in these settings~\cite{Burin2015_long_range,BlochMBL2D_2016,Tikhonov-Mirlin2018_power_law,Gopalakrishnan-Huse2019_instability}. Understanding the universal properties of such crossovers is an important open question, and our hope is that developing a more complete theory of the finite-size crossovers associated with the onset of Floquet heating and rare Floquet resonances will also prove illuminating for these other, arguably more challenging, cases. 

We note that the crossovers we have studied in this work are examples of boundaries between different types of thermal phases, since even the energy localized phase is described by a chaotic $H_{\rm eff}$. In contrast, the relevant processes for crossovers in to or between many-body localized regimes have extremely broadly distributed matrix elements that may need to be taken in to account~\cite{Morningstar-Huse2021_avalanches,Garratt-Roy2022_resonant}. Nonetheless, both Floquet thermalization crossovers studied in this work, and the observable crossover between thermal and MBL regimes, have some similarities. For example, these crossovers have \textit{apparent} critical exponents $\nu\cong 1$. In the case of Floquet thermalization, (setting some factors to $1$ for brevity) this is a result of the interzone relaxation time $\Gamma^{-1} \sim \euler^{\omega}=\euler^{(\omega/N)N}$ passing through the relevant Heisenberg time $\tau_\mathrm{H} \sim \euler^{s N}$ as a function of the control parameter $\omega/N$, as discussed above. In the case of the numerically observable MBL crossover, the apparent $\nu\cong 1$ in finite-size systems is similarly a result of a thermalization time that behaves as $\sim\euler^{k(W)N}$ near the crossover~\cite{Suntajs-Vidmar2020_quantum_chaos,Suntajs-Vidmar2020_ergodicity_breaking,Sels-Polkovnikov2021_obstruction,Sierant-Zakrzewski2020_thouless_time,Morningstar-Huse2021_avalanches,Crowley-Chandran2021, Long-Chandran2022_jacobi} passing through the Heisenberg time as a function of the control parameter $W$.  It may be that understanding the simpler cases of thermal-to-thermal crossovers, as we have done in this work, will provide some insights that are helpful for addressing the more challenging situations where the crossover involves nonthermal states. 

In future studies it would be interesting to realize some of the phenomenology explored in this work experimentally, as we have shown that it should be within reach of some current experimental platforms for quantum simulation. It would also be interesting to consider Floquet systems where $H_0$, the time-averaged Hamiltonian, is not fully chaotic, as an intermediate problem where the energy localized phase may thermalize only asymptotically slowly, or not at all. Another interesting direction is to consider an $H_0$ with quantum scars to explore the thermalization of the scar state across Floquet zones as a function of the drive frequency and energy of the scar state. Further studies of the dynamics throughout the phase diagram that we have mapped out here would also be interesting, as our work is primarily concerned with the final equilibrium that is achieved. Finally, many of our findings should have analogs in static settings where, for example, a system is governed by a Hamiltonian $\hat{H} = \hat{K} + \Delta \hat{M}$, where $\hat{M}$ has an integer spectrum, $\Delta$ is a large parameter, and $\hat{K}$ couples states within and across eigenspaces of $\hat{M}$~\cite{Abanin-Huveneers2017_rigorous}. Examples of such settings include systems with a weakly broken $U(1)$ symmetry corresponding to conservation of total charge, or scarred Hamiltonians derived from parent Hamiltonians with a spectrum generating algebra~\cite{Mark20a, Moudgalya20, O'Dea20}, or systems with approximate Hilbert space fragmentation due to weakly broken conservation of charge and dipole moment~\cite{Khemani20_shatter, Sala20}.

\begin{acknowledgments}
We thank Immanuel Bloch, Vir Bulchandani, Anushya Chandran, Sam Garratt, Sarang Gopalakrishnan, Anatoli Polkovnikov, Marcos Rigol,  David Wei and Johannes Zeiher for helpful discussions and/or previous collaborations. A.M. was supported in part by the DARPA DRINQS program, the Stanford Q-FARM Bloch Postdoctoral Fellowship in Quantum Science and Engineering, and the Gordon and Betty Moore Foundation’s EPiQS Initiative through Grant GBMF8686. D.A.H. and A.M. were supported in part by NSF QLCI grant OMA-2120757.  V.K. was supported by the US Department of Energy, Office of Science, Basic Energy Sciences, under Early Career Award Nos. DE-SC0021111, the Alfred P. Sloan Foundation through a Sloan Research Fellowship and the Packard Foundation through a Packard Fellowship in Science and Engineering.  Simulations presented in this work were performed on computational resources managed and supported by Princeton Research Computing. We also acknowledge the hospitality of the Kavli Institute for Theoretical Physics at the University of California, Santa Barbara (supported by NSF Grant PHY-1748958). 

\end{acknowledgments}

\bibliography{main}

\begin{thebibliography}{90}%
\makeatletter
\providecommand \@ifxundefined [1]{%
 \@ifx{#1\undefined}
}%
\providecommand \@ifnum [1]{%
 \ifnum #1\expandafter \@firstoftwo
 \else \expandafter \@secondoftwo
 \fi
}%
\providecommand \@ifx [1]{%
 \ifx #1\expandafter \@firstoftwo
 \else \expandafter \@secondoftwo
 \fi
}%
\providecommand \natexlab [1]{#1}%
\providecommand \enquote  [1]{``#1''}%
\providecommand \bibnamefont  [1]{#1}%
\providecommand \bibfnamefont [1]{#1}%
\providecommand \citenamefont [1]{#1}%
\providecommand \href@noop [0]{\@secondoftwo}%
\providecommand \href [0]{\begingroup \@sanitize@url \@href}%
\providecommand \@href[1]{\@@startlink{#1}\@@href}%
\providecommand \@@href[1]{\endgroup#1\@@endlink}%
\providecommand \@sanitize@url [0]{\catcode `\\12\catcode `\$12\catcode
  `\&12\catcode `\#12\catcode `\^12\catcode `\_12\catcode `\%12\relax}%
\providecommand \@@startlink[1]{}%
\providecommand \@@endlink[0]{}%
\providecommand \url  [0]{\begingroup\@sanitize@url \@url }%
\providecommand \@url [1]{\endgroup\@href {#1}{\urlprefix }}%
\providecommand \urlprefix  [0]{URL }%
\providecommand \Eprint [0]{\href }%
\providecommand \doibase [0]{http://dx.doi.org/}%
\providecommand \selectlanguage [0]{\@gobble}%
\providecommand \bibinfo  [0]{\@secondoftwo}%
\providecommand \bibfield  [0]{\@secondoftwo}%
\providecommand \translation [1]{[#1]}%
\providecommand \BibitemOpen [0]{}%
\providecommand \bibitemStop [0]{}%
\providecommand \bibitemNoStop [0]{.\EOS\space}%
\providecommand \EOS [0]{\spacefactor3000\relax}%
\providecommand \BibitemShut  [1]{\csname bibitem#1\endcsname}%
\let\auto@bib@innerbib\@empty
\bibitem [{\citenamefont {D'Alessio}\ \emph {et~al.}(2016)\citenamefont
  {D'Alessio}, \citenamefont {Kafri}, \citenamefont {Polkovnikov},\ and\
  \citenamefont {Rigol}}]{DAlessio-Rigol2016_review}%
  \BibitemOpen
  \bibfield  {author} {\bibinfo {author} {\bibfnamefont {Luca}\ \bibnamefont
  {D'Alessio}}, \bibinfo {author} {\bibfnamefont {Yariv}\ \bibnamefont
  {Kafri}}, \bibinfo {author} {\bibfnamefont {Anatoli}\ \bibnamefont
  {Polkovnikov}}, \ and\ \bibinfo {author} {\bibfnamefont {Marcos}\
  \bibnamefont {Rigol}},\ }\bibfield  {title} {\enquote {\bibinfo {title} {From
  quantum chaos and eigenstate thermalization to statistical mechanics and
  thermodynamics},}\ }\href {\doibase 10.1080/00018732.2016.1198134} {\bibfield
   {journal} {\bibinfo  {journal} {Advances in Physics}\ }\textbf {\bibinfo
  {volume} {65}},\ \bibinfo {pages} {239--362} (\bibinfo {year}
  {2016})}\BibitemShut {NoStop}%
\bibitem [{\citenamefont {Nandkishore}\ and\ \citenamefont
  {Huse}(2015)}]{Nandkishore-Huse2015_review}%
  \BibitemOpen
  \bibfield  {author} {\bibinfo {author} {\bibfnamefont {Rahul}\ \bibnamefont
  {Nandkishore}}\ and\ \bibinfo {author} {\bibfnamefont {David~A.}\
  \bibnamefont {Huse}},\ }\bibfield  {title} {\enquote {\bibinfo {title}
  {Many-body localization and thermalization in quantum statistical
  mechanics},}\ }\href {\doibase 10.1146/annurev-conmatphys-031214-014726}
  {\bibfield  {journal} {\bibinfo  {journal} {Annual Review of Condensed Matter
  Physics}\ }\textbf {\bibinfo {volume} {6}},\ \bibinfo {pages} {15--38}
  (\bibinfo {year} {2015})}\BibitemShut {NoStop}%
\bibitem [{\citenamefont {Abanin}\ \emph {et~al.}(2019)\citenamefont {Abanin},
  \citenamefont {Altman}, \citenamefont {Bloch},\ and\ \citenamefont
  {Serbyn}}]{Abanin-Serbyn2019_review}%
  \BibitemOpen
  \bibfield  {author} {\bibinfo {author} {\bibfnamefont {Dmitry~A.}\
  \bibnamefont {Abanin}}, \bibinfo {author} {\bibfnamefont {Ehud}\ \bibnamefont
  {Altman}}, \bibinfo {author} {\bibfnamefont {Immanuel}\ \bibnamefont
  {Bloch}}, \ and\ \bibinfo {author} {\bibfnamefont {Maksym}\ \bibnamefont
  {Serbyn}},\ }\bibfield  {title} {\enquote {\bibinfo {title} {Colloquium:
  Many-body localization, thermalization, and entanglement},}\ }\href {\doibase
  10.1103/RevModPhys.91.021001} {\bibfield  {journal} {\bibinfo  {journal}
  {Rev. Mod. Phys.}\ }\textbf {\bibinfo {volume} {91}},\ \bibinfo {pages}
  {021001} (\bibinfo {year} {2019})}\BibitemShut {NoStop}%
\bibitem [{\citenamefont {Alet}\ and\ \citenamefont
  {Laflorencie}(2018)}]{Alet-Laflorencie2018_review}%
  \BibitemOpen
  \bibfield  {author} {\bibinfo {author} {\bibfnamefont {Fabien}\ \bibnamefont
  {Alet}}\ and\ \bibinfo {author} {\bibfnamefont {Nicolas}\ \bibnamefont
  {Laflorencie}},\ }\bibfield  {title} {\enquote {\bibinfo {title} {Many-body
  localization: {An} introduction and selected topics},}\ }\href {\doibase
  10.1016/j.crhy.2018.03.003} {\bibfield  {journal} {\bibinfo  {journal}
  {Comptes Rendus Physique}\ }\bibinfo {series} {Quantum simulation /
  {Simulation} quantique},\ \textbf {\bibinfo {volume} {19}},\ \bibinfo {pages}
  {498--525} (\bibinfo {year} {2018})}\BibitemShut {NoStop}%
\bibitem [{\citenamefont {Serbyn}\ \emph {et~al.}(2021)\citenamefont {Serbyn},
  \citenamefont {Abanin},\ and\ \citenamefont
  {Papi{\'c}}}]{Serbyn-Papic2021_review}%
  \BibitemOpen
  \bibfield  {author} {\bibinfo {author} {\bibfnamefont {Maksym}\ \bibnamefont
  {Serbyn}}, \bibinfo {author} {\bibfnamefont {Dmitry~A.}\ \bibnamefont
  {Abanin}}, \ and\ \bibinfo {author} {\bibfnamefont {Zlatko}\ \bibnamefont
  {Papi{\'c}}},\ }\bibfield  {title} {\enquote {\bibinfo {title} {Quantum
  many-body scars and weak breaking of ergodicity},}\ }\href {\doibase
  10.1038/s41567-021-01230-2} {\bibfield  {journal} {\bibinfo  {journal}
  {Nature Physics}\ }\textbf {\bibinfo {volume} {17}},\ \bibinfo {pages}
  {675--685} (\bibinfo {year} {2021})}\BibitemShut {NoStop}%
\bibitem [{\citenamefont {Moudgalya}\ \emph {et~al.}(2022)\citenamefont
  {Moudgalya}, \citenamefont {Bernevig},\ and\ \citenamefont
  {Regnault}}]{Moudgalya-Regnault2021_review}%
  \BibitemOpen
  \bibfield  {author} {\bibinfo {author} {\bibfnamefont {Sanjay}\ \bibnamefont
  {Moudgalya}}, \bibinfo {author} {\bibfnamefont {B~Andrei}\ \bibnamefont
  {Bernevig}}, \ and\ \bibinfo {author} {\bibfnamefont {Nicolas}\ \bibnamefont
  {Regnault}},\ }\bibfield  {title} {\enquote {\bibinfo {title} {Quantum
  many-body scars and hilbert space fragmentation: a review of exact
  results},}\ }\href {\doibase 10.1088/1361-6633/ac73a0} {\bibfield  {journal}
  {\bibinfo  {journal} {Reports on Progress in Physics}\ }\textbf {\bibinfo
  {volume} {85}},\ \bibinfo {pages} {086501} (\bibinfo {year}
  {2022})}\BibitemShut {NoStop}%
\bibitem [{\citenamefont {Chandran}\ \emph {et~al.}(2023)\citenamefont
  {Chandran}, \citenamefont {Iadecola}, \citenamefont {Khemani},\ and\
  \citenamefont {Moessner}}]{Chandran-Moessner2022_review}%
  \BibitemOpen
  \bibfield  {author} {\bibinfo {author} {\bibfnamefont {Anushya}\ \bibnamefont
  {Chandran}}, \bibinfo {author} {\bibfnamefont {Thomas}\ \bibnamefont
  {Iadecola}}, \bibinfo {author} {\bibfnamefont {Vedika}\ \bibnamefont
  {Khemani}}, \ and\ \bibinfo {author} {\bibfnamefont {Roderich}\ \bibnamefont
  {Moessner}},\ }\bibfield  {title} {\enquote {\bibinfo {title} {Quantum
  many-body scars: A quasiparticle perspective},}\ }\href {\doibase
  10.1146/annurev-conmatphys-031620-101617} {\bibfield  {journal} {\bibinfo
  {journal} {Annual Review of Condensed Matter Physics}\ }\textbf {\bibinfo
  {volume} {14}},\ \bibinfo {pages} {443--469} (\bibinfo {year}
  {2023})}\BibitemShut {NoStop}%
\bibitem [{\citenamefont {Jensen}\ and\ \citenamefont
  {Shankar}(1985)}]{Jensen-Shankar1985_statistical}%
  \BibitemOpen
  \bibfield  {author} {\bibinfo {author} {\bibfnamefont {R.~V.}\ \bibnamefont
  {Jensen}}\ and\ \bibinfo {author} {\bibfnamefont {R.}~\bibnamefont
  {Shankar}},\ }\bibfield  {title} {\enquote {\bibinfo {title} {Statistical
  behavior in deterministic quantum systems with few degrees of freedom},}\
  }\href {\doibase 10.1103/PhysRevLett.54.1879} {\bibfield  {journal} {\bibinfo
   {journal} {Phys. Rev. Lett.}\ }\textbf {\bibinfo {volume} {54}},\ \bibinfo
  {pages} {1879--1882} (\bibinfo {year} {1985})}\BibitemShut {NoStop}%
\bibitem [{\citenamefont {Deutsch}(1991)}]{Deutsch1991_quantum}%
  \BibitemOpen
  \bibfield  {author} {\bibinfo {author} {\bibfnamefont {J.~M.}\ \bibnamefont
  {Deutsch}},\ }\bibfield  {title} {\enquote {\bibinfo {title} {{Quantum
  statistical mechanics in a closed system}},}\ }\href {\doibase
  10.1103/PhysRevA.43.2046} {\bibfield  {journal} {\bibinfo  {journal} {Phys.
  Rev. A}\ }\textbf {\bibinfo {volume} {43}},\ \bibinfo {pages} {2046--2049}
  (\bibinfo {year} {1991})}\BibitemShut {NoStop}%
\bibitem [{\citenamefont {Srednicki}(1994)}]{Srednicki1994_chaos}%
  \BibitemOpen
  \bibfield  {author} {\bibinfo {author} {\bibfnamefont {Mark}\ \bibnamefont
  {Srednicki}},\ }\bibfield  {title} {\enquote {\bibinfo {title} {Chaos and
  quantum thermalization},}\ }\href {\doibase 10.1103/PhysRevE.50.888}
  {\bibfield  {journal} {\bibinfo  {journal} {Phys. Rev. E}\ }\textbf {\bibinfo
  {volume} {50}},\ \bibinfo {pages} {888--901} (\bibinfo {year}
  {1994})}\BibitemShut {NoStop}%
\bibitem [{\citenamefont {Rigol}\ \emph {et~al.}(2008)\citenamefont {Rigol},
  \citenamefont {Dunjko},\ and\ \citenamefont
  {Olshanii}}]{Rigol-Olshanii2008_thermalization}%
  \BibitemOpen
  \bibfield  {author} {\bibinfo {author} {\bibfnamefont {Marcos}\ \bibnamefont
  {Rigol}}, \bibinfo {author} {\bibfnamefont {Vanja}\ \bibnamefont {Dunjko}}, \
  and\ \bibinfo {author} {\bibfnamefont {Maxim}\ \bibnamefont {Olshanii}},\
  }\bibfield  {title} {\enquote {\bibinfo {title} {Thermalization and its
  mechanism for generic isolated quantum systems},}\ }\href {\doibase
  10.1038/nature06838} {\bibfield  {journal} {\bibinfo  {journal} {Nature}\
  }\textbf {\bibinfo {volume} {452}},\ \bibinfo {pages} {854--858} (\bibinfo
  {year} {2008})}\BibitemShut {NoStop}%
\bibitem [{\citenamefont {Srednicki}(1996)}]{Srednicki1996_thermal}%
  \BibitemOpen
  \bibfield  {author} {\bibinfo {author} {\bibfnamefont {Mark}\ \bibnamefont
  {Srednicki}},\ }\bibfield  {title} {\enquote {\bibinfo {title} {Thermal
  fluctuations in quantized chaotic systems},}\ }\href {\doibase
  10.1088/0305-4470/29/4/003} {\bibfield  {journal} {\bibinfo  {journal}
  {Journal of Physics A: Mathematical and General}\ }\textbf {\bibinfo {volume}
  {29}},\ \bibinfo {pages} {L75--L79} (\bibinfo {year} {1996})}\BibitemShut
  {NoStop}%
\bibitem [{\citenamefont {Srednicki}(1999)}]{Srednicki1999_approach}%
  \BibitemOpen
  \bibfield  {author} {\bibinfo {author} {\bibfnamefont {Mark}\ \bibnamefont
  {Srednicki}},\ }\bibfield  {title} {\enquote {\bibinfo {title} {The approach
  to thermal equilibrium in quantized chaotic systems},}\ }\href {\doibase
  10.1088/0305-4470/32/7/007} {\bibfield  {journal} {\bibinfo  {journal}
  {Journal of Physics A: Mathematical and General}\ }\textbf {\bibinfo {volume}
  {32}},\ \bibinfo {pages} {1163--1175} (\bibinfo {year} {1999})}\BibitemShut
  {NoStop}%
\bibitem [{\citenamefont {Rigol}(2009{\natexlab{a}})}]{Rigol2009_breakdown}%
  \BibitemOpen
  \bibfield  {author} {\bibinfo {author} {\bibfnamefont {Marcos}\ \bibnamefont
  {Rigol}},\ }\bibfield  {title} {\enquote {\bibinfo {title} {Breakdown of
  thermalization in finite one-dimensional systems},}\ }\href {\doibase
  10.1103/PhysRevLett.103.100403} {\bibfield  {journal} {\bibinfo  {journal}
  {Phys. Rev. Lett.}\ }\textbf {\bibinfo {volume} {103}},\ \bibinfo {pages}
  {100403} (\bibinfo {year} {2009}{\natexlab{a}})}\BibitemShut {NoStop}%
\bibitem [{\citenamefont {Rigol}(2009{\natexlab{b}})}]{Rigol2009_quantum}%
  \BibitemOpen
  \bibfield  {author} {\bibinfo {author} {\bibfnamefont {Marcos}\ \bibnamefont
  {Rigol}},\ }\bibfield  {title} {\enquote {\bibinfo {title} {Quantum quenches
  and thermalization in one-dimensional fermionic systems},}\ }\href {\doibase
  10.1103/PhysRevA.80.053607} {\bibfield  {journal} {\bibinfo  {journal} {Phys.
  Rev. A}\ }\textbf {\bibinfo {volume} {80}},\ \bibinfo {pages} {053607}
  (\bibinfo {year} {2009}{\natexlab{b}})}\BibitemShut {NoStop}%
\bibitem [{\citenamefont {Beugeling}\ \emph {et~al.}(2014)\citenamefont
  {Beugeling}, \citenamefont {Moessner},\ and\ \citenamefont
  {Haque}}]{Beugeling-Haque2014_finite}%
  \BibitemOpen
  \bibfield  {author} {\bibinfo {author} {\bibfnamefont {W.}~\bibnamefont
  {Beugeling}}, \bibinfo {author} {\bibfnamefont {R.}~\bibnamefont {Moessner}},
  \ and\ \bibinfo {author} {\bibfnamefont {Masudul}\ \bibnamefont {Haque}},\
  }\bibfield  {title} {\enquote {\bibinfo {title} {Finite-size scaling of
  eigenstate thermalization},}\ }\href {\doibase 10.1103/PhysRevE.89.042112}
  {\bibfield  {journal} {\bibinfo  {journal} {Phys. Rev. E}\ }\textbf {\bibinfo
  {volume} {89}},\ \bibinfo {pages} {042112} (\bibinfo {year}
  {2014})}\BibitemShut {NoStop}%
\bibitem [{\citenamefont {Kim}\ \emph {et~al.}(2014)\citenamefont {Kim},
  \citenamefont {Ikeda},\ and\ \citenamefont {Huse}}]{Kim-Huse2014_testing}%
  \BibitemOpen
  \bibfield  {author} {\bibinfo {author} {\bibfnamefont {Hyungwon}\
  \bibnamefont {Kim}}, \bibinfo {author} {\bibfnamefont {Tatsuhiko~N.}\
  \bibnamefont {Ikeda}}, \ and\ \bibinfo {author} {\bibfnamefont {David~A.}\
  \bibnamefont {Huse}},\ }\bibfield  {title} {\enquote {\bibinfo {title}
  {Testing whether all eigenstates obey the eigenstate thermalization
  hypothesis},}\ }\href {\doibase 10.1103/PhysRevE.90.052105} {\bibfield
  {journal} {\bibinfo  {journal} {Phys. Rev. E}\ }\textbf {\bibinfo {volume}
  {90}},\ \bibinfo {pages} {052105} (\bibinfo {year} {2014})}\BibitemShut
  {NoStop}%
\bibitem [{\citenamefont {Mondaini}\ \emph {et~al.}(2016)\citenamefont
  {Mondaini}, \citenamefont {Fratus}, \citenamefont {Srednicki},\ and\
  \citenamefont {Rigol}}]{Mondaini-Rigol2016_eigenstate}%
  \BibitemOpen
  \bibfield  {author} {\bibinfo {author} {\bibfnamefont {Rubem}\ \bibnamefont
  {Mondaini}}, \bibinfo {author} {\bibfnamefont {Keith~R.}\ \bibnamefont
  {Fratus}}, \bibinfo {author} {\bibfnamefont {Mark}\ \bibnamefont
  {Srednicki}}, \ and\ \bibinfo {author} {\bibfnamefont {Marcos}\ \bibnamefont
  {Rigol}},\ }\bibfield  {title} {\enquote {\bibinfo {title} {Eigenstate
  thermalization in the two-dimensional transverse field ising model},}\ }\href
  {\doibase 10.1103/PhysRevE.93.032104} {\bibfield  {journal} {\bibinfo
  {journal} {Phys. Rev. E}\ }\textbf {\bibinfo {volume} {93}},\ \bibinfo
  {pages} {032104} (\bibinfo {year} {2016})}\BibitemShut {NoStop}%
\bibitem [{\citenamefont {Mondaini}\ and\ \citenamefont
  {Rigol}(2017)}]{Mondaini-Rigol2017_off_diagonal}%
  \BibitemOpen
  \bibfield  {author} {\bibinfo {author} {\bibfnamefont {Rubem}\ \bibnamefont
  {Mondaini}}\ and\ \bibinfo {author} {\bibfnamefont {Marcos}\ \bibnamefont
  {Rigol}},\ }\bibfield  {title} {\enquote {\bibinfo {title} {Eigenstate
  thermalization in the two-dimensional transverse field ising model. ii.
  off-diagonal matrix elements of observables},}\ }\href {\doibase
  10.1103/PhysRevE.96.012157} {\bibfield  {journal} {\bibinfo  {journal} {Phys.
  Rev. E}\ }\textbf {\bibinfo {volume} {96}},\ \bibinfo {pages} {012157}
  (\bibinfo {year} {2017})}\BibitemShut {NoStop}%
\bibitem [{\citenamefont {Lazarides}\ \emph
  {et~al.}(2014{\natexlab{a}})\citenamefont {Lazarides}, \citenamefont {Das},\
  and\ \citenamefont {Moessner}}]{Lazarides-Moessner2014_equilibrium}%
  \BibitemOpen
  \bibfield  {author} {\bibinfo {author} {\bibfnamefont {Achilleas}\
  \bibnamefont {Lazarides}}, \bibinfo {author} {\bibfnamefont {Arnab}\
  \bibnamefont {Das}}, \ and\ \bibinfo {author} {\bibfnamefont {Roderich}\
  \bibnamefont {Moessner}},\ }\bibfield  {title} {\enquote {\bibinfo {title}
  {Equilibrium states of generic quantum systems subject to periodic
  driving},}\ }\href {\doibase 10.1103/PhysRevE.90.012110} {\bibfield
  {journal} {\bibinfo  {journal} {Phys. Rev. E}\ }\textbf {\bibinfo {volume}
  {90}},\ \bibinfo {pages} {012110} (\bibinfo {year}
  {2014}{\natexlab{a}})}\BibitemShut {NoStop}%
\bibitem [{\citenamefont {D'Alessio}\ and\ \citenamefont
  {Rigol}(2014)}]{DAlessio-Rigol2014_long-time}%
  \BibitemOpen
  \bibfield  {author} {\bibinfo {author} {\bibfnamefont {Luca}\ \bibnamefont
  {D'Alessio}}\ and\ \bibinfo {author} {\bibfnamefont {Marcos}\ \bibnamefont
  {Rigol}},\ }\bibfield  {title} {\enquote {\bibinfo {title} {Long-time
  behavior of isolated periodically driven interacting lattice systems},}\
  }\href {\doibase 10.1103/PhysRevX.4.041048} {\bibfield  {journal} {\bibinfo
  {journal} {Phys. Rev. X}\ }\textbf {\bibinfo {volume} {4}},\ \bibinfo {pages}
  {041048} (\bibinfo {year} {2014})}\BibitemShut {NoStop}%
\bibitem [{\citenamefont {Prosen}(1998)}]{prosen1998time}%
  \BibitemOpen
  \bibfield  {author} {\bibinfo {author} {\bibfnamefont
  {Toma\ifmmode\check{z}\else\v{z}\fi{}}\ \bibnamefont {Prosen}},\ }\bibfield
  {title} {\enquote {\bibinfo {title} {Time evolution of a quantum many-body
  system: Transition from integrability to ergodicity in the thermodynamic
  limit},}\ }\href {\doibase 10.1103/PhysRevLett.80.1808} {\bibfield  {journal}
  {\bibinfo  {journal} {Phys. Rev. Lett.}\ }\textbf {\bibinfo {volume} {80}},\
  \bibinfo {pages} {1808--1811} (\bibinfo {year} {1998})}\BibitemShut {NoStop}%
\bibitem [{\citenamefont {D’Alessio}\ and\ \citenamefont
  {Polkovnikov}(2013)}]{DAlessio-Polkovnikov2013_manybody}%
  \BibitemOpen
  \bibfield  {author} {\bibinfo {author} {\bibfnamefont {Luca}\ \bibnamefont
  {D’Alessio}}\ and\ \bibinfo {author} {\bibfnamefont {Anatoli}\ \bibnamefont
  {Polkovnikov}},\ }\bibfield  {title} {\enquote {\bibinfo {title} {Many-body
  energy localization transition in periodically driven systems},}\ }\href
  {\doibase https://doi.org/10.1016/j.aop.2013.02.011} {\bibfield  {journal}
  {\bibinfo  {journal} {Annals of Physics}\ }\textbf {\bibinfo {volume}
  {333}},\ \bibinfo {pages} {19--33} (\bibinfo {year} {2013})}\BibitemShut
  {NoStop}%
\bibitem [{\citenamefont {Ponte}\ \emph {et~al.}(2015)\citenamefont {Ponte},
  \citenamefont {Papi\ifmmode~\acute{c}\else \'{c}\fi{}}, \citenamefont
  {Huveneers},\ and\ \citenamefont {Abanin}}]{Ponte-Abanin2015_mbl}%
  \BibitemOpen
  \bibfield  {author} {\bibinfo {author} {\bibfnamefont {Pedro}\ \bibnamefont
  {Ponte}}, \bibinfo {author} {\bibfnamefont {Z.}~\bibnamefont
  {Papi\ifmmode~\acute{c}\else \'{c}\fi{}}}, \bibinfo {author} {\bibfnamefont
  {Fran\ifmmode \mbox{\c{c}}\else~\c{c}\fi{}ois}\ \bibnamefont {Huveneers}}, \
  and\ \bibinfo {author} {\bibfnamefont {Dmitry~A.}\ \bibnamefont {Abanin}},\
  }\bibfield  {title} {\enquote {\bibinfo {title} {Many-body localization in
  periodically driven systems},}\ }\href {\doibase
  10.1103/PhysRevLett.114.140401} {\bibfield  {journal} {\bibinfo  {journal}
  {Phys. Rev. Lett.}\ }\textbf {\bibinfo {volume} {114}},\ \bibinfo {pages}
  {140401} (\bibinfo {year} {2015})}\BibitemShut {NoStop}%
\bibitem [{\citenamefont {Lazarides}\ \emph {et~al.}(2015)\citenamefont
  {Lazarides}, \citenamefont {Das},\ and\ \citenamefont
  {Moessner}}]{Lazarides-Moessner2015_fate}%
  \BibitemOpen
  \bibfield  {author} {\bibinfo {author} {\bibfnamefont {Achilleas}\
  \bibnamefont {Lazarides}}, \bibinfo {author} {\bibfnamefont {Arnab}\
  \bibnamefont {Das}}, \ and\ \bibinfo {author} {\bibfnamefont {Roderich}\
  \bibnamefont {Moessner}},\ }\bibfield  {title} {\enquote {\bibinfo {title}
  {Fate of many-body localization under periodic driving},}\ }\href {\doibase
  10.1103/PhysRevLett.115.030402} {\bibfield  {journal} {\bibinfo  {journal}
  {Phys. Rev. Lett.}\ }\textbf {\bibinfo {volume} {115}},\ \bibinfo {pages}
  {030402} (\bibinfo {year} {2015})}\BibitemShut {NoStop}%
\bibitem [{\citenamefont {Gritsev}\ and\ \citenamefont
  {Polkovnikov}(2017)}]{gritsev2017integrable}%
  \BibitemOpen
  \bibfield  {author} {\bibinfo {author} {\bibfnamefont {Vladimir}\
  \bibnamefont {Gritsev}}\ and\ \bibinfo {author} {\bibfnamefont {Anatoli}\
  \bibnamefont {Polkovnikov}},\ }\bibfield  {title} {\enquote {\bibinfo {title}
  {{Integrable Floquet dynamics}},}\ }\href {\doibase
  10.21468/SciPostPhys.2.3.021} {\bibfield  {journal} {\bibinfo  {journal}
  {SciPost Phys.}\ }\textbf {\bibinfo {volume} {2}},\ \bibinfo {pages} {021}
  (\bibinfo {year} {2017})}\BibitemShut {NoStop}%
\bibitem [{\citenamefont {Claeys}\ and\ \citenamefont
  {Caux}()}]{claeys2017breaking}%
  \BibitemOpen
  \bibfield  {author} {\bibinfo {author} {\bibfnamefont {Pieter~W.}\
  \bibnamefont {Claeys}}\ and\ \bibinfo {author} {\bibfnamefont
  {Jean-Sébastien}\ \bibnamefont {Caux}},\ }\bibfield  {title} {\enquote
  {\bibinfo {title} {Breaking the integrability of the heisenberg model through
  periodic driving},}\ }\href {\doibase arXiv:1708.07324} {\
  arXiv:1708.07324}\BibitemShut {NoStop}%
\bibitem [{\citenamefont {Lazarides}\ \emph
  {et~al.}(2014{\natexlab{b}})\citenamefont {Lazarides}, \citenamefont {Das},\
  and\ \citenamefont {Moessner}}]{Lazarides-Moessner2014_periodic}%
  \BibitemOpen
  \bibfield  {author} {\bibinfo {author} {\bibfnamefont {Achilleas}\
  \bibnamefont {Lazarides}}, \bibinfo {author} {\bibfnamefont {Arnab}\
  \bibnamefont {Das}}, \ and\ \bibinfo {author} {\bibfnamefont {Roderich}\
  \bibnamefont {Moessner}},\ }\bibfield  {title} {\enquote {\bibinfo {title}
  {Periodic thermodynamics of isolated quantum systems},}\ }\href {\doibase
  10.1103/PhysRevLett.112.150401} {\bibfield  {journal} {\bibinfo  {journal}
  {Phys. Rev. Lett.}\ }\textbf {\bibinfo {volume} {112}},\ \bibinfo {pages}
  {150401} (\bibinfo {year} {2014}{\natexlab{b}})}\BibitemShut {NoStop}%
\bibitem [{\citenamefont {Russomanno}\ \emph {et~al.}(2012)\citenamefont
  {Russomanno}, \citenamefont {Silva},\ and\ \citenamefont
  {Santoro}}]{Russomanno-Santoro2012_steady}%
  \BibitemOpen
  \bibfield  {author} {\bibinfo {author} {\bibfnamefont {Angelo}\ \bibnamefont
  {Russomanno}}, \bibinfo {author} {\bibfnamefont {Alessandro}\ \bibnamefont
  {Silva}}, \ and\ \bibinfo {author} {\bibfnamefont {Giuseppe~E.}\ \bibnamefont
  {Santoro}},\ }\bibfield  {title} {\enquote {\bibinfo {title} {Periodic steady
  regime and interference in a periodically driven quantum system},}\ }\href
  {\doibase 10.1103/PhysRevLett.109.257201} {\bibfield  {journal} {\bibinfo
  {journal} {Phys. Rev. Lett.}\ }\textbf {\bibinfo {volume} {109}},\ \bibinfo
  {pages} {257201} (\bibinfo {year} {2012})}\BibitemShut {NoStop}%
\bibitem [{\citenamefont {Khemani}\ \emph {et~al.}(2016)\citenamefont
  {Khemani}, \citenamefont {Lazarides}, \citenamefont {Moessner},\ and\
  \citenamefont {Sondhi}}]{Khemani2016}%
  \BibitemOpen
  \bibfield  {author} {\bibinfo {author} {\bibfnamefont {V.}~\bibnamefont
  {Khemani}}, \bibinfo {author} {\bibfnamefont {A.}~\bibnamefont {Lazarides}},
  \bibinfo {author} {\bibfnamefont {R.}~\bibnamefont {Moessner}}, \ and\
  \bibinfo {author} {\bibfnamefont {S.~L.}\ \bibnamefont {Sondhi}},\ }\bibfield
   {title} {\enquote {\bibinfo {title} {Phase structure of driven quantum
  systems},}\ }\href {\doibase 10.1103/PhysRevLett.116.250401} {\bibfield
  {journal} {\bibinfo  {journal} {Phys. Rev. Lett.}\ }\textbf {\bibinfo
  {volume} {116}},\ \bibinfo {pages} {250401} (\bibinfo {year}
  {2016})}\BibitemShut {NoStop}%
\bibitem [{\citenamefont {Else}\ \emph {et~al.}(2016)\citenamefont {Else},
  \citenamefont {Bauer},\ and\ \citenamefont {Nayak}}]{else_floquet_2016}%
  \BibitemOpen
  \bibfield  {author} {\bibinfo {author} {\bibfnamefont {D.}~\bibnamefont
  {Else}}, \bibinfo {author} {\bibfnamefont {B.}~\bibnamefont {Bauer}}, \ and\
  \bibinfo {author} {\bibfnamefont {C.}~\bibnamefont {Nayak}},\ }\bibfield
  {title} {\enquote {\bibinfo {title} {{F}loquet time crystals},}\ }\href
  {\doibase 10.1103/PhysRevLett.117.090402} {\bibfield  {journal} {\bibinfo
  {journal} {Phys. Rev. Lett.}\ }\textbf {\bibinfo {volume} {117}},\ \bibinfo
  {pages} {090402} (\bibinfo {year} {2016})}\BibitemShut {NoStop}%
\bibitem [{\citenamefont {von Keyserlingk}\ \emph {et~al.}(2016)\citenamefont
  {von Keyserlingk}, \citenamefont {Khemani},\ and\ \citenamefont
  {Sondhi}}]{CVS}%
  \BibitemOpen
  \bibfield  {author} {\bibinfo {author} {\bibfnamefont {C.~W.}\ \bibnamefont
  {von Keyserlingk}}, \bibinfo {author} {\bibfnamefont {V.}~\bibnamefont
  {Khemani}}, \ and\ \bibinfo {author} {\bibfnamefont {S.~L.}\ \bibnamefont
  {Sondhi}},\ }\bibfield  {title} {\enquote {\bibinfo {title} {Absolute
  stability and spatiotemporal long-range order in {F}loquet systems},}\ }\href
  {\doibase 10.1103/PhysRevB.94.085112} {\bibfield  {journal} {\bibinfo
  {journal} {Phys. Rev. B}\ }\textbf {\bibinfo {volume} {94}},\ \bibinfo
  {pages} {085112} (\bibinfo {year} {2016})}\BibitemShut {NoStop}%
\bibitem [{\citenamefont {Sacha}\ and\ \citenamefont
  {Zakrzewski}(2018)}]{SachaReview}%
  \BibitemOpen
  \bibfield  {author} {\bibinfo {author} {\bibfnamefont {K.}~\bibnamefont
  {Sacha}}\ and\ \bibinfo {author} {\bibfnamefont {J.}~\bibnamefont
  {Zakrzewski}},\ }\bibfield  {title} {\enquote {\bibinfo {title} {Time
  crystals: a review},}\ }\href {\doibase 10.1088/1361-6633/aa8b38} {\bibfield
  {journal} {\bibinfo  {journal} {Rep. Prog. Phys.}\ }\textbf {\bibinfo
  {volume} {81}},\ \bibinfo {pages} {016401} (\bibinfo {year}
  {2018})}\BibitemShut {NoStop}%
\bibitem [{\citenamefont {Khemani}\ \emph {et~al.}()\citenamefont {Khemani},
  \citenamefont {Moessner},\ and\ \citenamefont {Sondhi}}]{Khemani2019}%
  \BibitemOpen
  \bibfield  {author} {\bibinfo {author} {\bibfnamefont {Vedika}\ \bibnamefont
  {Khemani}}, \bibinfo {author} {\bibfnamefont {Roderich}\ \bibnamefont
  {Moessner}}, \ and\ \bibinfo {author} {\bibfnamefont {S.~L.}\ \bibnamefont
  {Sondhi}},\ }\bibfield  {title} {\enquote {\bibinfo {title} {A brief history
  of time crystals},}\ }\href {\doibase arXiv:1910.10745} {\
  arXiv:1910.10745}\BibitemShut {NoStop}%
\bibitem [{\citenamefont {Rudner}\ \emph {et~al.}(2013)\citenamefont {Rudner},
  \citenamefont {Lindner}, \citenamefont {Berg},\ and\ \citenamefont
  {Levin}}]{Rudner13}%
  \BibitemOpen
  \bibfield  {author} {\bibinfo {author} {\bibfnamefont {M.~S.}\ \bibnamefont
  {Rudner}}, \bibinfo {author} {\bibfnamefont {N.~H.}\ \bibnamefont {Lindner}},
  \bibinfo {author} {\bibfnamefont {E.}~\bibnamefont {Berg}}, \ and\ \bibinfo
  {author} {\bibfnamefont {M.}~\bibnamefont {Levin}},\ }\bibfield  {title}
  {\enquote {\bibinfo {title} {Anomalous edge states and the bulk-edge
  correspondence for periodically driven two-dimensional systems},}\ }\href
  {\doibase 10.1103/PhysRevX.3.031005} {\bibfield  {journal} {\bibinfo
  {journal} {Phys. Rev. X}\ }\textbf {\bibinfo {volume} {3}},\ \bibinfo {pages}
  {031005} (\bibinfo {year} {2013})}\BibitemShut {NoStop}%
\bibitem [{\citenamefont {Titum}\ \emph {et~al.}(2016)\citenamefont {Titum},
  \citenamefont {Berg}, \citenamefont {Rudner}, \citenamefont {Refael},\ and\
  \citenamefont {Lindner}}]{Titum16}%
  \BibitemOpen
  \bibfield  {author} {\bibinfo {author} {\bibfnamefont {P.}~\bibnamefont
  {Titum}}, \bibinfo {author} {\bibfnamefont {E.}~\bibnamefont {Berg}},
  \bibinfo {author} {\bibfnamefont {M.~S.}\ \bibnamefont {Rudner}}, \bibinfo
  {author} {\bibfnamefont {G.}~\bibnamefont {Refael}}, \ and\ \bibinfo {author}
  {\bibfnamefont {N.~H.}\ \bibnamefont {Lindner}},\ }\bibfield  {title}
  {\enquote {\bibinfo {title} {Anomalous {F}loquet-{A}nderson insulator as a
  nonadiabatic quantized charge pump},}\ }\href {\doibase
  10.1103/PhysRevX.6.021013} {\bibfield  {journal} {\bibinfo  {journal} {Phys.
  Rev. X}\ }\textbf {\bibinfo {volume} {6}},\ \bibinfo {pages} {021013}
  (\bibinfo {year} {2016})}\BibitemShut {NoStop}%
\bibitem [{\citenamefont {Abanin}\ \emph {et~al.}(2015)\citenamefont {Abanin},
  \citenamefont {De~Roeck},\ and\ \citenamefont
  {Huveneers}}]{Abanin-Huveneers2015_exponentially}%
  \BibitemOpen
  \bibfield  {author} {\bibinfo {author} {\bibfnamefont {Dmitry~A.}\
  \bibnamefont {Abanin}}, \bibinfo {author} {\bibfnamefont {Wojciech}\
  \bibnamefont {De~Roeck}}, \ and\ \bibinfo {author} {\bibfnamefont
  {Fran\ifmmode \mbox{\c{c}}\else~\c{c}\fi{}ois}\ \bibnamefont {Huveneers}},\
  }\bibfield  {title} {\enquote {\bibinfo {title} {Exponentially slow heating
  in periodically driven many-body systems},}\ }\href {\doibase
  10.1103/PhysRevLett.115.256803} {\bibfield  {journal} {\bibinfo  {journal}
  {Phys. Rev. Lett.}\ }\textbf {\bibinfo {volume} {115}},\ \bibinfo {pages}
  {256803} (\bibinfo {year} {2015})}\BibitemShut {NoStop}%
\bibitem [{\citenamefont {Mori}\ \emph {et~al.}(2016)\citenamefont {Mori},
  \citenamefont {Kuwahara},\ and\ \citenamefont
  {Saito}}]{Mori-Saito2016_rigorous}%
  \BibitemOpen
  \bibfield  {author} {\bibinfo {author} {\bibfnamefont {Takashi}\ \bibnamefont
  {Mori}}, \bibinfo {author} {\bibfnamefont {Tomotaka}\ \bibnamefont
  {Kuwahara}}, \ and\ \bibinfo {author} {\bibfnamefont {Keiji}\ \bibnamefont
  {Saito}},\ }\bibfield  {title} {\enquote {\bibinfo {title} {Rigorous bound on
  energy absorption and generic relaxation in periodically driven quantum
  systems},}\ }\href {\doibase 10.1103/PhysRevLett.116.120401} {\bibfield
  {journal} {\bibinfo  {journal} {Phys. Rev. Lett.}\ }\textbf {\bibinfo
  {volume} {116}},\ \bibinfo {pages} {120401} (\bibinfo {year}
  {2016})}\BibitemShut {NoStop}%
\bibitem [{\citenamefont {Kuwahara}\ \emph {et~al.}(2016)\citenamefont
  {Kuwahara}, \citenamefont {Mori},\ and\ \citenamefont
  {Saito}}]{Kuwahara-Saito2016_floquet}%
  \BibitemOpen
  \bibfield  {author} {\bibinfo {author} {\bibfnamefont {Tomotaka}\
  \bibnamefont {Kuwahara}}, \bibinfo {author} {\bibfnamefont {Takashi}\
  \bibnamefont {Mori}}, \ and\ \bibinfo {author} {\bibfnamefont {Keiji}\
  \bibnamefont {Saito}},\ }\bibfield  {title} {\enquote {\bibinfo {title}
  {Floquet–magnus theory and generic transient dynamics in periodically
  driven many-body quantum systems},}\ }\href {\doibase
  https://doi.org/10.1016/j.aop.2016.01.012} {\bibfield  {journal} {\bibinfo
  {journal} {Annals of Physics}\ }\textbf {\bibinfo {volume} {367}},\ \bibinfo
  {pages} {96--124} (\bibinfo {year} {2016})}\BibitemShut {NoStop}%
\bibitem [{\citenamefont {Abanin}\ \emph
  {et~al.}(2017{\natexlab{a}})\citenamefont {Abanin}, \citenamefont {De~Roeck},
  \citenamefont {Ho},\ and\ \citenamefont
  {Huveneers}}]{Abanin-Huveneers2017_effective}%
  \BibitemOpen
  \bibfield  {author} {\bibinfo {author} {\bibfnamefont {Dmitry~A.}\
  \bibnamefont {Abanin}}, \bibinfo {author} {\bibfnamefont {Wojciech}\
  \bibnamefont {De~Roeck}}, \bibinfo {author} {\bibfnamefont {Wen~Wei}\
  \bibnamefont {Ho}}, \ and\ \bibinfo {author} {\bibfnamefont {Fran\ifmmode
  \mbox{\c{c}}\else~\c{c}\fi{}ois}\ \bibnamefont {Huveneers}},\ }\bibfield
  {title} {\enquote {\bibinfo {title} {Effective hamiltonians,
  prethermalization, and slow energy absorption in periodically driven
  many-body systems},}\ }\href {\doibase 10.1103/PhysRevB.95.014112} {\bibfield
   {journal} {\bibinfo  {journal} {Phys. Rev. B}\ }\textbf {\bibinfo {volume}
  {95}},\ \bibinfo {pages} {014112} (\bibinfo {year}
  {2017}{\natexlab{a}})}\BibitemShut {NoStop}%
\bibitem [{\citenamefont {Abanin}\ \emph
  {et~al.}(2017{\natexlab{b}})\citenamefont {Abanin}, \citenamefont {De~Roeck},
  \citenamefont {Ho},\ and\ \citenamefont
  {Huveneers}}]{Abanin-Huveneers2017_rigorous}%
  \BibitemOpen
  \bibfield  {author} {\bibinfo {author} {\bibfnamefont {Dmitry}\ \bibnamefont
  {Abanin}}, \bibinfo {author} {\bibfnamefont {Wojciech}\ \bibnamefont
  {De~Roeck}}, \bibinfo {author} {\bibfnamefont {Wen~Wei}\ \bibnamefont {Ho}},
  \ and\ \bibinfo {author} {\bibfnamefont {Fran{\c{c}}ois}\ \bibnamefont
  {Huveneers}},\ }\bibfield  {title} {\enquote {\bibinfo {title} {A rigorous
  theory of many-body prethermalization for periodically driven and closed
  quantum systems},}\ }\href {\doibase 10.1007/s00220-017-2930-x} {\bibfield
  {journal} {\bibinfo  {journal} {Communications in Mathematical Physics}\
  }\textbf {\bibinfo {volume} {354}},\ \bibinfo {pages} {809--827} (\bibinfo
  {year} {2017}{\natexlab{b}})}\BibitemShut {NoStop}%
\bibitem [{\citenamefont {Rubio-Abadal}\ \emph {et~al.}(2020)\citenamefont
  {Rubio-Abadal}, \citenamefont {Ippoliti}, \citenamefont {Hollerith},
  \citenamefont {Wei}, \citenamefont {Rui}, \citenamefont {Sondhi},
  \citenamefont {Khemani}, \citenamefont {Gross},\ and\ \citenamefont
  {Bloch}}]{RubioAbadal-Bloch2020}%
  \BibitemOpen
  \bibfield  {author} {\bibinfo {author} {\bibfnamefont {Antonio}\ \bibnamefont
  {Rubio-Abadal}}, \bibinfo {author} {\bibfnamefont {Matteo}\ \bibnamefont
  {Ippoliti}}, \bibinfo {author} {\bibfnamefont {Simon}\ \bibnamefont
  {Hollerith}}, \bibinfo {author} {\bibfnamefont {David}\ \bibnamefont {Wei}},
  \bibinfo {author} {\bibfnamefont {Jun}\ \bibnamefont {Rui}}, \bibinfo
  {author} {\bibfnamefont {S.~L.}\ \bibnamefont {Sondhi}}, \bibinfo {author}
  {\bibfnamefont {Vedika}\ \bibnamefont {Khemani}}, \bibinfo {author}
  {\bibfnamefont {Christian}\ \bibnamefont {Gross}}, \ and\ \bibinfo {author}
  {\bibfnamefont {Immanuel}\ \bibnamefont {Bloch}},\ }\bibfield  {title}
  {\enquote {\bibinfo {title} {Floquet prethermalization in a bose-hubbard
  system},}\ }\href {\doibase 10.1103/PhysRevX.10.021044} {\bibfield  {journal}
  {\bibinfo  {journal} {Phys. Rev. X}\ }\textbf {\bibinfo {volume} {10}},\
  \bibinfo {pages} {021044} (\bibinfo {year} {2020})}\BibitemShut {NoStop}%
\bibitem [{\citenamefont {De~Roeck}\ and\ \citenamefont
  {Verreet}()}]{DeRoeck-Verreet2019_very_slow}%
  \BibitemOpen
  \bibfield  {author} {\bibinfo {author} {\bibfnamefont {Wojciech}\
  \bibnamefont {De~Roeck}}\ and\ \bibinfo {author} {\bibfnamefont {Victor}\
  \bibnamefont {Verreet}},\ }\bibfield  {title} {\enquote {\bibinfo {title}
  {Very slow heating for weakly driven quantum many-body systems},}\ }\href
  {\doibase arXiv:1911.01998} {\ arXiv:1911.01998}\BibitemShut {NoStop}%
\bibitem [{\citenamefont {Luitz}\ \emph {et~al.}(2020)\citenamefont {Luitz},
  \citenamefont {Moessner}, \citenamefont {Sondhi},\ and\ \citenamefont
  {Khemani}}]{Luitz-Khemani2020}%
  \BibitemOpen
  \bibfield  {author} {\bibinfo {author} {\bibfnamefont {David~J.}\
  \bibnamefont {Luitz}}, \bibinfo {author} {\bibfnamefont {Roderich}\
  \bibnamefont {Moessner}}, \bibinfo {author} {\bibfnamefont {S.~L.}\
  \bibnamefont {Sondhi}}, \ and\ \bibinfo {author} {\bibfnamefont {Vedika}\
  \bibnamefont {Khemani}},\ }\bibfield  {title} {\enquote {\bibinfo {title}
  {Prethermalization without temperature},}\ }\href {\doibase
  10.1103/PhysRevX.10.021046} {\bibfield  {journal} {\bibinfo  {journal} {Phys.
  Rev. X}\ }\textbf {\bibinfo {volume} {10}},\ \bibinfo {pages} {021046}
  (\bibinfo {year} {2020})}\BibitemShut {NoStop}%
\bibitem [{\citenamefont {Ho}\ and\ \citenamefont
  {De~Roeck}()}]{Ho-DeRoeck2020_rigorous_without}%
  \BibitemOpen
  \bibfield  {author} {\bibinfo {author} {\bibfnamefont {Wen~Wei}\ \bibnamefont
  {Ho}}\ and\ \bibinfo {author} {\bibfnamefont {Wojciech}\ \bibnamefont
  {De~Roeck}},\ }\bibfield  {title} {\enquote {\bibinfo {title} {A rigorous
  theory of prethermalization without temperature},}\ }\href {\doibase
  arXiv:2011.14583} {\ arXiv:2011.14583}\BibitemShut {NoStop}%
\bibitem [{\citenamefont {Else}\ \emph
  {et~al.}(2017{\natexlab{a}})\citenamefont {Else}, \citenamefont {Bauer},\
  and\ \citenamefont {Nayak}}]{Else-Nayak2017_time_translation}%
  \BibitemOpen
  \bibfield  {author} {\bibinfo {author} {\bibfnamefont {Dominic~V.}\
  \bibnamefont {Else}}, \bibinfo {author} {\bibfnamefont {Bela}\ \bibnamefont
  {Bauer}}, \ and\ \bibinfo {author} {\bibfnamefont {Chetan}\ \bibnamefont
  {Nayak}},\ }\bibfield  {title} {\enquote {\bibinfo {title} {Prethermal phases
  of matter protected by time-translation symmetry},}\ }\href {\doibase
  10.1103/PhysRevX.7.011026} {\bibfield  {journal} {\bibinfo  {journal} {Phys.
  Rev. X}\ }\textbf {\bibinfo {volume} {7}},\ \bibinfo {pages} {011026}
  (\bibinfo {year} {2017}{\natexlab{a}})}\BibitemShut {NoStop}%
\bibitem [{\citenamefont {Lerose}\ \emph {et~al.}(2019)\citenamefont {Lerose},
  \citenamefont {Marino}, \citenamefont {Gambassi},\ and\ \citenamefont
  {Silva}}]{Lerose-Silva2019_kapitza}%
  \BibitemOpen
  \bibfield  {author} {\bibinfo {author} {\bibfnamefont {Alessio}\ \bibnamefont
  {Lerose}}, \bibinfo {author} {\bibfnamefont {Jamir}\ \bibnamefont {Marino}},
  \bibinfo {author} {\bibfnamefont {Andrea}\ \bibnamefont {Gambassi}}, \ and\
  \bibinfo {author} {\bibfnamefont {Alessandro}\ \bibnamefont {Silva}},\
  }\bibfield  {title} {\enquote {\bibinfo {title} {Prethermal quantum many-body
  kapitza phases of periodically driven spin systems},}\ }\href {\doibase
  10.1103/PhysRevB.100.104306} {\bibfield  {journal} {\bibinfo  {journal}
  {Phys. Rev. B}\ }\textbf {\bibinfo {volume} {100}},\ \bibinfo {pages}
  {104306} (\bibinfo {year} {2019})}\BibitemShut {NoStop}%
\bibitem [{\citenamefont {Machado}\ \emph {et~al.}(2020)\citenamefont
  {Machado}, \citenamefont {Else}, \citenamefont {Kahanamoku-Meyer},
  \citenamefont {Nayak},\ and\ \citenamefont
  {Yao}}]{Machado-Yao2020_long_range}%
  \BibitemOpen
  \bibfield  {author} {\bibinfo {author} {\bibfnamefont {Francisco}\
  \bibnamefont {Machado}}, \bibinfo {author} {\bibfnamefont {Dominic~V.}\
  \bibnamefont {Else}}, \bibinfo {author} {\bibfnamefont {Gregory~D.}\
  \bibnamefont {Kahanamoku-Meyer}}, \bibinfo {author} {\bibfnamefont {Chetan}\
  \bibnamefont {Nayak}}, \ and\ \bibinfo {author} {\bibfnamefont {Norman~Y.}\
  \bibnamefont {Yao}},\ }\bibfield  {title} {\enquote {\bibinfo {title}
  {Long-range prethermal phases of nonequilibrium matter},}\ }\href {\doibase
  10.1103/PhysRevX.10.011043} {\bibfield  {journal} {\bibinfo  {journal} {Phys.
  Rev. X}\ }\textbf {\bibinfo {volume} {10}},\ \bibinfo {pages} {011043}
  (\bibinfo {year} {2020})}\BibitemShut {NoStop}%
\bibitem [{\citenamefont {Collura}\ \emph {et~al.}(2022)\citenamefont
  {Collura}, \citenamefont {De~Luca}, \citenamefont {Rossini},\ and\
  \citenamefont {Lerose}}]{Collura-Lerose2022_confinement}%
  \BibitemOpen
  \bibfield  {author} {\bibinfo {author} {\bibfnamefont {Mario}\ \bibnamefont
  {Collura}}, \bibinfo {author} {\bibfnamefont {Andrea}\ \bibnamefont
  {De~Luca}}, \bibinfo {author} {\bibfnamefont {Davide}\ \bibnamefont
  {Rossini}}, \ and\ \bibinfo {author} {\bibfnamefont {Alessio}\ \bibnamefont
  {Lerose}},\ }\bibfield  {title} {\enquote {\bibinfo {title} {Discrete
  time-crystalline response stabilized by domain-wall confinement},}\ }\href
  {\doibase 10.1103/PhysRevX.12.031037} {\bibfield  {journal} {\bibinfo
  {journal} {Phys. Rev. X}\ }\textbf {\bibinfo {volume} {12}},\ \bibinfo
  {pages} {031037} (\bibinfo {year} {2022})}\BibitemShut {NoStop}%
\bibitem [{\citenamefont {Else}\ \emph
  {et~al.}(2017{\natexlab{b}})\citenamefont {Else}, \citenamefont {Fendley},
  \citenamefont {Kemp},\ and\ \citenamefont
  {Nayak}}]{Else-Nayak2017_prethermal_zero}%
  \BibitemOpen
  \bibfield  {author} {\bibinfo {author} {\bibfnamefont {Dominic~V.}\
  \bibnamefont {Else}}, \bibinfo {author} {\bibfnamefont {Paul}\ \bibnamefont
  {Fendley}}, \bibinfo {author} {\bibfnamefont {Jack}\ \bibnamefont {Kemp}}, \
  and\ \bibinfo {author} {\bibfnamefont {Chetan}\ \bibnamefont {Nayak}},\
  }\bibfield  {title} {\enquote {\bibinfo {title} {Prethermal strong zero modes
  and topological qubits},}\ }\href {\doibase 10.1103/PhysRevX.7.041062}
  {\bibfield  {journal} {\bibinfo  {journal} {Phys. Rev. X}\ }\textbf {\bibinfo
  {volume} {7}},\ \bibinfo {pages} {041062} (\bibinfo {year}
  {2017}{\natexlab{b}})}\BibitemShut {NoStop}%
\bibitem [{\citenamefont {Altshuler}\ \emph {et~al.}(1997)\citenamefont
  {Altshuler}, \citenamefont {Gefen}, \citenamefont {Kamenev},\ and\
  \citenamefont {Levitov}}]{Altshuler-Levitov1997_quasiparticle}%
  \BibitemOpen
  \bibfield  {author} {\bibinfo {author} {\bibfnamefont {Boris~L.}\
  \bibnamefont {Altshuler}}, \bibinfo {author} {\bibfnamefont {Yuval}\
  \bibnamefont {Gefen}}, \bibinfo {author} {\bibfnamefont {Alex}\ \bibnamefont
  {Kamenev}}, \ and\ \bibinfo {author} {\bibfnamefont {Leonid~S.}\ \bibnamefont
  {Levitov}},\ }\bibfield  {title} {\enquote {\bibinfo {title} {Quasiparticle
  lifetime in a finite system: A nonperturbative approach},}\ }\href {\doibase
  10.1103/PhysRevLett.78.2803} {\bibfield  {journal} {\bibinfo  {journal}
  {Phys. Rev. Lett.}\ }\textbf {\bibinfo {volume} {78}},\ \bibinfo {pages}
  {2803--2806} (\bibinfo {year} {1997})}\BibitemShut {NoStop}%
\bibitem [{\citenamefont {Burin}(2015)}]{Burin2015_long_range}%
  \BibitemOpen
  \bibfield  {author} {\bibinfo {author} {\bibfnamefont {Alexander~L.}\
  \bibnamefont {Burin}},\ }\bibfield  {title} {\enquote {\bibinfo {title}
  {Many-body delocalization in a strongly disordered system with long-range
  interactions: Finite-size scaling},}\ }\href {\doibase
  10.1103/PhysRevB.91.094202} {\bibfield  {journal} {\bibinfo  {journal} {Phys.
  Rev. B}\ }\textbf {\bibinfo {volume} {91}},\ \bibinfo {pages} {094202}
  (\bibinfo {year} {2015})}\BibitemShut {NoStop}%
\bibitem [{\citenamefont {Tikhonov}\ and\ \citenamefont
  {Mirlin}(2018)}]{Tikhonov-Mirlin2018_power_law}%
  \BibitemOpen
  \bibfield  {author} {\bibinfo {author} {\bibfnamefont {K.~S.}\ \bibnamefont
  {Tikhonov}}\ and\ \bibinfo {author} {\bibfnamefont {A.~D.}\ \bibnamefont
  {Mirlin}},\ }\bibfield  {title} {\enquote {\bibinfo {title} {Many-body
  localization transition with power-law interactions: Statistics of
  eigenstates},}\ }\href {\doibase 10.1103/PhysRevB.97.214205} {\bibfield
  {journal} {\bibinfo  {journal} {Phys. Rev. B}\ }\textbf {\bibinfo {volume}
  {97}},\ \bibinfo {pages} {214205} (\bibinfo {year} {2018})}\BibitemShut
  {NoStop}%
\bibitem [{\citenamefont {Gopalakrishnan}\ and\ \citenamefont
  {Huse}(2019)}]{Gopalakrishnan-Huse2019_instability}%
  \BibitemOpen
  \bibfield  {author} {\bibinfo {author} {\bibfnamefont {Sarang}\ \bibnamefont
  {Gopalakrishnan}}\ and\ \bibinfo {author} {\bibfnamefont {David~A.}\
  \bibnamefont {Huse}},\ }\bibfield  {title} {\enquote {\bibinfo {title}
  {Instability of many-body localized systems as a phase transition in a
  nonstandard thermodynamic limit},}\ }\href {\doibase
  10.1103/PhysRevB.99.134305} {\bibfield  {journal} {\bibinfo  {journal} {Phys.
  Rev. B}\ }\textbf {\bibinfo {volume} {99}},\ \bibinfo {pages} {134305}
  (\bibinfo {year} {2019})}\BibitemShut {NoStop}%
\bibitem [{\citenamefont {Bulchandani}\ \emph {et~al.}(2022)\citenamefont
  {Bulchandani}, \citenamefont {Huse},\ and\ \citenamefont
  {Gopalakrishnan}}]{Bulchandani-Gopalakrishnan2021_onset}%
  \BibitemOpen
  \bibfield  {author} {\bibinfo {author} {\bibfnamefont {Vir~B.}\ \bibnamefont
  {Bulchandani}}, \bibinfo {author} {\bibfnamefont {David~A.}\ \bibnamefont
  {Huse}}, \ and\ \bibinfo {author} {\bibfnamefont {Sarang}\ \bibnamefont
  {Gopalakrishnan}},\ }\bibfield  {title} {\enquote {\bibinfo {title} {Onset of
  many-body quantum chaos due to breaking integrability},}\ }\href {\doibase
  10.1103/PhysRevB.105.214308} {\bibfield  {journal} {\bibinfo  {journal}
  {Phys. Rev. B}\ }\textbf {\bibinfo {volume} {105}},\ \bibinfo {pages}
  {214308} (\bibinfo {year} {2022})}\BibitemShut {NoStop}%
\bibitem [{\citenamefont {Preskill}(2018)}]{Preskill2018_nisq}%
  \BibitemOpen
  \bibfield  {author} {\bibinfo {author} {\bibfnamefont {John}\ \bibnamefont
  {Preskill}},\ }\bibfield  {title} {\enquote {\bibinfo {title} {Quantum
  {C}omputing in the {NISQ} era and beyond},}\ }\href {\doibase
  10.22331/q-2018-08-06-79} {\bibfield  {journal} {\bibinfo  {journal}
  {{Quantum}}\ }\textbf {\bibinfo {volume} {2}},\ \bibinfo {pages} {79}
  (\bibinfo {year} {2018})}\BibitemShut {NoStop}%
\bibitem [{\citenamefont {Morningstar}\ \emph
  {et~al.}(2022{\natexlab{a}})\citenamefont {Morningstar}, \citenamefont
  {Hauru}, \citenamefont {Beall}, \citenamefont {Ganahl}, \citenamefont
  {Lewis}, \citenamefont {Khemani},\ and\ \citenamefont
  {Vidal}}]{Morningstar-Vidal2021_simulation}%
  \BibitemOpen
  \bibfield  {author} {\bibinfo {author} {\bibfnamefont {Alan}\ \bibnamefont
  {Morningstar}}, \bibinfo {author} {\bibfnamefont {Markus}\ \bibnamefont
  {Hauru}}, \bibinfo {author} {\bibfnamefont {Jackson}\ \bibnamefont {Beall}},
  \bibinfo {author} {\bibfnamefont {Martin}\ \bibnamefont {Ganahl}}, \bibinfo
  {author} {\bibfnamefont {Adam~G.M.}\ \bibnamefont {Lewis}}, \bibinfo {author}
  {\bibfnamefont {Vedika}\ \bibnamefont {Khemani}}, \ and\ \bibinfo {author}
  {\bibfnamefont {Guifre}\ \bibnamefont {Vidal}},\ }\bibfield  {title}
  {\enquote {\bibinfo {title} {Simulation of quantum many-body dynamics with
  tensor processing units: Floquet prethermalization},}\ }\href {\doibase
  10.1103/PRXQuantum.3.020331} {\bibfield  {journal} {\bibinfo  {journal} {PRX
  Quantum}\ }\textbf {\bibinfo {volume} {3}},\ \bibinfo {pages} {020331}
  (\bibinfo {year} {2022}{\natexlab{a}})}\BibitemShut {NoStop}%
\bibitem [{\citenamefont {Seetharam}\ \emph {et~al.}(2018)\citenamefont
  {Seetharam}, \citenamefont {Titum}, \citenamefont {Kolodrubetz},\ and\
  \citenamefont {Refael}}]{Seetheram-Refael2018_absence}%
  \BibitemOpen
  \bibfield  {author} {\bibinfo {author} {\bibfnamefont {Karthik}\ \bibnamefont
  {Seetharam}}, \bibinfo {author} {\bibfnamefont {Paraj}\ \bibnamefont
  {Titum}}, \bibinfo {author} {\bibfnamefont {Michael}\ \bibnamefont
  {Kolodrubetz}}, \ and\ \bibinfo {author} {\bibfnamefont {Gil}\ \bibnamefont
  {Refael}},\ }\bibfield  {title} {\enquote {\bibinfo {title} {Absence of
  thermalization in finite isolated interacting floquet systems},}\ }\href
  {\doibase 10.1103/PhysRevB.97.014311} {\bibfield  {journal} {\bibinfo
  {journal} {Phys. Rev. B}\ }\textbf {\bibinfo {volume} {97}},\ \bibinfo
  {pages} {014311} (\bibinfo {year} {2018})}\BibitemShut {NoStop}%
\bibitem [{\citenamefont {Cao}(2021)}]{Cao2021_statistical}%
  \BibitemOpen
  \bibfield  {author} {\bibinfo {author} {\bibfnamefont {Xiangyu}\ \bibnamefont
  {Cao}},\ }\bibfield  {title} {\enquote {\bibinfo {title} {A statistical
  mechanism for operator growth},}\ }\href {\doibase 10.1088/1751-8121/abe77c}
  {\bibfield  {journal} {\bibinfo  {journal} {Journal of Physics A:
  Mathematical and Theoretical}\ }\textbf {\bibinfo {volume} {54}},\ \bibinfo
  {pages} {144001} (\bibinfo {year} {2021})}\BibitemShut {NoStop}%
\bibitem [{\citenamefont {Sels}\ and\ \citenamefont
  {Polkovnikov}(2023)}]{Sels-Polkovnikov2022_dilute}%
  \BibitemOpen
  \bibfield  {author} {\bibinfo {author} {\bibfnamefont {Dries}\ \bibnamefont
  {Sels}}\ and\ \bibinfo {author} {\bibfnamefont {Anatoli}\ \bibnamefont
  {Polkovnikov}},\ }\bibfield  {title} {\enquote {\bibinfo {title}
  {Thermalization of dilute impurities in one-dimensional spin chains},}\
  }\href {\doibase 10.1103/PhysRevX.13.011041} {\bibfield  {journal} {\bibinfo
  {journal} {Phys. Rev. X}\ }\textbf {\bibinfo {volume} {13}},\ \bibinfo
  {pages} {011041} (\bibinfo {year} {2023})}\BibitemShut {NoStop}%
\bibitem [{\citenamefont {Parker}\ \emph {et~al.}(2019)\citenamefont {Parker},
  \citenamefont {Cao}, \citenamefont {Avdoshkin}, \citenamefont {Scaffidi},\
  and\ \citenamefont {Altman}}]{Parker-Altman2019_hypothesis}%
  \BibitemOpen
  \bibfield  {author} {\bibinfo {author} {\bibfnamefont {Daniel~E.}\
  \bibnamefont {Parker}}, \bibinfo {author} {\bibfnamefont {Xiangyu}\
  \bibnamefont {Cao}}, \bibinfo {author} {\bibfnamefont {Alexander}\
  \bibnamefont {Avdoshkin}}, \bibinfo {author} {\bibfnamefont {Thomas}\
  \bibnamefont {Scaffidi}}, \ and\ \bibinfo {author} {\bibfnamefont {Ehud}\
  \bibnamefont {Altman}},\ }\bibfield  {title} {\enquote {\bibinfo {title} {A
  universal operator growth hypothesis},}\ }\href {\doibase
  10.1103/PhysRevX.9.041017} {\bibfield  {journal} {\bibinfo  {journal} {Phys.
  Rev. X}\ }\textbf {\bibinfo {volume} {9}},\ \bibinfo {pages} {041017}
  (\bibinfo {year} {2019})}\BibitemShut {NoStop}%
\bibitem [{\citenamefont {Mallayya}\ and\ \citenamefont
  {Rigol}(2019)}]{Mallayya-Rigol2019_heating}%
  \BibitemOpen
  \bibfield  {author} {\bibinfo {author} {\bibfnamefont {Krishnanand}\
  \bibnamefont {Mallayya}}\ and\ \bibinfo {author} {\bibfnamefont {Marcos}\
  \bibnamefont {Rigol}},\ }\bibfield  {title} {\enquote {\bibinfo {title}
  {Heating rates in periodically driven strongly interacting quantum many-body
  systems},}\ }\href {\doibase 10.1103/PhysRevLett.123.240603} {\bibfield
  {journal} {\bibinfo  {journal} {Phys. Rev. Lett.}\ }\textbf {\bibinfo
  {volume} {123}},\ \bibinfo {pages} {240603} (\bibinfo {year}
  {2019})}\BibitemShut {NoStop}%
\bibitem [{\citenamefont {Mori}(2022)}]{Mori2022_heating}%
  \BibitemOpen
  \bibfield  {author} {\bibinfo {author} {\bibfnamefont {Takashi}\ \bibnamefont
  {Mori}},\ }\bibfield  {title} {\enquote {\bibinfo {title} {Heating rates
  under fast periodic driving beyond linear response},}\ }\href {\doibase
  10.1103/PhysRevLett.128.050604} {\bibfield  {journal} {\bibinfo  {journal}
  {Phys. Rev. Lett.}\ }\textbf {\bibinfo {volume} {128}},\ \bibinfo {pages}
  {050604} (\bibinfo {year} {2022})}\BibitemShut {NoStop}%
\bibitem [{\citenamefont {Ikeda}\ and\ \citenamefont
  {Polkovnikov}(2021)}]{Ikeda-Polkovnikov2021_fgr}%
  \BibitemOpen
  \bibfield  {author} {\bibinfo {author} {\bibfnamefont {Tatsuhiko~N.}\
  \bibnamefont {Ikeda}}\ and\ \bibinfo {author} {\bibfnamefont {Anatoli}\
  \bibnamefont {Polkovnikov}},\ }\bibfield  {title} {\enquote {\bibinfo {title}
  {Fermi's golden rule for heating in strongly driven floquet systems},}\
  }\href {\doibase 10.1103/PhysRevB.104.134308} {\bibfield  {journal} {\bibinfo
   {journal} {Phys. Rev. B}\ }\textbf {\bibinfo {volume} {104}},\ \bibinfo
  {pages} {134308} (\bibinfo {year} {2021})}\BibitemShut {NoStop}%
\bibitem [{\citenamefont {Rakcheev}\ and\ \citenamefont
  {L\"auchli}(2022)}]{Rakcheev-Lauchli2020_estimating}%
  \BibitemOpen
  \bibfield  {author} {\bibinfo {author} {\bibfnamefont {Artem}\ \bibnamefont
  {Rakcheev}}\ and\ \bibinfo {author} {\bibfnamefont {Andreas~M.}\ \bibnamefont
  {L\"auchli}},\ }\bibfield  {title} {\enquote {\bibinfo {title} {Estimating
  heating times in periodically driven quantum many-body systems via avoided
  crossing spectroscopy},}\ }\href {\doibase 10.1103/PhysRevResearch.4.043174}
  {\bibfield  {journal} {\bibinfo  {journal} {Phys. Rev. Res.}\ }\textbf
  {\bibinfo {volume} {4}},\ \bibinfo {pages} {043174} (\bibinfo {year}
  {2022})}\BibitemShut {NoStop}%
\bibitem [{\citenamefont {Micklitz}\ \emph {et~al.}(2022)\citenamefont
  {Micklitz}, \citenamefont {Morningstar}, \citenamefont {Altland},\ and\
  \citenamefont {Huse}}]{Micklitz-Huse2022_emergence}%
  \BibitemOpen
  \bibfield  {author} {\bibinfo {author} {\bibfnamefont {Tobias}\ \bibnamefont
  {Micklitz}}, \bibinfo {author} {\bibfnamefont {Alan}\ \bibnamefont
  {Morningstar}}, \bibinfo {author} {\bibfnamefont {Alexander}\ \bibnamefont
  {Altland}}, \ and\ \bibinfo {author} {\bibfnamefont {David~A.}\ \bibnamefont
  {Huse}},\ }\bibfield  {title} {\enquote {\bibinfo {title} {Emergence of
  fermi's golden rule},}\ }\href {\doibase 10.1103/PhysRevLett.129.140402}
  {\bibfield  {journal} {\bibinfo  {journal} {Phys. Rev. Lett.}\ }\textbf
  {\bibinfo {volume} {129}},\ \bibinfo {pages} {140402} (\bibinfo {year}
  {2022})}\BibitemShut {NoStop}%
\bibitem [{\citenamefont {Bukov}\ \emph {et~al.}(2016)\citenamefont {Bukov},
  \citenamefont {Heyl}, \citenamefont {Huse},\ and\ \citenamefont
  {Polkovnikov}}]{Bukov-Polkovnikov2016_heating}%
  \BibitemOpen
  \bibfield  {author} {\bibinfo {author} {\bibfnamefont {Marin}\ \bibnamefont
  {Bukov}}, \bibinfo {author} {\bibfnamefont {Markus}\ \bibnamefont {Heyl}},
  \bibinfo {author} {\bibfnamefont {David~A.}\ \bibnamefont {Huse}}, \ and\
  \bibinfo {author} {\bibfnamefont {Anatoli}\ \bibnamefont {Polkovnikov}},\
  }\bibfield  {title} {\enquote {\bibinfo {title} {Heating and many-body
  resonances in a periodically driven two-band system},}\ }\href {\doibase
  10.1103/PhysRevB.93.155132} {\bibfield  {journal} {\bibinfo  {journal} {Phys.
  Rev. B}\ }\textbf {\bibinfo {volume} {93}},\ \bibinfo {pages} {155132}
  (\bibinfo {year} {2016})}\BibitemShut {NoStop}%
\bibitem [{\citenamefont {Serbyn}\ \emph {et~al.}(2015)\citenamefont {Serbyn},
  \citenamefont {Papi\ifmmode~\acute{c}\else \'{c}\fi{}},\ and\ \citenamefont
  {Abanin}}]{Serbyn-Abanin2015_criterion}%
  \BibitemOpen
  \bibfield  {author} {\bibinfo {author} {\bibfnamefont {Maksym}\ \bibnamefont
  {Serbyn}}, \bibinfo {author} {\bibfnamefont {Z.}~\bibnamefont
  {Papi\ifmmode~\acute{c}\else \'{c}\fi{}}}, \ and\ \bibinfo {author}
  {\bibfnamefont {Dmitry~A.}\ \bibnamefont {Abanin}},\ }\bibfield  {title}
  {\enquote {\bibinfo {title} {Criterion for many-body
  localization-delocalization phase transition},}\ }\href {\doibase
  10.1103/PhysRevX.5.041047} {\bibfield  {journal} {\bibinfo  {journal} {Phys.
  Rev. X}\ }\textbf {\bibinfo {volume} {5}},\ \bibinfo {pages} {041047}
  (\bibinfo {year} {2015})}\BibitemShut {NoStop}%
\bibitem [{\citenamefont {Morningstar}\ \emph
  {et~al.}(2022{\natexlab{b}})\citenamefont {Morningstar}, \citenamefont
  {Colmenarez}, \citenamefont {Khemani}, \citenamefont {Luitz},\ and\
  \citenamefont {Huse}}]{Morningstar-Huse2021_avalanches}%
  \BibitemOpen
  \bibfield  {author} {\bibinfo {author} {\bibfnamefont {Alan}\ \bibnamefont
  {Morningstar}}, \bibinfo {author} {\bibfnamefont {Luis}\ \bibnamefont
  {Colmenarez}}, \bibinfo {author} {\bibfnamefont {Vedika}\ \bibnamefont
  {Khemani}}, \bibinfo {author} {\bibfnamefont {David~J.}\ \bibnamefont
  {Luitz}}, \ and\ \bibinfo {author} {\bibfnamefont {David~A.}\ \bibnamefont
  {Huse}},\ }\bibfield  {title} {\enquote {\bibinfo {title} {Avalanches and
  many-body resonances in many-body localized systems},}\ }\href {\doibase
  10.1103/PhysRevB.105.174205} {\bibfield  {journal} {\bibinfo  {journal}
  {Phys. Rev. B}\ }\textbf {\bibinfo {volume} {105}},\ \bibinfo {pages}
  {174205} (\bibinfo {year} {2022}{\natexlab{b}})}\BibitemShut {NoStop}%
\bibitem [{\citenamefont {Garratt}\ \emph {et~al.}(2021)\citenamefont
  {Garratt}, \citenamefont {Roy},\ and\ \citenamefont
  {Chalker}}]{Garratt-Chalker2021_local}%
  \BibitemOpen
  \bibfield  {author} {\bibinfo {author} {\bibfnamefont {S.~J.}\ \bibnamefont
  {Garratt}}, \bibinfo {author} {\bibfnamefont {Sthitadhi}\ \bibnamefont
  {Roy}}, \ and\ \bibinfo {author} {\bibfnamefont {J.~T.}\ \bibnamefont
  {Chalker}},\ }\bibfield  {title} {\enquote {\bibinfo {title} {Local
  resonances and parametric level dynamics in the many-body localized phase},}\
  }\href {\doibase 10.1103/PhysRevB.104.184203} {\bibfield  {journal} {\bibinfo
   {journal} {Phys. Rev. B}\ }\textbf {\bibinfo {volume} {104}},\ \bibinfo
  {pages} {184203} (\bibinfo {year} {2021})}\BibitemShut {NoStop}%
\bibitem [{\citenamefont {Garratt}\ and\ \citenamefont
  {Roy}(2022)}]{Garratt-Roy2022_resonant}%
  \BibitemOpen
  \bibfield  {author} {\bibinfo {author} {\bibfnamefont {Samuel~J.}\
  \bibnamefont {Garratt}}\ and\ \bibinfo {author} {\bibfnamefont {Sthitadhi}\
  \bibnamefont {Roy}},\ }\bibfield  {title} {\enquote {\bibinfo {title}
  {Resonant energy scales and local observables in the many-body localized
  phase},}\ }\href {\doibase 10.1103/PhysRevB.106.054309} {\bibfield  {journal}
  {\bibinfo  {journal} {Phys. Rev. B}\ }\textbf {\bibinfo {volume} {106}},\
  \bibinfo {pages} {054309} (\bibinfo {year} {2022})}\BibitemShut {NoStop}%
\bibitem [{\citenamefont {Wei}\ \emph {et~al.}(2022)\citenamefont {Wei},
  \citenamefont {Rubio-Abadal}, \citenamefont {Ye}, \citenamefont {Machado},
  \citenamefont {Kemp}, \citenamefont {Srakaew}, \citenamefont {Hollerith},
  \citenamefont {Rui}, \citenamefont {Gopalakrishnan}, \citenamefont {Yao},
  \citenamefont {Bloch},\ and\ \citenamefont {Zeiher}}]{Wei-Zeiher2021_kpz}%
  \BibitemOpen
  \bibfield  {author} {\bibinfo {author} {\bibfnamefont {David}\ \bibnamefont
  {Wei}}, \bibinfo {author} {\bibfnamefont {Antonio}\ \bibnamefont
  {Rubio-Abadal}}, \bibinfo {author} {\bibfnamefont {Bingtian}\ \bibnamefont
  {Ye}}, \bibinfo {author} {\bibfnamefont {Francisco}\ \bibnamefont {Machado}},
  \bibinfo {author} {\bibfnamefont {Jack}\ \bibnamefont {Kemp}}, \bibinfo
  {author} {\bibfnamefont {Kritsana}\ \bibnamefont {Srakaew}}, \bibinfo
  {author} {\bibfnamefont {Simon}\ \bibnamefont {Hollerith}}, \bibinfo {author}
  {\bibfnamefont {Jun}\ \bibnamefont {Rui}}, \bibinfo {author} {\bibfnamefont
  {Sarang}\ \bibnamefont {Gopalakrishnan}}, \bibinfo {author} {\bibfnamefont
  {Norman~Y.}\ \bibnamefont {Yao}}, \bibinfo {author} {\bibfnamefont
  {Immanuel}\ \bibnamefont {Bloch}}, \ and\ \bibinfo {author} {\bibfnamefont
  {Johannes}\ \bibnamefont {Zeiher}},\ }\bibfield  {title} {\enquote {\bibinfo
  {title} {Quantum gas microscopy of kardar-parisi-zhang superdiffusion},}\
  }\href {\doibase 10.1126/science.abk2397} {\bibfield  {journal} {\bibinfo
  {journal} {Science}\ }\textbf {\bibinfo {volume} {376}},\ \bibinfo {pages}
  {716--720} (\bibinfo {year} {2022})}\BibitemShut {NoStop}%
\bibitem [{\citenamefont {Levitov}\ and\ \citenamefont
  {Lesovik}(1993)}]{levitov-Lesovik1993_charge}%
  \BibitemOpen
  \bibfield  {author} {\bibinfo {author} {\bibfnamefont {Leonid~S}\
  \bibnamefont {Levitov}}\ and\ \bibinfo {author} {\bibfnamefont {Gordey~B}\
  \bibnamefont {Lesovik}},\ }\bibfield  {title} {\enquote {\bibinfo {title}
  {Charge distribution in quantum shot noise},}\ }\href
  {http://jetpletters.ru/ps/1186/article_17907.shtml} {\bibfield  {journal}
  {\bibinfo  {journal} {JETP Letters}\ }\textbf {\bibinfo {volume} {58}},\
  \bibinfo {pages} {225} (\bibinfo {year} {1993})}\BibitemShut {NoStop}%
\bibitem [{\citenamefont {Groha}\ \emph {et~al.}(2018)\citenamefont {Groha},
  \citenamefont {Essler},\ and\ \citenamefont
  {Calabrese}}]{Groha-Calabrese2018_fcs}%
  \BibitemOpen
  \bibfield  {author} {\bibinfo {author} {\bibfnamefont {Stefan}\ \bibnamefont
  {Groha}}, \bibinfo {author} {\bibfnamefont {Fabian H.~L.}\ \bibnamefont
  {Essler}}, \ and\ \bibinfo {author} {\bibfnamefont {Pasquale}\ \bibnamefont
  {Calabrese}},\ }\bibfield  {title} {\enquote {\bibinfo {title} {{Full
  counting statistics in the transverse field Ising chain}},}\ }\href {\doibase
  10.21468/SciPostPhys.4.6.043} {\bibfield  {journal} {\bibinfo  {journal}
  {SciPost Phys.}\ }\textbf {\bibinfo {volume} {4}},\ \bibinfo {pages} {43}
  (\bibinfo {year} {2018})}\BibitemShut {NoStop}%
\bibitem [{\citenamefont {Gopalakrishnan}\ \emph {et~al.}()\citenamefont
  {Gopalakrishnan}, \citenamefont {Morningstar}, \citenamefont {Vasseur},\ and\
  \citenamefont {Khemani}}]{Gopalakrishnan-Khemani2022_xxz}%
  \BibitemOpen
  \bibfield  {author} {\bibinfo {author} {\bibfnamefont {Sarang}\ \bibnamefont
  {Gopalakrishnan}}, \bibinfo {author} {\bibfnamefont {Alan}\ \bibnamefont
  {Morningstar}}, \bibinfo {author} {\bibfnamefont {Romain}\ \bibnamefont
  {Vasseur}}, \ and\ \bibinfo {author} {\bibfnamefont {Vedika}\ \bibnamefont
  {Khemani}},\ }\bibfield  {title} {\enquote {\bibinfo {title} {Distinct
  universality classes of diffusive transport from full counting statistics},}\
  }\href {\doibase arXiv:2203.09526} {\ arXiv:2203.09526}\BibitemShut {NoStop}%
\bibitem [{\citenamefont {Heyl}\ \emph {et~al.}(2019)\citenamefont {Heyl},
  \citenamefont {Hauke},\ and\ \citenamefont
  {Zoller}}]{Heyl-Zoller2019_trotter}%
  \BibitemOpen
  \bibfield  {author} {\bibinfo {author} {\bibfnamefont {Markus}\ \bibnamefont
  {Heyl}}, \bibinfo {author} {\bibfnamefont {Philipp}\ \bibnamefont {Hauke}}, \
  and\ \bibinfo {author} {\bibfnamefont {Peter}\ \bibnamefont {Zoller}},\
  }\bibfield  {title} {\enquote {\bibinfo {title} {Quantum localization bounds
  trotter errors in digital quantum simulation},}\ }\href {\doibase
  10.1126/sciadv.aau8342} {\bibfield  {journal} {\bibinfo  {journal} {Science
  Advances}\ }\textbf {\bibinfo {volume} {5}},\ \bibinfo {pages} {eaau8342}
  (\bibinfo {year} {2019})}\BibitemShut {NoStop}%
\bibitem [{\citenamefont {Sieberer}\ \emph {et~al.}(2019)\citenamefont
  {Sieberer}, \citenamefont {Olsacher}, \citenamefont {Elben}, \citenamefont
  {Heyl}, \citenamefont {Hauke}, \citenamefont {Haake},\ and\ \citenamefont
  {Zoller}}]{Sieberer-Zoller2019_kicked}%
  \BibitemOpen
  \bibfield  {author} {\bibinfo {author} {\bibfnamefont {Lukas~M.}\
  \bibnamefont {Sieberer}}, \bibinfo {author} {\bibfnamefont {Tobias}\
  \bibnamefont {Olsacher}}, \bibinfo {author} {\bibfnamefont {Andreas}\
  \bibnamefont {Elben}}, \bibinfo {author} {\bibfnamefont {Markus}\
  \bibnamefont {Heyl}}, \bibinfo {author} {\bibfnamefont {Philipp}\
  \bibnamefont {Hauke}}, \bibinfo {author} {\bibfnamefont {Fritz}\ \bibnamefont
  {Haake}}, \ and\ \bibinfo {author} {\bibfnamefont {Peter}\ \bibnamefont
  {Zoller}},\ }\bibfield  {title} {\enquote {\bibinfo {title} {Digital quantum
  simulation, trotter errors, and quantum chaos of the kicked top},}\ }\href
  {\doibase 10.1038/s41534-019-0192-5} {\bibfield  {journal} {\bibinfo
  {journal} {npj Quantum Information}\ }\textbf {\bibinfo {volume} {5}},\
  \bibinfo {pages} {78} (\bibinfo {year} {2019})}\BibitemShut {NoStop}%
\bibitem [{\citenamefont {Kargi}\ \emph {et~al.}()\citenamefont {Kargi},
  \citenamefont {Dehollain}, \citenamefont {Henriques}, \citenamefont
  {Sieberer}, \citenamefont {Olsacher}, \citenamefont {Hauke}, \citenamefont
  {Heyl}, \citenamefont {Zoller},\ and\ \citenamefont
  {Langford}}]{Kargi-Landford2021_performance}%
  \BibitemOpen
  \bibfield  {author} {\bibinfo {author} {\bibfnamefont {Cahit}\ \bibnamefont
  {Kargi}}, \bibinfo {author} {\bibfnamefont {Juan~Pablo}\ \bibnamefont
  {Dehollain}}, \bibinfo {author} {\bibfnamefont {Fabio}\ \bibnamefont
  {Henriques}}, \bibinfo {author} {\bibfnamefont {Lukas~M.}\ \bibnamefont
  {Sieberer}}, \bibinfo {author} {\bibfnamefont {Tobias}\ \bibnamefont
  {Olsacher}}, \bibinfo {author} {\bibfnamefont {Philipp}\ \bibnamefont
  {Hauke}}, \bibinfo {author} {\bibfnamefont {Markus}\ \bibnamefont {Heyl}},
  \bibinfo {author} {\bibfnamefont {Peter}\ \bibnamefont {Zoller}}, \ and\
  \bibinfo {author} {\bibfnamefont {Nathan~K.}\ \bibnamefont {Langford}},\
  }\bibfield  {title} {\enquote {\bibinfo {title} {Quantum chaos and universal
  trotterisation behaviours in digital quantum simulations},}\ }\href {\doibase
  arXiv:2110.11113} {\ arXiv:2110.11113}\BibitemShut {NoStop}%
\bibitem [{\citenamefont {Crowley}\ and\ \citenamefont
  {Chandran}(2022)}]{Crowley-Chandran2021}%
  \BibitemOpen
  \bibfield  {author} {\bibinfo {author} {\bibfnamefont {Philip J~D}\
  \bibnamefont {Crowley}}\ and\ \bibinfo {author} {\bibfnamefont {Anushya}\
  \bibnamefont {Chandran}},\ }\bibfield  {title} {\enquote {\bibinfo {title} {A
  constructive theory of the numerically accessible many-body localized to
  thermal crossover},}\ }\href {\doibase 10.21468/SciPostPhys.12.6.201}
  {\bibfield  {journal} {\bibinfo  {journal} {SciPost Phys.}\ }\textbf
  {\bibinfo {volume} {12}},\ \bibinfo {pages} {201} (\bibinfo {year}
  {2022})}\BibitemShut {NoStop}%
\bibitem [{\citenamefont {Long}\ \emph {et~al.}()\citenamefont {Long},
  \citenamefont {Crowley}, \citenamefont {Khemani},\ and\ \citenamefont
  {Chandran}}]{Long-Chandran2022_jacobi}%
  \BibitemOpen
  \bibfield  {author} {\bibinfo {author} {\bibfnamefont {David~M.}\
  \bibnamefont {Long}}, \bibinfo {author} {\bibfnamefont {Philip J.~D.}\
  \bibnamefont {Crowley}}, \bibinfo {author} {\bibfnamefont {Vedika}\
  \bibnamefont {Khemani}}, \ and\ \bibinfo {author} {\bibfnamefont {Anushya}\
  \bibnamefont {Chandran}},\ }\bibfield  {title} {\enquote {\bibinfo {title}
  {Phenomenology of the prethermal many-body localized regime},}\ }\href
  {\doibase arXiv:2207.05761} {\ arXiv:2207.05761}\BibitemShut {NoStop}%
\bibitem [{\citenamefont {yoon Choi}\ \emph {et~al.}(2016)\citenamefont {yoon
  Choi}, \citenamefont {Hild}, \citenamefont {Zeiher}, \citenamefont {Schauss},
  \citenamefont {Rubio-Abadal}, \citenamefont {Yefsah}, \citenamefont
  {Khemani}, \citenamefont {Huse}, \citenamefont {Bloch},\ and\ \citenamefont
  {Gross}}]{BlochMBL2D_2016}%
  \BibitemOpen
  \bibfield  {author} {\bibinfo {author} {\bibfnamefont {Jae}\ \bibnamefont
  {yoon Choi}}, \bibinfo {author} {\bibfnamefont {Sebastian}\ \bibnamefont
  {Hild}}, \bibinfo {author} {\bibfnamefont {Johannes}\ \bibnamefont {Zeiher}},
  \bibinfo {author} {\bibfnamefont {Peter}\ \bibnamefont {Schauss}}, \bibinfo
  {author} {\bibfnamefont {Antonio}\ \bibnamefont {Rubio-Abadal}}, \bibinfo
  {author} {\bibfnamefont {Tarik}\ \bibnamefont {Yefsah}}, \bibinfo {author}
  {\bibfnamefont {Vedika}\ \bibnamefont {Khemani}}, \bibinfo {author}
  {\bibfnamefont {David~A.}\ \bibnamefont {Huse}}, \bibinfo {author}
  {\bibfnamefont {Immanuel}\ \bibnamefont {Bloch}}, \ and\ \bibinfo {author}
  {\bibfnamefont {Christian}\ \bibnamefont {Gross}},\ }\bibfield  {title}
  {\enquote {\bibinfo {title} {Exploring the many-body localization transition
  in two dimensions},}\ }\href {\doibase 10.1126/science.aaf8834} {\bibfield
  {journal} {\bibinfo  {journal} {Science}\ }\textbf {\bibinfo {volume}
  {352}},\ \bibinfo {pages} {1547--1552} (\bibinfo {year} {2016})}\BibitemShut
  {NoStop}%
\bibitem [{\citenamefont {\ifmmode~\check{S}\else \v{S}\fi{}untajs}\ \emph
  {et~al.}(2020{\natexlab{a}})\citenamefont {\ifmmode~\check{S}\else
  \v{S}\fi{}untajs}, \citenamefont {Bon\ifmmode~\check{c}\else \v{c}\fi{}a},
  \citenamefont {Prosen},\ and\ \citenamefont
  {Vidmar}}]{Suntajs-Vidmar2020_quantum_chaos}%
  \BibitemOpen
  \bibfield  {author} {\bibinfo {author} {\bibfnamefont {Jan}\ \bibnamefont
  {\ifmmode~\check{S}\else \v{S}\fi{}untajs}}, \bibinfo {author} {\bibfnamefont
  {Janez}\ \bibnamefont {Bon\ifmmode~\check{c}\else \v{c}\fi{}a}}, \bibinfo
  {author} {\bibfnamefont {Toma\ifmmode \check{z}\else~\v{z}\fi{}}\
  \bibnamefont {Prosen}}, \ and\ \bibinfo {author} {\bibfnamefont {Lev}\
  \bibnamefont {Vidmar}},\ }\bibfield  {title} {\enquote {\bibinfo {title}
  {Quantum chaos challenges many-body localization},}\ }\href {\doibase
  10.1103/PhysRevE.102.062144} {\bibfield  {journal} {\bibinfo  {journal}
  {Phys. Rev. E}\ }\textbf {\bibinfo {volume} {102}},\ \bibinfo {pages}
  {062144} (\bibinfo {year} {2020}{\natexlab{a}})}\BibitemShut {NoStop}%
\bibitem [{\citenamefont {\ifmmode~\check{S}\else \v{S}\fi{}untajs}\ \emph
  {et~al.}(2020{\natexlab{b}})\citenamefont {\ifmmode~\check{S}\else
  \v{S}\fi{}untajs}, \citenamefont {Bon\ifmmode~\check{c}\else \v{c}\fi{}a},
  \citenamefont {Prosen},\ and\ \citenamefont
  {Vidmar}}]{Suntajs-Vidmar2020_ergodicity_breaking}%
  \BibitemOpen
  \bibfield  {author} {\bibinfo {author} {\bibfnamefont {Jan}\ \bibnamefont
  {\ifmmode~\check{S}\else \v{S}\fi{}untajs}}, \bibinfo {author} {\bibfnamefont
  {Janez}\ \bibnamefont {Bon\ifmmode~\check{c}\else \v{c}\fi{}a}}, \bibinfo
  {author} {\bibfnamefont {Toma\ifmmode \check{z}\else~\v{z}\fi{}}\
  \bibnamefont {Prosen}}, \ and\ \bibinfo {author} {\bibfnamefont {Lev}\
  \bibnamefont {Vidmar}},\ }\bibfield  {title} {\enquote {\bibinfo {title}
  {Ergodicity breaking transition in finite disordered spin chains},}\ }\href
  {\doibase 10.1103/PhysRevB.102.064207} {\bibfield  {journal} {\bibinfo
  {journal} {Phys. Rev. B}\ }\textbf {\bibinfo {volume} {102}},\ \bibinfo
  {pages} {064207} (\bibinfo {year} {2020}{\natexlab{b}})}\BibitemShut
  {NoStop}%
\bibitem [{\citenamefont {Sels}\ and\ \citenamefont
  {Polkovnikov}(2021)}]{Sels-Polkovnikov2021_obstruction}%
  \BibitemOpen
  \bibfield  {author} {\bibinfo {author} {\bibfnamefont {Dries}\ \bibnamefont
  {Sels}}\ and\ \bibinfo {author} {\bibfnamefont {Anatoli}\ \bibnamefont
  {Polkovnikov}},\ }\bibfield  {title} {\enquote {\bibinfo {title} {Dynamical
  obstruction to localization in a disordered spin chain},}\ }\href {\doibase
  10.1103/PhysRevE.104.054105} {\bibfield  {journal} {\bibinfo  {journal}
  {Phys. Rev. E}\ }\textbf {\bibinfo {volume} {104}},\ \bibinfo {pages}
  {054105} (\bibinfo {year} {2021})}\BibitemShut {NoStop}%
\bibitem [{\citenamefont {Sierant}\ \emph {et~al.}(2020)\citenamefont
  {Sierant}, \citenamefont {Delande},\ and\ \citenamefont
  {Zakrzewski}}]{Sierant-Zakrzewski2020_thouless_time}%
  \BibitemOpen
  \bibfield  {author} {\bibinfo {author} {\bibfnamefont {Piotr}\ \bibnamefont
  {Sierant}}, \bibinfo {author} {\bibfnamefont {Dominique}\ \bibnamefont
  {Delande}}, \ and\ \bibinfo {author} {\bibfnamefont {Jakub}\ \bibnamefont
  {Zakrzewski}},\ }\bibfield  {title} {\enquote {\bibinfo {title} {Thouless
  time analysis of anderson and many-body localization transitions},}\ }\href
  {\doibase 10.1103/PhysRevLett.124.186601} {\bibfield  {journal} {\bibinfo
  {journal} {Phys. Rev. Lett.}\ }\textbf {\bibinfo {volume} {124}},\ \bibinfo
  {pages} {186601} (\bibinfo {year} {2020})}\BibitemShut {NoStop}%
\bibitem [{\citenamefont {Mark}\ \emph {et~al.}(2020)\citenamefont {Mark},
  \citenamefont {Lin},\ and\ \citenamefont {Motrunich}}]{Mark20a}%
  \BibitemOpen
  \bibfield  {author} {\bibinfo {author} {\bibfnamefont {Daniel~K.}\
  \bibnamefont {Mark}}, \bibinfo {author} {\bibfnamefont {Cheng-Ju}\
  \bibnamefont {Lin}}, \ and\ \bibinfo {author} {\bibfnamefont {Olexei~I.}\
  \bibnamefont {Motrunich}},\ }\bibfield  {title} {\enquote {\bibinfo {title}
  {Unified structure for exact towers of scar states in the
  affleck-kennedy-lieb-tasaki and other models},}\ }\href {\doibase
  10.1103/PhysRevB.101.195131} {\bibfield  {journal} {\bibinfo  {journal}
  {Phys. Rev. B}\ }\textbf {\bibinfo {volume} {101}},\ \bibinfo {pages}
  {195131} (\bibinfo {year} {2020})}\BibitemShut {NoStop}%
\bibitem [{\citenamefont {Moudgalya}\ \emph {et~al.}(2020)\citenamefont
  {Moudgalya}, \citenamefont {Regnault},\ and\ \citenamefont
  {Bernevig}}]{Moudgalya20}%
  \BibitemOpen
  \bibfield  {author} {\bibinfo {author} {\bibfnamefont {Sanjay}\ \bibnamefont
  {Moudgalya}}, \bibinfo {author} {\bibfnamefont {Nicolas}\ \bibnamefont
  {Regnault}}, \ and\ \bibinfo {author} {\bibfnamefont {B.~Andrei}\
  \bibnamefont {Bernevig}},\ }\bibfield  {title} {\enquote {\bibinfo {title}
  {$\ensuremath{\eta}$-pairing in hubbard models: From spectrum generating
  algebras to quantum many-body scars},}\ }\href {\doibase
  10.1103/PhysRevB.102.085140} {\bibfield  {journal} {\bibinfo  {journal}
  {Phys. Rev. B}\ }\textbf {\bibinfo {volume} {102}},\ \bibinfo {pages}
  {085140} (\bibinfo {year} {2020})}\BibitemShut {NoStop}%
\bibitem [{\citenamefont {O'Dea}\ \emph {et~al.}(2020)\citenamefont {O'Dea},
  \citenamefont {Burnell}, \citenamefont {Chandran},\ and\ \citenamefont
  {Khemani}}]{O'Dea20}%
  \BibitemOpen
  \bibfield  {author} {\bibinfo {author} {\bibfnamefont {Nicholas}\
  \bibnamefont {O'Dea}}, \bibinfo {author} {\bibfnamefont {Fiona}\ \bibnamefont
  {Burnell}}, \bibinfo {author} {\bibfnamefont {Anushya}\ \bibnamefont
  {Chandran}}, \ and\ \bibinfo {author} {\bibfnamefont {Vedika}\ \bibnamefont
  {Khemani}},\ }\bibfield  {title} {\enquote {\bibinfo {title} {From tunnels to
  towers: Quantum scars from lie algebras and $q$-deformed lie algebras},}\
  }\href {\doibase 10.1103/PhysRevResearch.2.043305} {\bibfield  {journal}
  {\bibinfo  {journal} {Phys. Rev. Research}\ }\textbf {\bibinfo {volume}
  {2}},\ \bibinfo {pages} {043305} (\bibinfo {year} {2020})}\BibitemShut
  {NoStop}%
\bibitem [{\citenamefont {Khemani}\ \emph {et~al.}(2020)\citenamefont
  {Khemani}, \citenamefont {Hermele},\ and\ \citenamefont
  {Nandkishore}}]{Khemani20_shatter}%
  \BibitemOpen
  \bibfield  {author} {\bibinfo {author} {\bibfnamefont {Vedika}\ \bibnamefont
  {Khemani}}, \bibinfo {author} {\bibfnamefont {Michael}\ \bibnamefont
  {Hermele}}, \ and\ \bibinfo {author} {\bibfnamefont {Rahul}\ \bibnamefont
  {Nandkishore}},\ }\bibfield  {title} {\enquote {\bibinfo {title}
  {{Localization from Hilbert space shattering: From theory to physical
  realizations}},}\ }\href {\doibase 10.1103/PhysRevB.101.174204} {\bibfield
  {journal} {\bibinfo  {journal} {Phys. Rev. B}\ }\textbf {\bibinfo {volume}
  {101}},\ \bibinfo {pages} {174204} (\bibinfo {year} {2020})}\BibitemShut
  {NoStop}%
\bibitem [{\citenamefont {Sala}\ \emph {et~al.}(2020)\citenamefont {Sala},
  \citenamefont {Rakovszky}, \citenamefont {Verresen}, \citenamefont {Knap},\
  and\ \citenamefont {Pollmann}}]{Sala20}%
  \BibitemOpen
  \bibfield  {author} {\bibinfo {author} {\bibfnamefont {Pablo}\ \bibnamefont
  {Sala}}, \bibinfo {author} {\bibfnamefont {Tibor}\ \bibnamefont {Rakovszky}},
  \bibinfo {author} {\bibfnamefont {Ruben}\ \bibnamefont {Verresen}}, \bibinfo
  {author} {\bibfnamefont {Michael}\ \bibnamefont {Knap}}, \ and\ \bibinfo
  {author} {\bibfnamefont {Frank}\ \bibnamefont {Pollmann}},\ }\bibfield
  {title} {\enquote {\bibinfo {title} {{Ergodicity Breaking Arising from
  Hilbert Space Fragmentation in Dipole-Conserving Hamiltonians}},}\ }\href
  {\doibase 10.1103/PhysRevX.10.011047} {\bibfield  {journal} {\bibinfo
  {journal} {Phys. Rev. X}\ }\textbf {\bibinfo {volume} {10}},\ \bibinfo
  {pages} {011047} (\bibinfo {year} {2020})}\BibitemShut {NoStop}%
\end{thebibliography}%

\appendix

\section{The effective Hamiltonian $H_\mathrm{eff}$ \label{app:Heff}}
In Sec.~\ref{sec:theory} we use the concept of an effective Hamiltonian $H_\mathrm{eff}$. This is a concept that is helpful in discussions of Floquet prethermalization and thermalization, but in later sections we use $H_0$ as a sufficient definition of energy. Here we want to clarify the difference between $H_\mathrm{eff}$ and an approximation to it, like $H_0$.

A good $H_\mathrm{eff}$ is one that captures the processes that do not result in resonant transitions across Floquet zones, e.g., interzone transitions that are mediated by virtual transitions to other Floquet zones. A nearly optimal $H_\mathrm{eff}$ therefore approximates the dynamics of the system on timescales shorter than $\Gamma^{-1}$, the long timescale on which quanta of energy $\sim \omega$ are resonantly exchanged with the drive. For nearly optimal $H_\mathrm{eff}$ and large enough $\omega$, $H_\mathrm{eff} \mod \omega$ will be approximately conserved for all times, up to a precision $\Gamma$ (Fig.~\ref{fig:ladder}).
For our considerations, we can think of $H_\mathrm{eff}$ as being ``obtained" perturbatively until the series stops improving the approximation $e^{-iH_\mathrm{eff}T} \approx U_F$, although we do not actually need to have a strict definition of $H_\mathrm{eff}$ or a method for constructing it. The perturbative series is in the small parameter $(J/\omega)$, where $J$ is a microscopic energy scale and it goes up to order $n\sim O(\omega/J)$. When $\omega/J \sim N$, there are terms with support on $O(N)$ qubits, but those terms are suppressed by coefficients that are $\sim [1/O(N)]^{O(N)}$. Due to this extreme suppression, as we scale $\omega$ up faster with $N$, the effective Hamiltonian gets \textit{more} local, not less local, in the sense that all terms but $H_0$ go to zero in the limit of large $N$ due to their coefficients vanishing.

Thus, for our purposes, using $H_0$ as a notion of energy in the system is sufficient due to the large overlap between $H_0$ and $H_\mathrm{eff}$. $H_0$ is approximately conserved on timescales shorter than $\Gamma^{-1}$, too, but the precision of that approximate conservation law is not good enough for $H_0$ to describe the full dynamics of the system on long timescales, because it does not capture the perturbative processes mediated by virtual transitions to other Floquet zones. Importantly, this means that the widths of the peaks of the distribution of $H_0$ in eigenstates of $U_F$ will not be $\sim\Gamma$, but will be substantially broader due to intrazone dynamics that are captured by $H_\mathrm{eff}$ but not by $H_0$. More specifically, due to the correction $H_\mathrm{eff} - H_0 = \frac {1}{\omega} H_1 + ...$, we expect an additional broadening of the energy peaks (in energy $H_0$) of $\sim \frac{\sqrt{N}}{\omega}$.

\section{Estimating $s(e)$ and $f(\Delta e, \bar{e})$ from finite-size data \label{app:estimate_s_f}}

Near $E=0$, where most of the spectral weight is, the density of states is Gaussian with variance $\sigma^2 = 2N$ (from Sec.~\ref{sec:num}). Thus we expect $D(0) = (4\pi N)^{-\frac{1}{2}} \euler^{N \log 2}$ and so we extract $s$ from small-$N$ numerical data via
\begin{align}
    s\left(\frac{\Delta e}{2}\right) = \frac{1}{N} \log \left( \sqrt{4 \pi N} D\left(\frac{\Delta e}{2}\right)\right).
\end{align}

For the matrix elements, we assume the form $V\left(-\frac{\Delta e}{2}, \frac{\Delta e}{2}\right) = A \euler^{-N f(\Delta e, 0)}$ and take the $N$ and $\Delta e$ dependence of $A$~\cite{DAlessio-Rigol2016_review} to be subleading within the small window of $N$ we have data for. Thus we extract $f$ from numerical data via
\begin{align}
    f(\Delta e, 0) = -\frac{1}{N} \log \left[ \frac{1}{A} V\left(-\frac{\Delta e}{2}, \frac{\Delta e}{2}\right) \right],
\end{align}
and fit $A$ in order to collapse the data beyond $\Delta e \sim \frac{1}{N}$ [$\omega \sim O(1)$]. 

\section{The two-zone approximation \label{app:two_zone}}
The Floquet thermalization/resonances phase diagram of Fig.~\ref{fig:phase_diagram_one_way_two_way} was generated using the two-zone approximation. In this approximation we are ignoring the Floquet zones outside of the two being considered (at energy densities $e_1$ and $e_2$). Minor quantitative modifications to the phase diagram can occur due to considering the other zones. The concern is that, when considering the point $(e_1,e_2)$, if there is another $e_3$ in the energy ladder that has a larger density of states than $e_1$ and $e_2$, it may be relevant. More specifically, if $(e_1, e_2)$ is in the one-way Floquet thermalization phase, and $e_3$ has a larger density of states and $(e_2, e_3)$ are in the two-way phase, then we should update the DOS of $e_2$ to include the states at $e_3$, and that may cause $(e_1, e_2)$ to become part of the two-way phase.

We will first state which areas of the phase diagram are completely safe from the influence of extra zones, i.e., the rest of the energy ladder. Since the phase diagram is symmetric, we consider the tile that has positive $\Delta e$ and negative $\bar{e}$ (see Fig.~\ref{fig:phase_diagram_one_way_two_way}), so $D(e_1)<D(e_2)$. Now as long as $e_2\in [-\frac{\Delta e}{2}, \frac{\Delta e}{2}]$, then it is the rung in the ladder with the highest density of states. This means that the phase diagram (in the tile considered) could only be modified by third-zone effects below the line $e_2 = \frac{e_1}{3}$.

But this region can be constrained even further, because the danger comes from other pairs of rungs in the ladder that are in the two-way phase. Thus a line of constant $\Delta e$ that is tangent to the top of the two-way phase also bounds the region of concern.

From these considerations, we can deduce that the two-way phase may bulge out into the one-way phase a little more than is shown in Fig.~\ref{fig:phase_diagram_one_way_two_way} in the lower left and upper right corners of that phase diagram, but that is all that can be modified by going beyond the two-zone approximation.

\section{Finite-time Floquet thermalization crossover in infinite systems \label{app:finite_time}}
For completeness, in this appendix we note a few aspects of the Floquet thermalization crossover that occurs in systems of effectively infinite size when tuning the drive frequency $\omega$ and observing at fixed time $t$. Large enough systems do eventually thermalize to infinite temperature (see Sec.~\ref{sec:theory}), and the energy density changes at a rate $\Gamma \sim \euler^{-\omega/\omega_0}$. As we increase $\omega$, most of the heating that occurs in the system will go from happening before time $t$ to after time $t$. Thus a crossover in the energy density $e$ at time $t$ will occur, and here we show that this crossover sharpens in a sense as the observation time $t$ is increased.

Consider a system initialized with $e=O(-1)$ energy density and driven to heat up to $e=0$ eventually. We have a fixed observation time $t$ and our tuning parameter is $\omega$.  The change in the energy density by the observation time is $\delta e(\omega) \sim t \euler^{-\omega/\omega_0}$ as long as the heating is far from complete. This hits a set threshold value when $\omega \sim \omega_0 \log t$. The width of the crossover through that threshold value is 
\begin{align}
    \left| \left. \frac{d\delta e}{d\omega} \right|_{\omega = \omega_0 \log t} \right|^{-1} \sim \omega_0. 
\end{align}
Therefore, the width of the Floquet thermalization crossover is down by a factor of $\log t$ from the location of the crossover, and this means that the crossover sharpens up logarithmically as a function of increasing observation time $t$. 

If we exit the regime where $N$ is effectively infinite and allow the observation time to approach the inverse many-body level spacing of the system, then $\log t \propto N$ so the crossover sharpens linearly in $N$. This is consistent with statements in the main text that the width of the crossovers (in $\omega/N$) shown on the phase diagram scale as $N^{-1}$.

\end{document}